\RequirePackage{lineno} 

\documentclass[twocolumn,showpacs,preprintnumbers,amsmath,amssymb,nofootinbib]{revtex4}

\usepackage{graphicx}
\usepackage{dcolumn}
\usepackage{bm}
 
\usepackage{epstopdf}
\def\bea{\begin{eqnarray}}
\def\eea{\end{eqnarray}}

\def\pp{\mbox{$p$-$p$}}

\def\pa{\mbox{$p$-A}}
\def\da{\mbox{$d$-A}}
\def\auau{\mbox{Au-Au}}

\def\pbpb{\mbox{Pb-Pb}}
\def\ppb{\mbox{$p$-Pb}}
\def\pn{\mbox{$p$-N}}
\def\aa{\mbox{A-A}}
\def\nn{\mbox{N-N}}

\def\pt{$p_t$}
\def\mt{$m_t$}
\def\yt{$y_t$}

\def\nch{$n_{ch}$}
\def\mmpt{$\bar p_t$}

\begin{document} 

\setlength{\pdfpagewidth}{8.5in}
\setlength{\pdfpageheight}{11in}

\setpagewiselinenumbers
\modulolinenumbers[5]

\preprint{version 2.0}

\title{Glauber-model analysis of 5 TeV $\bf p$-Pb centrality compared to a two-component (soft + hard) model of hadron production in high-energy nuclear collisions
}

\author{Thomas A.\ Trainor}\affiliation{CENPA 354290, University of Washington, Seattle, WA 98195}


\date{\today}

\begin{abstract}
A recent study of 5 TeV $p$-Pb centrality combined a Glauber model of  $p$-Pb collision geometry with an assumption of linear scaling between $n_{ch}$ (charge) integrated within some $\eta$ acceptance and the number of nucleon participants $N_{part}$. The study concluded that $N_{part}$ increases to nearly 16 in central collisions, and the high-$p_t$ region of $p$-Pb $p_t$ spectra rescaled by the Glauber-estimated number of $p$-N binary collisions remains consistent with a $p$-$p$ spectrum for the same energy, independent of \ppb\ centrality. However, the relation between $N_{part}$ and $n_{ch}$ derived from a two-component (soft + hard) model (TCM) study of ensemble-mean $\bar p_t$ data for the same system is quite different. This article reports a detailed analysis of the Glauber study and the question of centrality in $p$-A collisions. The Glauber centrality model is compared with the $\bar p_t$ TCM to understand  the sources of major differences. The assumption of linear proportionality between $n_{ch}$ and $N_{part}$ is found to be inconsistent with $\bar p_t$ data. Properties of the convolution integral relating a differential cross section and hadron production model to an event distribution on $n_{ch}$ are examined. An alternative differential-cross-section distribution is inferred from charge-multiplicity data, and the upper limit on $N_{part}$ is estimated to be near 8. The TCM centrality model is then applied to $p_t$ spectrum ratios to predict results for $p$-Pb spectra. The spectrum TCM is tested with identified-pion spectra from 5 TeV $p$-Pb collisions and the result is consistent with previous $p$-$p$ TCM results. A TCM prediction that the spectrum ratio at high $p_t$ should {\em increase to 14} for central $p$-Pb collisions due to quadratic dependence of dijet production on $n_{ch}$ is consistent with $\bar p_t$ data from the same system.
\end{abstract}


\maketitle

 \section{Introduction}

Estimation of collision centrality (e.g.\ impact parameter $b$) for nucleus-nucleus (\aa) collisions according to a Glauber Monte Carlo (MC) model based on the eikonal approximation seems well established and reasonably accurate~\cite{glauber,powerlaw}. The concept  of centrality (or impact parameter) for \pp\ collisions has been invoked within some Monte Carlo models (e.g.\ PYTHIA~\cite{pythia}), but its relevance may be questioned~\cite{pptheory,ppquad}.  The centrality of asymmetric \pa\ or \da\ systems is important because of the role such data are expected to play in verifying the formation of quark-gluon plasma (QGP) in more-central \aa\ collisions. However, centrality estimation in asymmetric systems is more difficult as revealed in the present study.

Centrality determination requires definition of a quantitative relation between geometry parameters such as participant nucleon number $N_{part}$ or number of nucleon-nucleon (\nn) binary collisions $N_{bin}$ and a measured quantity such as integrated charge multiplicity \nch\ within some angular acceptance $\Delta \eta$. A hadron production model is required for such definition, and the accuracy of centrality determination depends on the validity of the model, especially its basis in various forms of data.

The two-component (soft + hard) model (TCM) of hadron production near midrapidity ($\eta \approx 0$)~\cite{kn} has been applied to hadron yield, spectrum and correlation data from \pp, \pa\ and \aa\ collisions at the relativistic heavy ion collider (RHIC)~\cite{ppprd,porter2,porter3,hardspec,jetspec,anomalous,ppquad} and the large hadron collider (LHC)~\cite{alicetomspec,alicetommpt,tommpt}. Based on substantial evidence from data the soft component is interpreted to represent participant-nucleon dissociation to charge-neutral hadron pairs, and the hard component is interpreted to represent fragmentation of large-angle-scattered low-$x$ partons (gluons) to minimum-bias (MB) dijets. 

The TCM has been applied recently to ensemble-mean \mmpt\ data from 5 TeV \ppb\ collisions (as well as \pp\ and \pbpb\ data)~\cite{alicempt} and describes those data within their uncertainties~\cite{alicetommpt,tommpt}. The TCM thereby determines the relation between $N_{part}$ and \nch\ for that system which, in effect, relates \ppb\ centrality to a measured quantity.

A recent study of \ppb\ centrality adopted an alternative strategy based on a Glauber MC model of the \ppb\ system including certain assumptions about collision geometry and hadron production, especially assumption of {\em linear scaling} between $N_{part}$ and \nch~\cite{aliceppbprod}. The Glauber study also reported the systematics of \ppb\ \pt\ spectra, concluding that at higher \pt\ (e.g.\ above 10 GeV/c) the spectra exhibit binary-collision scaling: spectra divided by $N_{bin} = N_{part} - 1$ are consistent with \pp\ spectra in that \pt\ interval independent of \nch\ or centrality.  The inferred relation between $N_{part}$ (and other Glauber parameters) and \nch\ differs greatly from TCM-based \mmpt\ studies reported in Refs.~\cite{alicetommpt,tommpt}. The conclusion about binary-collision scaling of the high-\pt\ region of \ppb\ spectra is also at odds with TCM results as demonstrated below. It is essential to determine the reasons for disagreement and which centrality method, if either, is correct. 

This article reports a detailed study of the Glauber centrality method and its relation to the TCM. The internal consistency of the model is examined. Glauber results are compared step-by-step with TCM results. As noted, major differences emerge concerning the assumed relation between $N_{part}$ and \nch, between what is adopted for the Glauber study and what is inferred from \mmpt\ data for the TCM. The geometric Glauber MC used to describe \ppb\ collisions and predict the cross-section distribution on $N_{part}$ is a major issue. The centrality trend for \ppb\ spectrum ratios inferred via the Glauber model conflicts with measured identified-pion spectra that precisely follow TCM predictions.

This article is arranged as follows:
Section~\ref{glauber3}  briefly introduces the Glauber-model study of \ppb\ centrality from Ref.~\cite{aliceppbprod}.
Section~\ref{ppb} describes the TCM for \mmpt\ data from 5 TeV \ppb\ collisions as reported in Ref.~\cite{tommpt}.
Section~\ref{glaubermod} provides a detailed description of the Glauber-model analysis with some implications and possible inconsistencies.
Section~\ref{glaubervstcm} compares the Glauber analysis to TCM results and itemizes differences.
Section~\ref{oops} presents an alternative TCM centrality analysis and suggests basic reasons for disagreement between Glauber analysis and TCM.
Section~\ref{ratios} describes TCM predictions for spectrum-ratio trends compared to those reported in Ref.~\cite{aliceppbprod}. Identified-pion data from the \ppb\ collision system are introduced to test the predictions.
Section~\ref{sys} discusses systematic uncertainties.
Sections~\ref{disc} and~\ref{summ} include discussion and summary.
Appendix~\ref{specapp} presents the TCM for \pp\ \pt\ spectrum data used as a basis for the \pp\ and \ppb\ \mmpt\ TCMs.
Appendix~\ref{ppmptapp} describes the TCM for \mmpt\ data from \pp\ collisions.

\section{Glauber analysis of $\bf p$-$\bf Pb$ collisions} \label{glauber3}


Reference~\cite{aliceppbprod} motivates the study of \ppb\ centrality with two issues: (a) \pa\ is a null hypothesis for QGP formation in \aa\ collisions, and (b) recent claims of ``collectivity'' in \pa\ and \pp\ systems~\cite{dusling} should be evaluated. 

 \pa\ collisions are intended to serve as a control or reference system ``...to disentangle hot nuclear matter effects which are characteristic of the formation of the quark-gluon plasma (QGP) from cold nuclear matter effects'' -- for instance, by comparisons of spectrum ratio $R_{AA}$ from \aa\ collisions with ratio $R_{pA}$ from \pa\ collisions. Whereas $R_{AA} \ll 1$ at higher \pt\ for more-central collisions is interpreted to indicate strong jet quenching consistent with QGP formation $R_{pA} \approx 1$ (binary-collision scaling) in the same \pt\ interval and more-central \pa\ collisions would indicate no jet quenching -- a preferred null result.


However, the \aa--\pa\ $\sim$ QGP--no-QGP dichotomy is inconsistent with recent interpretations of data to indicate collective manifestations (flows) in smaller systems, e.g.\ \pa\ or even \pp\ collisions~\cite{dusling}. Evidence is cited (e.g.\ the \mmpt\ data reported in Ref.~\cite{alicempt}) to claim  that data for lower \pt\ in  ``...p–Pb collisions cannot be explained by an incoherent [i.e.\ linear] superposition of pp [\pn] collisions,'' any deviations interpreted to signal ``collectivity'' i.e.\ hydrodynamic flows. The strength of collective effects is said to increase with \nch\ and therefore with \pa\ centrality, implying a strong collision-geometry dependence. Determining \pa\ collision geometry is therefore essential.

Accurate estimation of collision geometry is also required to evaluate spectrum ratio $R_{pA}$ that includes number of binary collisions $N_{bin}$ in the ratio. Reference~\cite{aliceppbprod} concludes that ``...particle production at high pT in p–Pb collisions indeed can be approximated by an incoherent [linear] superposition of pp collisions'' because ratio estimates appear to be independent of \ppb\ centrality in that \pt\ interval. But the conclusion depends on accurate estimation of $N_{bin}$ in relation to a measured quantity.

Centrality estimation is based on a Glauber MC used to relate $N_{part}$ to fractional cross section $\sigma/\sigma_0$ and an assumption that serves as a hadron production model: Charge multiplicity $n_{ch}$ integrated within some pseudorapidity $\eta$ interval is proportional to the number of participant nucleons $N_{part}$. The two elements are combined in a convolution integral to predict an event-frequency distribution on \nch\ which is then compared to data. Centrality classes are defined by cuts on \nch\ and corresponding parameter values for each class determined from the MC: ``For a given centrality class, defined by selections in the measured [\nch] distribution, the information from the Glauber MC in the corresponding generated distribution is used to calculate [means of several Glauber parameters, e.g.\ $N_{part}$].'' The method is applied to the same 5 TeV \ppb\ data sample that appears in Refs.~\cite{alicempt,alicetommpt,tommpt}.

The present study reviews the Glauber methods and compares them to results from a TCM analysis~\cite{tommpt} of \ppb\ \mmpt\ data from Ref.~\cite{alicempt} wherein certain contradictions emerge. An alternative centrality analysis derived from the \mmpt\ TCM leads to quite different results.

\section{$\bf \bar p_t$ TCM for $\bf p$-$\bf Pb$  collisions} \label{ppb}

Appendix~\ref{specapp} describes a TCM for \pp\ \pt\ spectra and reviews systematic evolution with event multiplicity and collision energy. \pt\ spectrum structure is directly related to \mmpt\ trends. Appendix~\ref{ppmptapp} describes a TCM for \mmpt\ data from \pp\ collisions as a basis for \ppb\ \mmpt\ analysis. With the dominant role of MB jets established for \pp\ (\pn, \nn) collisions  and elements of the \pp\ \mmpt\ TCM introduced the \ppb\ \mmpt\ TCM is presented here in greater detail. Note that the TCM is a {\em linear-superposition model}.

The TCM for complex A-B collisions relies on participant-pair number $N_{part}/2$, number of \nn\ binary collisions $N_{bin}$ and mean number of binary collisions per participant pair $\nu = 2N_{bin} / N_{part}$. In addition, hard/soft ratio $x(n_s) \equiv \bar \rho_{hNN} / \bar \rho_{sNN}$ averaged over participant \nn\ pairs within individual A-B collisions is a generalization of $x(n_s) = \bar \rho_h / \bar \rho_s$ for \pp\ collisions~\cite{ppprd}. Reference~\cite{alicetommpt} reported a preliminary TCM analysis of \mmpt\ data from Ref.~\cite{alicempt}. Reference~\cite{tommpt} describes an updated TCM based on jet systematics in Refs.~\cite{jetspec2,alicetomspec}.
As noted in Ref.~\cite{tommpt} the $\bar p_t$ trend for \ppb\ collisions is identical to the \pp\ trend at lower \nch\ but deviates substantially at higher \nch, suggesting a formulation of the TCM for the \ppb\ collision system based on generalization of the product $x(n_s) \nu(n_s)$, with soft multiplicity $n_s$ ($\propto$ total number of {\em participant} low-$x$ gluons) as the independent variable. For \pp\ collisions $\nu \equiv 1$ and $x(n_s) \approx \alpha \bar \rho_s$ based on spectrum studies~\cite{ppprd,ppquad}. For \aa\ collisions $\nu$ is defined by a Glauber MC~\cite{powerlaw}, and $x(\nu)$ is inferred from a measured trend of per-participant hadron yields~\cite{alicemultpbpb}. Corresponding elements for \ppb\ data are determined below.

\subsection{Formulating a $\bf p$-Pb TCM} \label{patcm}

A universal TCM for hadron \pt\ spectrum $\bar \rho_0(p_t)$ is expressed for any A-B system as
\bea \label{ppspectcm}
\bar \rho_0(p_t) &\approx& \frac{1}{p_t}\frac{d^2 n_{ch}}{dp_t d\eta}
\\ \nonumber
 &=& \bar \rho_s(p_t) + \bar \rho_h(p_t)
\\ \nonumber
&=&\frac{N_{part}}{2} \bar \rho_{sNN} \hat S_0(p_t) + N_{bin} \bar \rho_{hNN} \hat H_0(p_t) 
\\ \nonumber
\frac{\bar \rho_0(p_t)}{\bar \rho_s} &=& \hat S_0(p_t) + x(n_s)\nu(n_s) \hat H_0(p_t),
\eea
where $\bar \rho_x = n_x / \Delta \eta$  are averaged over acceptance $\Delta \eta$ and hats denote unit-integral (on \pt\ or \yt) quantities.
For \pp\ collisions $N_{part}/2 = N_{bin} \equiv 1$. For composite \mbox{A-B} systems hard and soft components factorize as shown. With integration over \pt\ or \yt\ the mean charge density is
\bea \label{nchppb}
\bar \rho_0 &=& \bar \rho_s + \bar \rho_h
\\ \nonumber
&=& \frac{N_{part}}{2} \bar \rho_{sNN}(n_s) + N_{bin} \bar \rho_{hNN}(n_s)
\\ \nonumber
\frac{2}{N_{part}}  \bar \rho_0 &=& \bar \rho_{sNN}(n_s)  \left[1 + x(n_s)\nu(n_s) \right]
\\ \nonumber
\frac{\bar \rho_0}{\bar \rho_{s}} &=& 1 + x(n_s)\nu(n_s) 
\\ \nonumber
\frac{\bar \rho_0'}{\bar \rho_{s}} &=& \frac{n_{ch}'}{n_{s}} ~=~ \xi + x(n_s)\nu(n_s),
\eea
where $x(n_s) \equiv \bar \rho_{hNN} / \bar \rho_{sNN} \approx \alpha \bar \rho_{sNN}$ for \pp\ or \pa\ collisions, $\bar \rho_{s} = [N_{part}(n_s)/2]\, \bar \rho_{sNN}(n_s)$ is a factorized soft-component density for any system and $\bar \rho_h(n_s) = N_{bin}(n_s) \bar \rho_{hNN}(n_s)$ is a factorized hard component.   

The ensemble-mean {\em total} \pt\ integrated over some angular acceptance $\Delta \eta$ is
\bea
\bar P_t &=& \Delta \eta \int_0^\infty dp_t\, p_t^2\, \bar \rho_0(p_t)
~=~ \bar P_{ts} + \bar P_{th}
\\ \nonumber
&=& \frac{N_{part}}{2}  n_{sNN}(n_s) \bar p_{tsNN} + N_{bin}  n_{hNN}(n_s) \bar p_{thNN}.
\eea 
 Data indicate that $\bar p_{tsNN} \rightarrow \bar p_{ts}$ is a universal quantity. The corresponding TCM for $\bar p_t'$ with nonzero $p_{t,cut}$ is
\bea \label{pampttcm}
\frac{\bar P_t'}{n_{ch}'} &\equiv&  \bar p_t' ~\approx~ \frac{\bar p_{ts} + x(n_s) \nu(n_s) \, \bar p_{thNN}(n_s)}{\xi + x(n_s)\, \nu(n_s)}
\\ \nonumber
\frac{n_{ch}'}{n_{s}}\, \bar p_t' &\approx& \frac{\bar P_t}{n_{s}}
~=~ \bar p_{ts} + x(n_s)\nu(n_s) \, \bar p_{thNN}(n_s),
\eea
where $\xi$ is the fraction of $\hat S_0(p_t)$ admitted by a low-\pt\ acceptance cut $p_{t,cut}$ and primes indicate corresponding biased quantities.
For \pa\ data evolution of factors $x(n_s)\, \nu(n_s)$ from strictly \pp--like to alternative behavior is observed near a transition point  $\bar \rho_{s0}$, but $\bar p_{thNN}(n_s) \rightarrow \bar p_{th0}$ is assumed to maintain a fixed \pp\ (\nn) value in the \pa\ system (i.e.\ no jet modification).


Soft density $\bar \rho_s$, interpreted to represent participant low-$x$ gluons, is adopted as a universal TCM parameter for all collision systems. 
The factorization
\bea \label{rhosnn}
\bar \rho_s &=&  [N_{part}(n_s)/2] \, \bar \rho_{sNN}(n_s)
\eea
then defines $\bar \rho_{sNN}(n_s)$ for any collision system wherein $N_{part}(n_s)/2$ is defined, and  $x(n_s) \approx \alpha \bar \rho_{sNN}(n_s)$ for \pn\ collisions within \pa\ collisions~\cite{ppprd,ppquad}.  It follows that $N_{part}(n_s)/2 = \alpha \bar \rho_s / x(n_s)$ given a model for $x(n_s)$. For \pa\ collisions $N_{bin} = N_{part} - 1$, and $\nu(n_s) = 2N_{bin} / N_{part}$ are then determined by $x(n_s)$. 
Based on previous analysis in Ref.~\cite{tommpt} \ppb\ data indicate that \mmpt\ increases with \nch\ according to a \pp\ trend for lower \nch\ but less rapidly above a transition point, suggesting a similar structure for $x(n_s)$. The simplest form is linear increase with $\bar \rho_s$ also above the  transition point but with reduced slope. 

\begin{figure}[h]
  \includegraphics[width=1.65in]{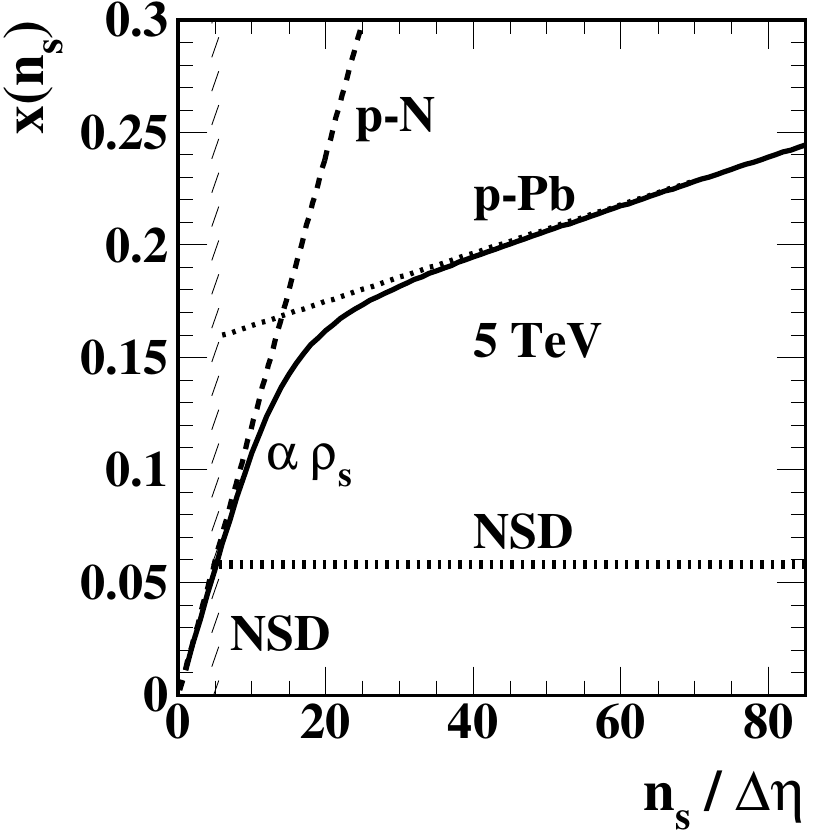}
  \includegraphics[width=1.65in]{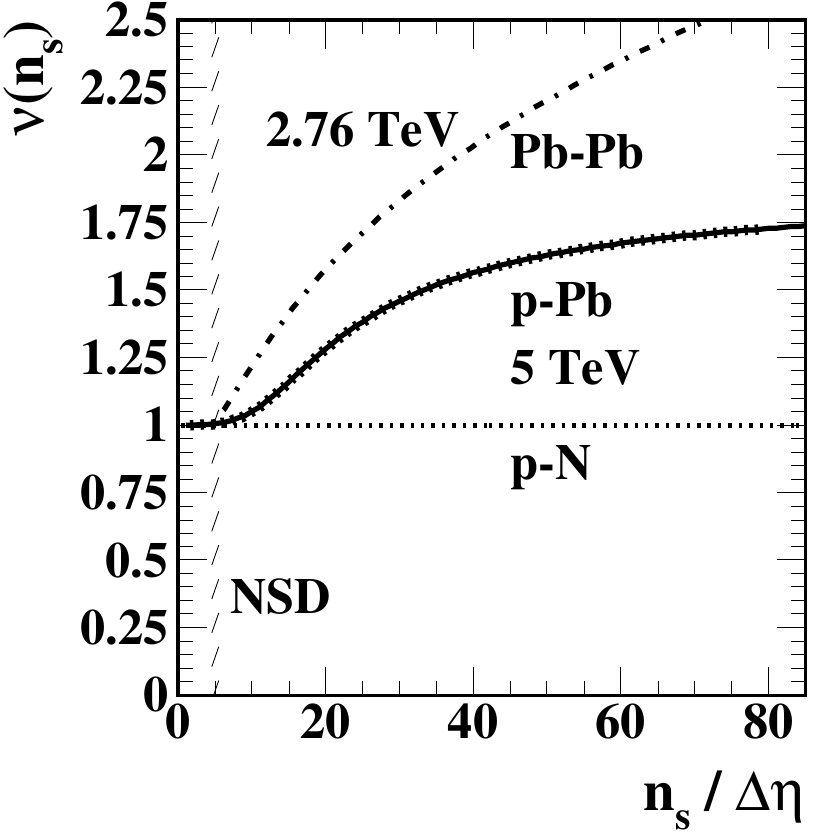}
  \caption{\label{paforms}
  Left: Evolution of TCM hard/soft parameter $x(n_s)$ with mean soft charge density $\bar \rho_s = n_s / \Delta \eta$ following a linear \pn\ (\pp) trend (dashed) for lower multiplicities and a trend with ten-fold reduced slope for higher multiplicities (dotted) to describe \ppb\ \mmpt\ data.
  Right: Mean participant pathlength $\nu \equiv 2N_{bin} / N_{part}$ vs $\bar \rho_s$ (solid) as determined by the $x(n_s)$ trend in the left panel (see text). The $\nu$ trend for \pbpb\ collisions (dash-dotted) is included for comparison.
   } 
 \end{figure}

Figure~\ref{paforms} (left) shows a model for $x(n_s)$ in the form
\bea \label{xmodel}
x(n_s) &=& \frac{\alpha}{\left\{[1/ \bar \rho_s]^{n_1} + [1/f(n_s)]^{n_1}\right\}^{1/n_1}},
\eea
where $f(n_s) = \bar \rho_{s0} + m_0(\bar \rho_s - \bar \rho_{s0})$.
Below the transition at $\bar \rho_{s0}$, $x(n_s) \approx \alpha \bar \rho_s$ as for \pp\ collisions (dashed line). Above the transition $x(n_s)$ increases with slope $m_0 < 1$ (dotted line). Exponent $n_1$ controls the transition width. Specific parameter values for $x(n_s)$ are noted below. The dotted line and hatched band indicate estimates for $x(n_s)$ and $\bar \rho_s$ from non-single-diffractive (NSD) \pp\ collisions.

Figure~\ref{paforms} (right) shows $\nu \equiv 2N_{bin} / N_{part}$ for \ppb\ data (solid curve) based on $N_{part}(n_s)/2 = \alpha \bar \rho_s / x(n_s)$ and $N_{bin} = N_{part} - 1$ as noted above, with $x(n_s)$ as described in the left panel (solid curve). 
The dash-dotted curve indicates the $\nu = 2N_{bin} / N_{part} \approx (N_{part}/2)^{1/3}$ trend for \pbpb\ collisions for comparison, consistent with the eikonal approximation assumed for the \aa\ Glauber model. For \pbpb\ $\nu \in [1,6]$ whereas for \ppb\ $\nu \in [1,2]$.


\subsection{TCM description of p-A data}

\begin{figure}[h]
  \includegraphics[width=1.65in]{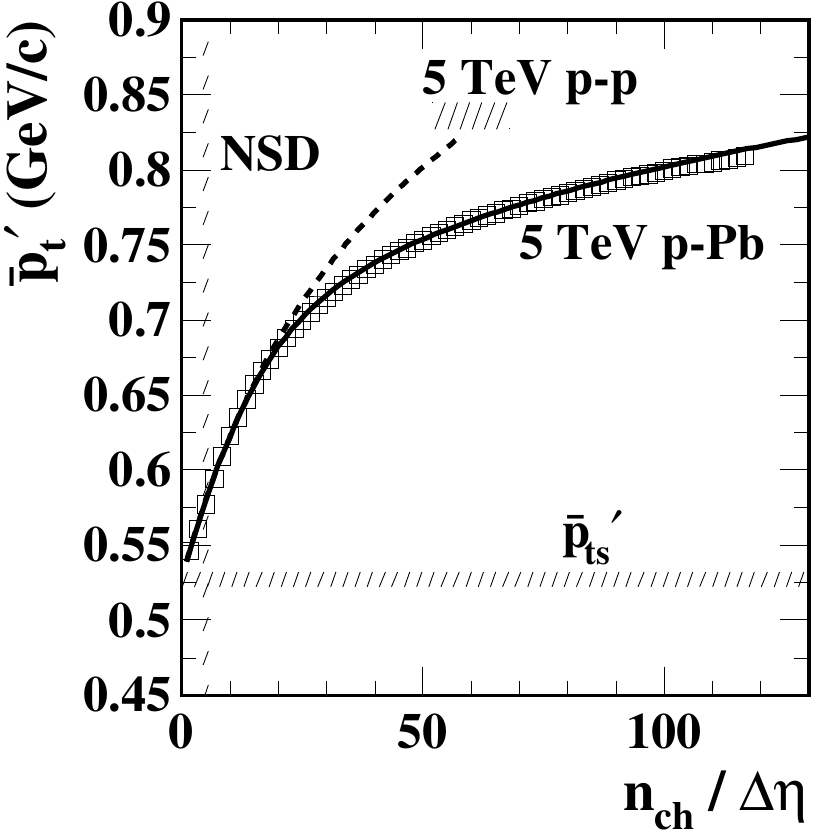}
  \includegraphics[width=1.65in]{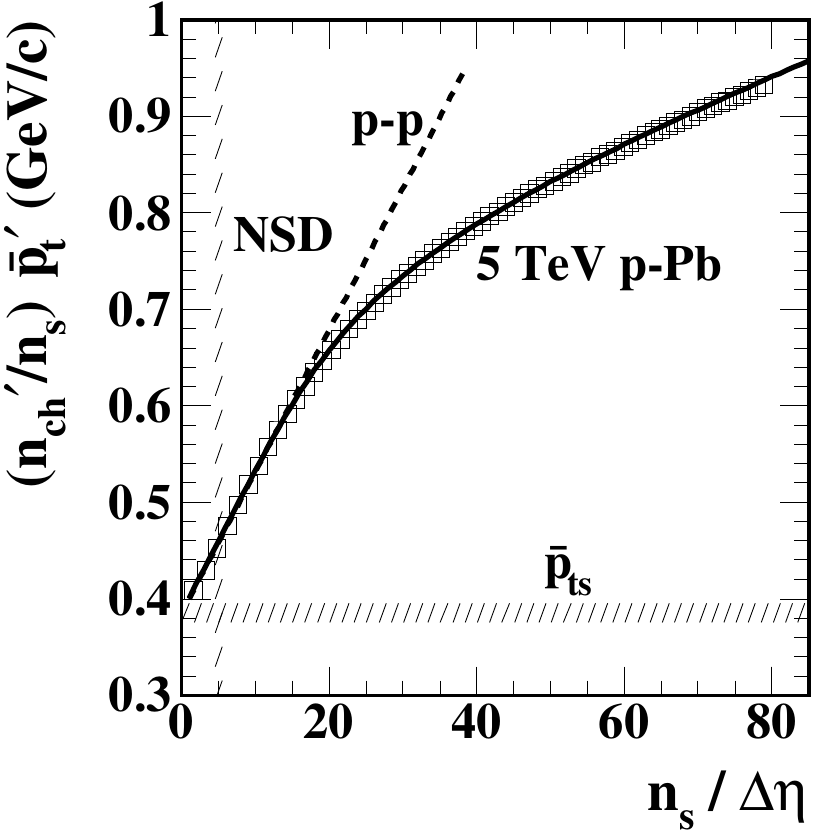}
  \caption{\label{padata}
  Left: Uncorrected ensemble-mean \mmpt\ data from 5 TeV \ppb\ collisions (open squares) vs corrected charge density $\bar \rho_0 = n_{ch} / \Delta \eta$ from Ref.~\cite{alicempt}. The solid and dashed curves are TCM data descriptions from Ref.~\cite{tommpt}.
Right: Curves and data in the left panel transformed by factor $\bar \rho_0'/\bar \rho_s = n_{ch}' / n_s$ defined by Eq.~(\ref{nchppb}) (fifth line).
   }  
\end{figure}

Figure~\ref{padata} (left) shows uncorrected $\bar p_t'$ data for 106 million 5 TeV \ppb\ collisions vs corrected \nch\ (points) from Ref.~\cite{alicempt}. The dashed curve is the TCM for 5 TeV \pp\ collisions given by Eq.~(\ref{ppmpttcm}) with $\alpha = 0.0113$ derived from the parametrization in Fig.~\ref{enrat3} (right), $\bar p_{ts} \approx 0.4$ GeV/c, $\bar p_{th0} = 1.3$ GeV/c and $\xi = 0.73$. The solid curve through points is the TCM described by Eqs.~(\ref{pampttcm}) and (\ref{xmodel})  with parameters  $\alpha = 0.0113$ and $\bar p_{th0} = 1.3$ GeV/c held fixed as for 5 TeV \pp\ collisions (assuming no jet modification).  Parameters $\bar \rho_{s0} \approx 3 \bar \rho_{sNSD} \approx 15$ and $m_0 \approx 0.10$  are adjusted to accommodate the \ppb\ data. Exponent $n_1=5$ affects the TCM shape only near $\bar \rho_{s0}$.  

Figure~\ref{padata} (right) shows data in the left panel converted to $(n_{ch}' / n_s)\, \bar p_t' \approx \bar P_t / n_{s}$ by factor $\xi + x(n_s) \nu (n_s)$ as in Eq.~(\ref{nchppb}) (fifth line). The dashed line is the TCM for 5 TeV \pp\ collisions defined by Eq.~(\ref{niceeq}). The solid curve is the \ppb\ TCM defined by Eq.~(\ref{pampttcm})  (second line) corresponding to the solid curve in the left panel. Transforming data from left to right panels requires an estimate of $n_s$ for the data to evaluate the required conversion factor $\xi + x(n_s) \nu(n_s)$. The map $n_s \rightarrow n_{ch}$ for the TCM from Eq.~(\ref{nchppb}) (second line) is inverted  via linear interpolation to provide the map $n_{ch} \rightarrow n_s$ for data. 



\ppb\ \mmpt\ data provide understanding of the transition from isolated \pp\ or \nn\ collisions to the geometry of compound A-B systems, from noneikonal \pp\ to eikonal \aa\ Glauber model. The central element is factorization of the soft density $\bar \rho_s = \bar \rho_{sNN}(n_s) \, N_{part}(n_s)/2 $ combining \nn\ internal structure ($\bar \rho_{sNN}$) and A-B geometry ($N_{part}$). The \ppb\ TCM is based on the key assumption that the dijet production trend $\bar \rho_{hNN} = \alpha \bar \rho_{sNN}^2$ (averaged over all \nn\ collisions) is universal, in which case
\bea \label{nparmodel}
x(n_s) &\equiv& \frac{\bar \rho_{hNN}}{\bar \rho_{sNN}} = \frac{\alpha \bar \rho_{s}}{N_{part}/2}
\eea
determines $N_{part}(n_s)$ given a model for $x(n_s)$. The \pa\ $x(n_s)$ model is the simplest extrapolation of the \pp\ $x(n_s) \approx \alpha \bar \rho_s$ linear trend possible: a continuing linear trend but with reduced slope beyond a transition point as in Eq.~(\ref{xmodel}). For \ppb\ collisions the other Glauber parameters are immediately determined as in Sec.~\ref{patcm} above.
TCM $x(n_s)$ and $\nu(n_s)$ parameters are portable across all A-B systems, although their details may vary. 



\subsection{p-Pb $\bf \bar p_t$ data and the Glauber model}

The \ppb\ \mmpt\ analysis described above has direct bearing on the Glauber analysis of Ref.~\cite{aliceppbprod} through the TCM relation of \ppb\ centrality to observed \nch\ in the form of inferred relation $N_{part}(\bar \rho_0)$ and hadron production in the form of the mean number of hadrons per participant pair.

\begin{figure}[h]
  \includegraphics[width=1.65in]{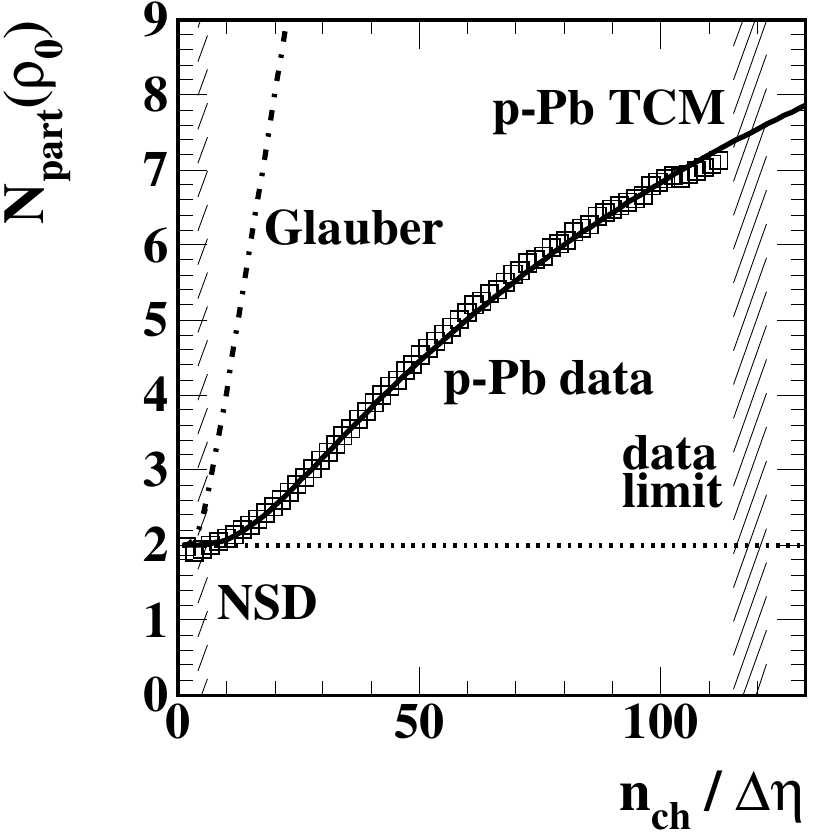}
  \includegraphics[width=1.65in]{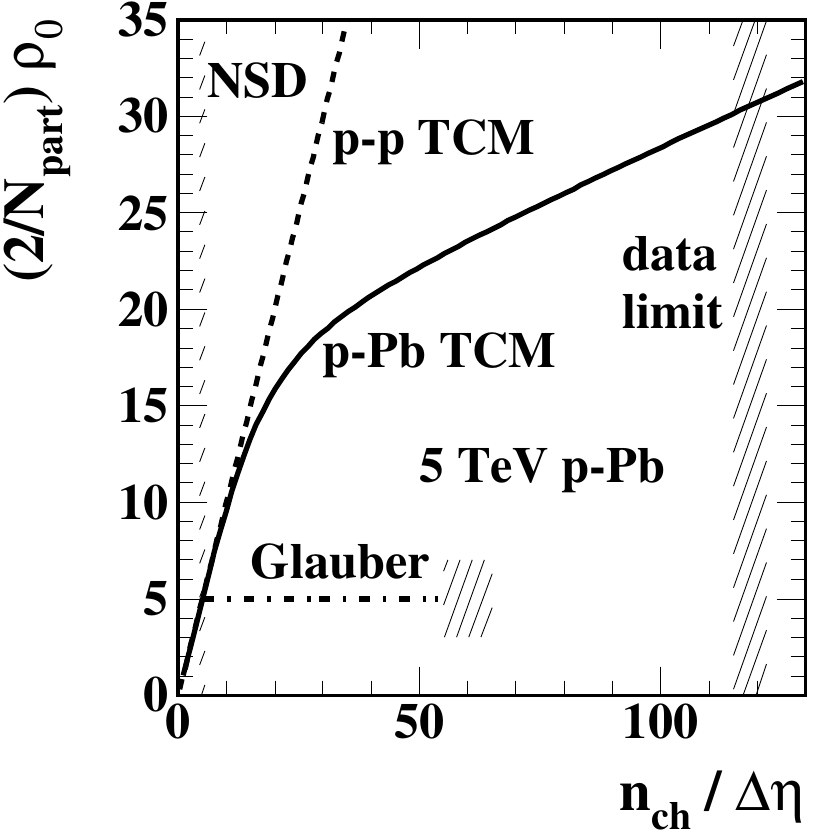}
  \caption{\label{paforms2}
Left: $N_{part}(\bar \rho_0)$ (points) inferred from \mmpt\ data in Fig.~\ref{padata} (right) via Eq.~(\ref{pampttcm}) (second line) and the curves in Fig.~\ref{paforms}. The corresponding TCM is the solid curve.
  Right: The TCM trends for number of hadrons per participant pair inferred for 5 TeV \ppb\ data (solid) and \pp\ data (dashed) compared to the assumed relation for the Glauber study (dash-dotted).
   }  
 \end{figure}

Figure~\ref{paforms2} (left) shows data (points) in the form of Eq.~(\ref{pampttcm}) (second line) and Fig.~\ref{padata} (right panel) inverted to solve for $\nu$ and thus $N_{part} = 2/(2-\nu)$ using $\bar p_{ts} \approx 0.4$, $\bar p_{thNN} \approx 1.3$ GeV/c and $x(n_s)$ as defined in Eq.~(\ref{xmodel}). The corresponding TCM is the solid curve. Glauber-assumed linear scaling of $N_{part}$ with \nch\ is represented by the dash-dotted curve.
The Glauber analysis of Ref.~\cite{aliceppbprod} emphasizes a joint distribution on $(\bar \rho_0,N_{part})$ simulated by a MC based on the same linear scaling.  The  TCM  \mmpt\ analysis defines a ``locus of modes'' defined by the solid curve $N_{part}(\bar \rho_0)$---the most-probable points on the space $(\bar \rho_0,N_{part})$ or approximately the mean values on $\bar \rho_0$ with $N_{part}$ fixed or on $N_{part}$ with $\bar \rho_0$ fixed.

Figure~\ref{paforms2} (right) shows hadron production per participant pair in the form $(2/N_{part})\bar \rho_0$ vs $\bar \rho_0$ (solid curve) as inferred from the \ppb\ \mmpt\ analysis in Ref.~\cite{tommpt}. The \pp\ trend is the dashed line with $N_{part}/2 \equiv 1$. The hadron production model for the Glauber analysis assumes that this quantity remains close to the NSD \pp\ value $\approx 5$ (dash-dotted). TCM and Glauber descriptions of \ppb\ centrality are thus likely to be very different.

To provide context for what follows some limiting parameter values may be useful. The \mmpt\ data for 7 TeV \pp\ collisions in Fig.~\ref{alice5a} (left) extend to $\bar \rho_{0pp} \approx 10\, \bar \rho_{0NSD} \approx 60$. The \mmpt\ data for 5 TeV \ppb\ collisions extend to $(2/N_{part}) \bar \rho_0 \approx 6\, \bar \rho_{0NSD} \approx 30$ as in Fig.~\ref{paforms2} (right). At that upper limit  $N_{part} \approx 7.5$ as in  Fig.~\ref{paforms2} (left) compared to $N_{part}\approx 16$ for $(2/N_{part}) \bar \rho_0 \approx 5$ inferred from the Glauber MC as reported in Ref.~\cite{aliceppbprod} and Table~\ref{ppbparams1} below.  If placed tangent along a Pb nucleus ($r \approx 7.1$ fm) diameter the number of nucleons ($r \approx 0.85$ fm) would be about 8. The mean number of \nn\ binary collisions per participant in \auau\ or \pbpb\ collisions is $\nu < 6$~\cite{anomalous}.

\section{Glauber model of $\bf p$-$\bf Pb$ centrality} \label{glaubermod}

Reference~\cite{aliceppbprod} presents a Glauber centrality model for 5 TeV \ppb\ collisions that may be compare with results from the TCM analysis of \ppb\ \mmpt\ data summarized above. Glauber parameters $N_{part}$, $N_{bin}$ and $b$ are obtained from Table 2, and mean charge densities at midrapidity in the form $\bar \rho_0 = n_{ch} / \Delta \eta$ are estimated from Fig.~16 for seven centrality classes of \ppb\ collisions. Parameter values are summarized in Table~\ref{ppbparams1} of this article.
The particle data sample for Ref.~\cite{aliceppbprod} is the same 106 million 5 TeV NSD \ppb\ collisions reported in Ref.~\cite{alicempt} as described in Sec.~\ref{ppb}. Thus, the TCM obtained from \ppb\ \mmpt\ analysis is directly applicable to the Glauber analysis.


\subsection{$\bf p$-$\bf Pb$ Glauber analysis strategy} \label{strategy}

The Glauber model of A-B collision geometry is used to relate fractional cross section $\sigma / \sigma_0$ ($\sigma_0$ is a measured total cross section) to collision geometry parameters such as impact parameter $b$, participant-nucleon number $N_{part}$, \nn\ binary-collision number $N_{bin}$ and the derived collision number per participant pair $\nu \equiv 2N_{bin} / N_{part}$. A Monte Carlo simulation including  geometric models of the collision partners and relevant cross sections is used to relate those quantities on a statistical basis~\cite{powerlaw}.
``The Glauber-MC determines on an event-by-event basis the properties of the collision geometry, such as ...$N_{part}...,$'' and an event ensemble determines statistical parameter mean values and probability distributions, e.g.\ $P(N_{part})$.


Some definition of a particle-production model denoted by conditional probability $P(n_{ch}|N_{part})$ is an essential requirement and must include an assumed \nch\ vs $N_{part}$ trend. The model in Ref.~\cite{aliceppbprod} is a NBD with parameters  $\mu$ and $k$ and the assumption that those parameters {\em scale linearly} with $N_{part}$ (e.g.\ $n_{ch} \propto N_{part}$). Convolution of $P(n_{ch}|N_{part})$ with $P(N_{part})$ yields a distribution on some observable \nch: $P(n_{ch}) = \sum_{N_{part}} P(n_{ch}|N_{part}) P(N_{part})$. Model $P(n_{ch}|N_{part})$ is ``validated'' by comparing the convolution integral to a measured $P(n_{ch})$ distribution. It is concluded that fitted ``values of parameters $\mu$ and $k$ are similar to those obtained by fitting  the corresponding multiplicity distributions in pp collisions at 7 TeV'' (but see Sec.~\ref{details}).

Once $P(n_{ch}|N_{part})$ has been so validated  it is used to determine mean $N_{part}$ for a given \nch\ (centrality) interval: ``The collision geometry is determined by fitting the measured [e.g.\ V0A $P(n_{ch})$ defined below] distribution with [the convolution integral]'' referred to as a NBD-Glauber fit.  The average $N_{part}$ (or other Glauber parameter) value for each of several defined event classes (based on \nch\ bins) are obtained: ``For a given centrality class, defined by selections in the measured [e.g.\ V0A] distribution, the information from the Glauber MC in the {\em corresponding generated distribution} [emphasis added] is used to calculate the mean number of participants $\langle N_{part} \rangle$ [here simply denoted $N_{part}$]....'' An alternative procedure based on running integrals is described below.



\subsection{p-Pb Glauber-model results} \label{glaubresults}

\begin{figure}[h]
  \includegraphics[width=1.65in]{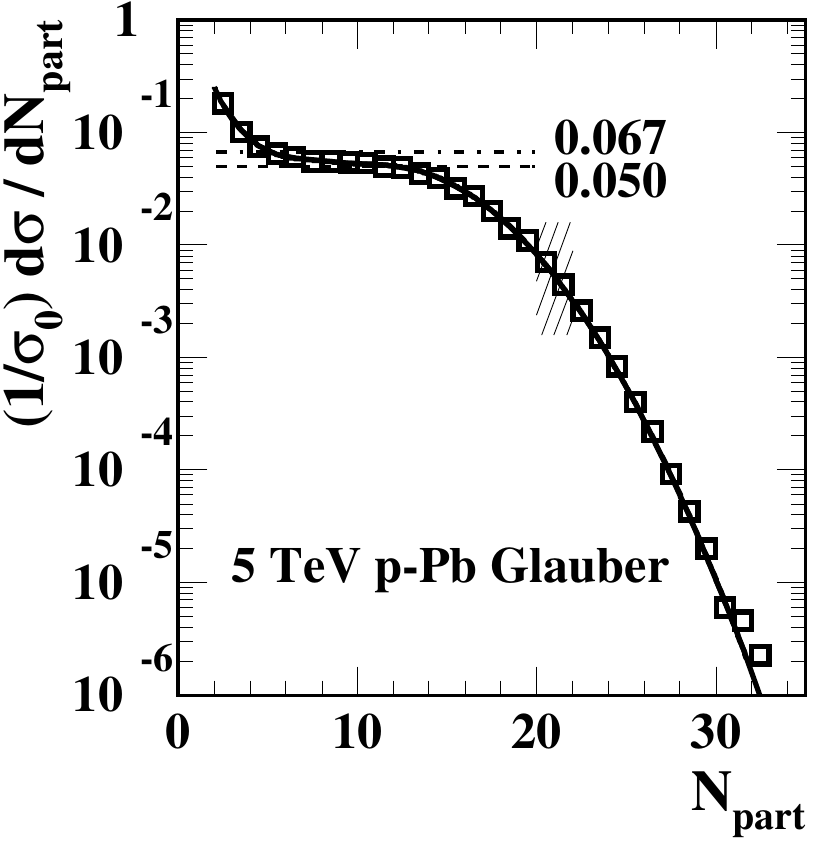}
 \includegraphics[width=1.65in]{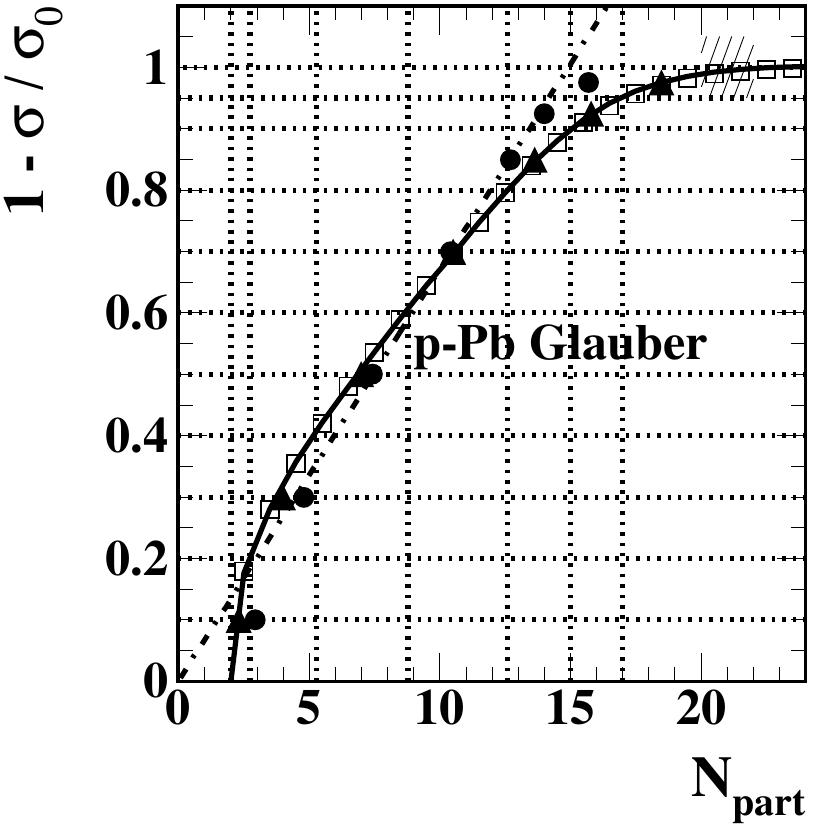}
  \caption{\label{glauber1}
  Left:  Monte Carlo data from Fig.~3 (left, Std Glauber) of Ref.~\cite{aliceppbprod} (open squares) representing a Glauber  simulation of 5 TeV \ppb\ collisions. A function describing the data (curve) is given by Eq.~(\ref{glauberdata}).
  Right: Running integrals of data from Ref.~\cite{aliceppbprod} (open squares) and Eq.~(\ref{glauberdata}) (solid curve) in the left panel. The solid points are discussed in the text.
   }  
 \end{figure}

Figure~\ref{glauber1} (left) shows Monte Carlo data obtained from Fig.~3 (left, Std-Glauber) of Ref.~\cite{aliceppbprod} (open squares) representing a Glauber  simulation of 5 TeV \ppb\ collisions. A function describing the Monte Carlo data (curve) is
\bea \label{glauberdata}
p(N_{part}) &=& 0.05 \exp\{-[(N_{part}-11.4)/6.4]^2\}
\\ \nonumber
q(N_{part}) &=& 0.84/N_{part}^{1.75} [1 + (N_{part}/9.3)^{10}]
\\ \nonumber
\frac{1}{\sigma_0} \frac{d\sigma}{dN_{part}} &=& p(N_{part}) + q(N_{part})
\eea
employed below for data description. Beyond its inflection point near 10 the distribution is Gaussian: $p(N_{part})$. The left panel in Fig.~3 of  Ref.~\cite{aliceppbprod} is labeled ``Events (arb units)'' but is consistent with a unit-normal probability distribution modeling the differential cross section. Tail structure for $N_{part} > 20$ is not relevant to measurements.


If \pn\ cross section $\sigma_{pN}$ represented all \pn\ encounters then for  a minimum-bias 5 TeV \ppb\ event ensemble $\bar N_{bin} \equiv A \sigma_{pN} / \sigma_{pA}$ with measured $\sigma_{pA} \approx 2.1$ b and $\sigma_{pN} = 70\pm 5$ mb~\cite{aliceppbprod}. $\bar N_{bin} = 208 \times 70 / 2100 = 6.9 \pm 0.5$ compares to mean $\bar N_{part} = \bar N_{bin} + 1 =  8.3$ obtained from the Glauber distribution in Fig.~\ref{glauber1} (left). The MC value is consistent with the measured cross sections within the 7\% uncertainty in $\sigma_{pN}$, but the measured cross sections were inputs to the Glauber MC. 


Figure~\ref{glauber1} (right) shows running integrals of data from Ref.~\cite{aliceppbprod} (open squares) and Eq.~(\ref{glauberdata}) (solid curve) in the left panel. Both integrals go to unity asymptotically without further normalization. The horizontal dotted lines are centrality bin edges defined in Ref.~\cite{aliceppbprod}. The vertical dotted lines are centrality bin edges on $N_{part}$ determined by the solid curve and horizontal dotted lines. Centrality bin {\em centers} on  $N_{part}$ are denoted by the solid triangles, also on the solid curve, with values denoted $N_{part}$ (unprimed) in Table~\ref{ppbparams1}. The solid dots, representing $N_{part}'$ entries in Table~\ref{ppbparams1} obtained from Table 2  of Ref.~\cite{aliceppbprod}, deviate substantially from the running integrals. Whereas the integrals are consistent with a value 0.050 at the inflection point in the left panel the solid dots are consistent with a value 0.067 (slope of the dash-dotted line through solid dots). The Glauber parameter values from  Ref.~\cite{aliceppbprod} thus appear to be inconsistent with the basic Monte Carlo data.

For experimental determination of collision centralities a Glauber simulation must be related to a measurable quantity (e.g.\ a particle multiplicity), of which several are considered in Ref.~\cite{aliceppbprod}. For this study results from a VZERO-A or V0A scintillator detector covering acceptance $2.8 < \eta < 5.1$ in the Pb hemisphere provide an example. The signal amplitude from the V0A detector is denoted here by $n_x$ (to distinguish from \nch\ or $\bar \rho_0$ relating to midrapidity). The main issue is the relation between simulated $N_{part}$ and measured $n_x$ determined through the intermediary of a differential cross section.

\begin{figure}[h]
  \includegraphics[width=1.62in]{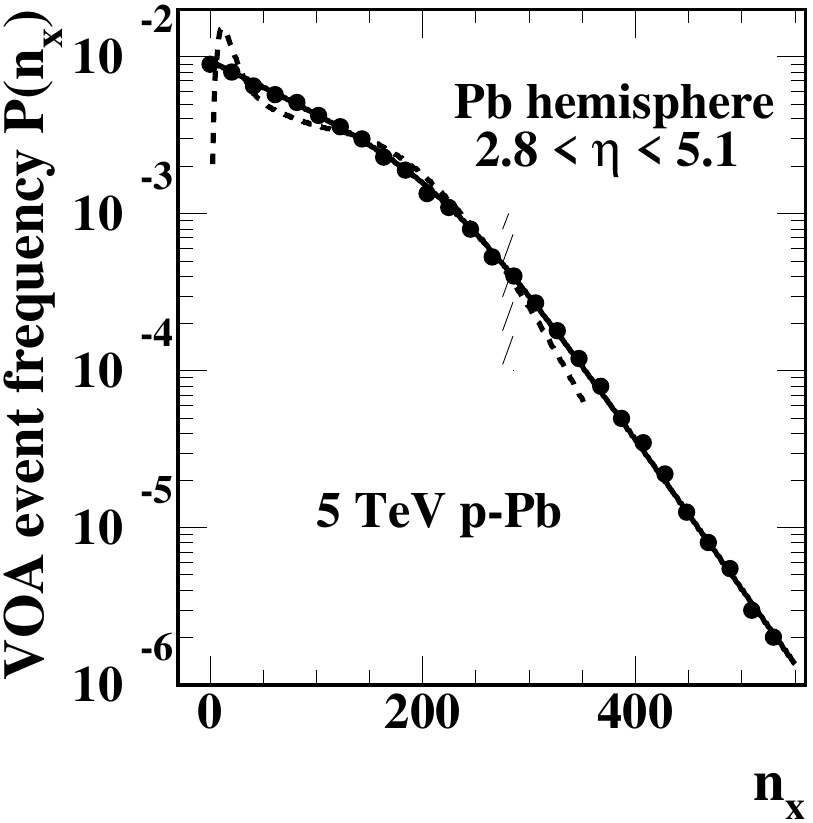}
  \includegraphics[width=1.69in]{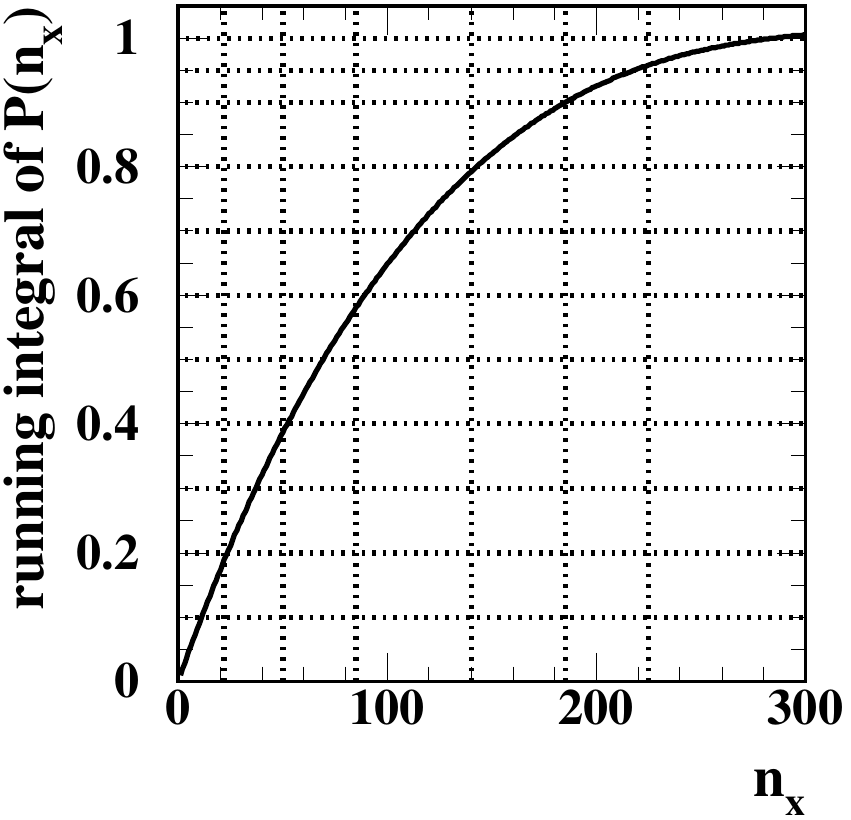}
  \caption{\label{ppbvoa}
  Left:  Normalized event-frequency distribution $P(n_x)$ on V0A amplitude $n_x$ (points) for 5 TeV \ppb\ collisions from Fig.~1 of Ref.~\cite{aliceppbprod}. A function describing the data (curve) is given by Eq.~(\ref{nxprob}).
  Right:  The running integral on $n_x$ of the $P(n_x)$ expression in Eq.~(\ref{nxprob}) with centrality bins (dotted).
   }  
 \end{figure}

Figure~\ref{ppbvoa} (left) shows normalized event-frequency (probability) distribution $P(n_x)$ on V0A multiplicity $n_x$ (points) for 5 TeV \ppb\ collisions from Fig.~1 of Ref.~\cite{aliceppbprod}. A model function (curve) describing the V0A data is
\bea \label{nxprob}
A(n_x) &=&0.0075 + 0.0145 \{1 + \tanh[(n_x - 225)/90]\}/2
\nonumber \\ 
P(n_x)  &=& \frac{1}{N_0} \frac{dN_{evt}}{dn_x} =  \exp\left[- \int_0^{n_x} \hspace{-.1in} dn_x'\, A(n_x') - 4.67\right]
\eea
which is accurate to a few percent. To establish a connection between $N_{part}$ and $n_x$ it is apparently assumed in Ref.~\cite{aliceppbprod} that $P(n_x) \rightarrow (1/\sigma_0) d\sigma/dn_x$ or $dN_{evt} \propto d\sigma$ -- centrality classes $\sigma / \sigma_0$ are defined by binning $P(n_x)$. For comparison the dashed curve is the Glauber differential cross section in Fig.~\ref{glauber1} (left) transformed from $N_{part}$ to $n_x$ with a Jacobian derived from the curve in Fig.~\ref{v0anch} (left).

Figure~\ref{ppbvoa} (right) shows a running integral on $n_x$ of the $P(n_x)$ expression in Eq.~(\ref{nxprob}) with asymptotic limit 1.02. The horizontal dotted lines are centrality bin edges and the vertical dotted lines are the corresponding bin edges on $n_x$ from Fig.~1 of  Ref.~\cite{aliceppbprod}. Correspondence with the running integral supports the assertion that $P(n_x) \rightarrow (1/\sigma_0) d\sigma/dn_x$ is assumed for the \ppb\ Glauber analysis.

Table~\ref{ppbparams1} summarizes some results of the \ppb\ Glauber analysis from Table 2 of Ref.~\cite{aliceppbprod} relating to the V0A detector. Primed quantities are found to be biased in the present study. The primed Glauber parameters correspond to the solid dots in Fig.~\ref{glauber1} (right) whereas the unprimed $N_{part}$ values correspond to the triangles on the solid curve. The $\bar \rho_0' = n_{ch}' / \Delta \eta$ values, obtained from bin edges on V0A multiplicity $n_x$, do not correspond to the defined centrality bins, as established in Sec.~\ref{oops} below.

\begin{table}[h]
  \caption{V0A Glauber parameters for 5 TeV \ppb\ collisions are from Table 2 and $\bar \rho_0' = n_{ch}'/ \Delta \eta$ densities are from Fig.~16 of Ref.~\cite{aliceppbprod}. Primes indicate biased entries as determined in the present study. The event sample is approximately NSD.
}
  \label{ppbparams1}
\begin{center}
\begin{tabular}{|c|c|c|c|c|c|} \hline
 centrality (\%) & $b'$ (fm) & $N_{part}'$ & $N_{bin}'$ & $n_{ch}' / \Delta \eta$ & $N_{part}$  \\ \hline
0 - 5   & 3.12 & 15.7  & 14.7 &  44.6 & 18.5  \\ \hline
 ~\,5 - 10  & 3.50 &  14.0 & 13.0 & 35.9 &  15.8 \\ \hline
 10 - 20  & 3.85 &  12.7 &11.7  & 30.0  & 13.65  \\ \hline
  20 - 40 & 4.54 &  10.4 & 9.36 & 23.0 &  10.5   \\ \hline
40 - 60   & 5.57 & 7.42  & 6.42 & 15.8 &  7.0  \\ \hline
 60 - 80  & 6.63 & 4.81  & 3.81 & 9.7 &  4.0  \\ \hline
 ~\,80 - 100  & 7.51 &  2.94 &1.94  & 4.2 & 2.3  \\ \hline
\end{tabular}
\end{center}
\end{table}

Based on the key assumption that $N_{part} \propto n_{x}$ the V0A data in Fig.~\ref{ppbvoa} (left) were fitted with a convolution integral including the Glauber-model result as in Fig.~\ref{glauber1} (left) described by Eq.~(\ref{glauberdata}) and a parametrized \pp\ (\pn) negative binomial distribution (NBD) $P(n_{x};\mu,k)$, where $\mu = \bar n_x$ and $\mu / k$ is a fluctuation measure. If V0A multiplicity $n_x$ is assumed proportional to Glauber $N_{part}$ then $P(n_{x};\mu,k) \rightarrow P(n_{x};N_{part}\mu,N_{part}k)$ and $\mu$ and $k$ values are determined by fitting the V0A distribution.

\begin{figure}[h]
  \includegraphics[width=3.3in]{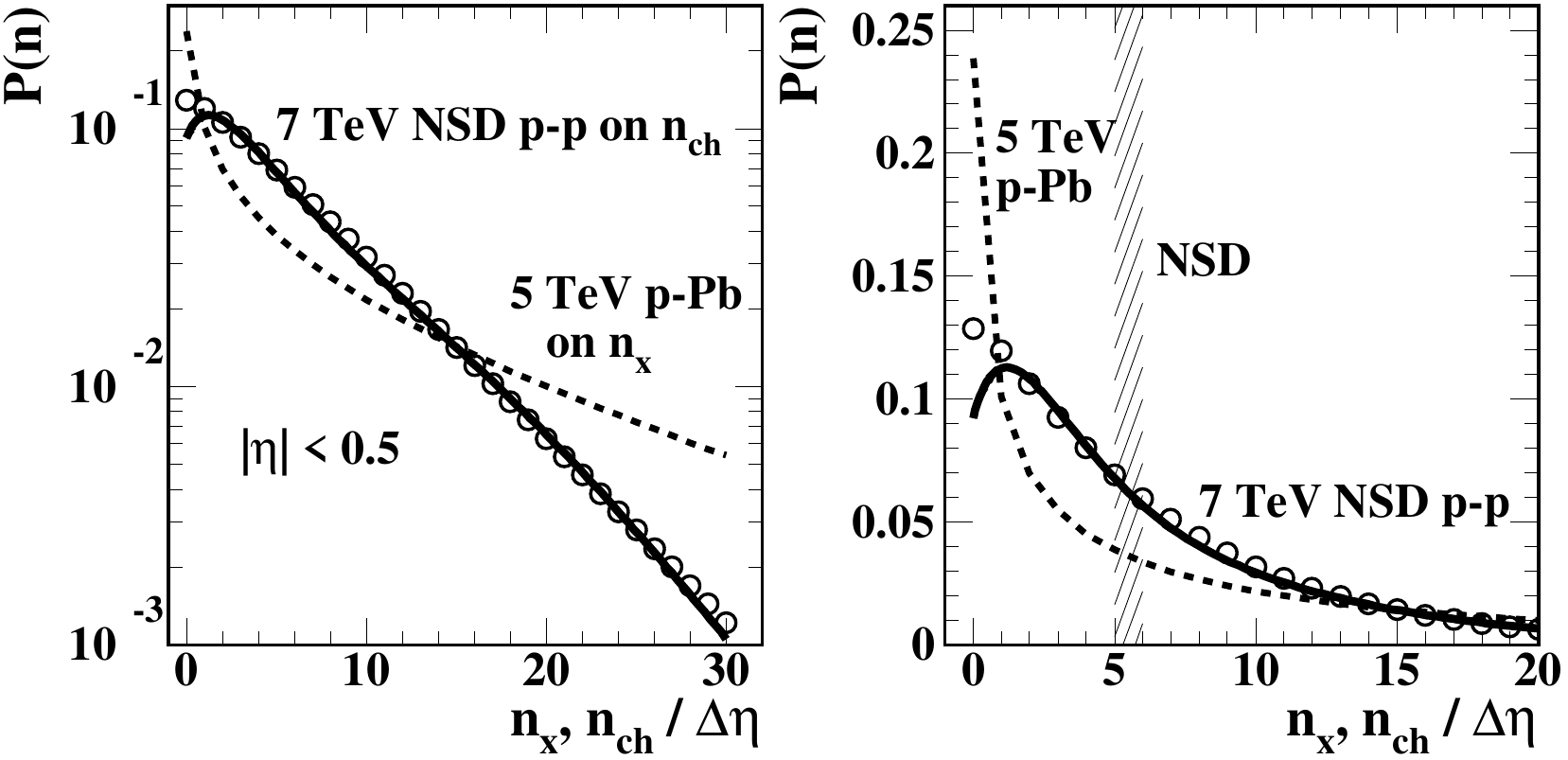}
  \caption{\label{ppbmult1}
  Left:  \pp\ NBD on $n_x$ inferred from a 5 TeV \ppb\ V0A fit (dashed).  The solid curve is a double-NBD direct fit to 7 TeV NSD \pp\ data as in Ref.~\cite{alicemult}. The open points are discussed in the text.
  Right: Those curves and data in linear format with NSD means $\approx 5$ for 5 TeV and $\approx 6$ for 7 TeV.
   } 
 \end{figure}

\begin{figure}[h]
\includegraphics[width=1.60in]{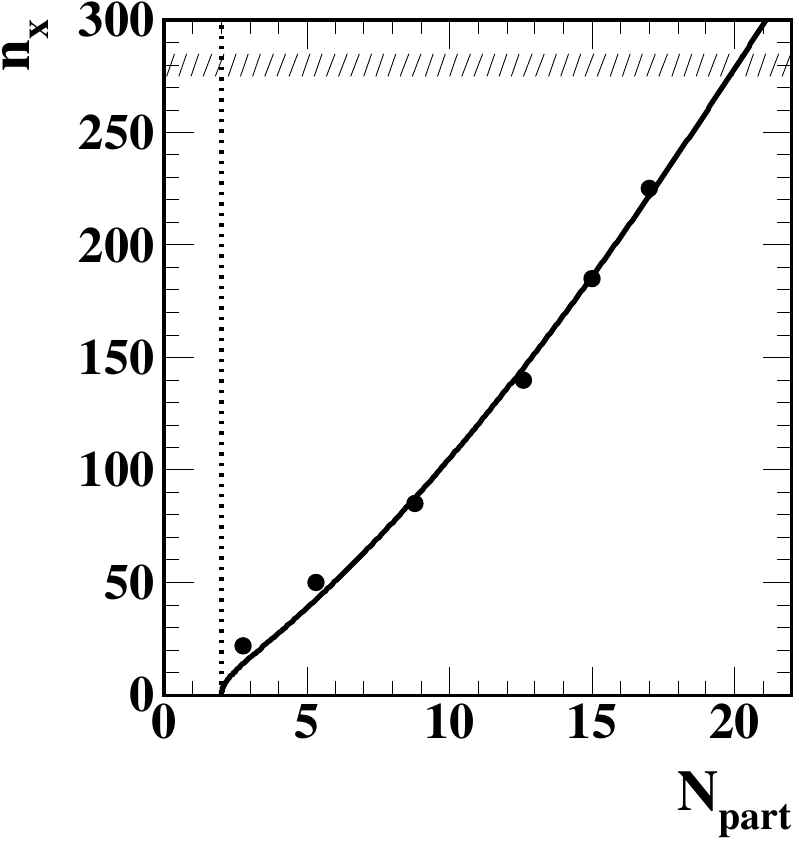}
  \includegraphics[width=1.70in]{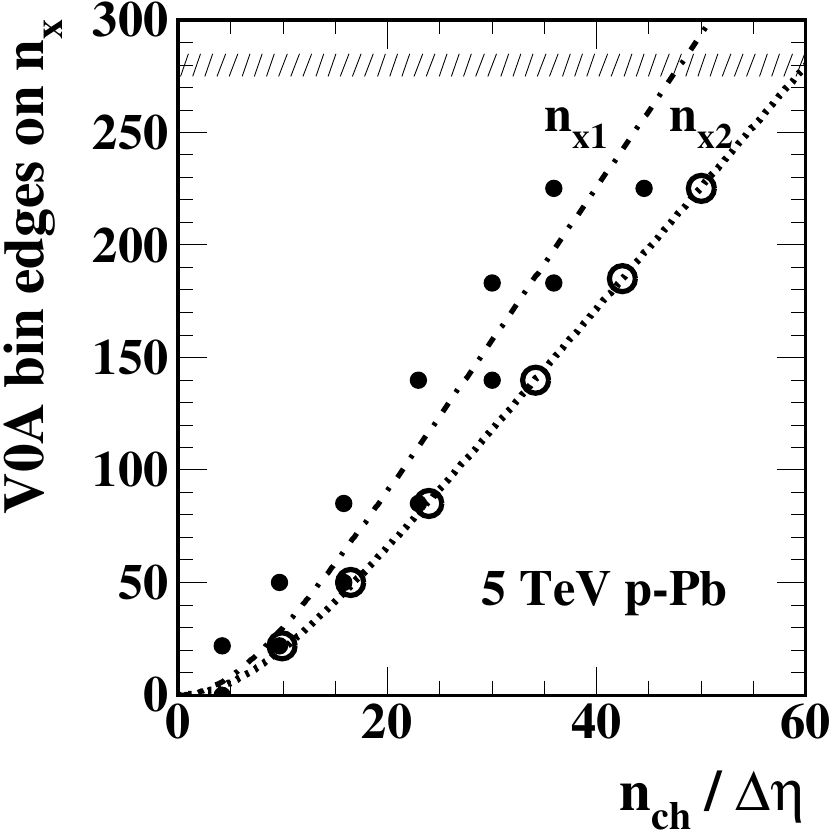}
  \caption{\label{v0anch}
Left:   Centrality bin edges on $N_{part}$ and $n_x$ (points) from Figs.~\ref{glauber1} (right) and \ref{ppbvoa} (right) that relate $n_x$ to $N_{part}$.  The solid curve is derived from Eqs.~(\ref{npartnx}) and (\ref{npart2}).
 Right:   pairs of bin edges on $n_x$ vs corresponding midrapidity centrality-mean values of $\bar \rho_0' \equiv n_{ch}' / \Delta \eta$ (solid points) for seven V0A centrality bins. The dash-dotted and dotted curves are Eqs.~(\ref{nchnx}). The open circles are explained in the  text.
   }  
 \end{figure}

Figure~\ref{ppbmult1} (left) shows the \pp\ NBD on $n_x$ inferred from the 5 TeV \ppb\ V0A fit (dashed curve) with $(\mu,k) = (11.0,0.44)$ from Table 1 of Ref.~\cite{aliceppbprod}. 
To provide a reference from \pp\ data the solid curve is a double-NBD direct fit to 7 TeV NSD \pp\ data (150 million events) on \nch\ as in Ref.~\cite{alicemult}. The open circles represent a single-NBD fit to the solid curve with parameters $(\mu,k) = (6.0,1.1)$ for direct comparison with the V0A results.

Figure~\ref{ppbmult1} (right) shows the same results on a linear plotting format. The hatched band encompasses the measured NSD \pp\ values on \nch\ for 5 and 7 TeV. The 7 TeV NSD mean is $\bar \rho_{0NSD} \approx 6.17$ in agreement with NBD $\mu = 6$ from Ref.~\cite{alicemult}.  The fitted \pn\ NBD on $n_x$ is determined by combining the Glauber distribution on $N_{part}$ with the $N_{part} \propto n_x$ assumption. The NBD mean $\mu = 11$ on $n_x$ is consistent with the 5 TeV NSD value $\bar \rho_{0NSD} \approx 5$ according to the curve in Fig.~\ref{v0anch} (right) below. The shapes of the two NBD distributions are very different, but the difference is explained in Sec.~\ref{oops}. What follows is a determination of the relations among $N_{part}$, $n_x$ and $n_{ch}$ that result from the Ref.~\cite{aliceppbprod} Glauber \ppb\ analysis.



Figure~\ref{v0anch} (left) shows the correspondence (points) between centrality bin edges on $N_{part}$ and on $n_x$ from Figs.~\ref{glauber1} (right) and \ref{ppbvoa} (right). The solid curve is
\bea \label{npartnx}
N_{part}(n_x) &=& (n_x / 4.2)^{1/1.4}
\eea
except that $N_{part} \geq 2$ is imposed as a constraint via
\bea \label{npart2}
{N_{part}} &\rightarrow & [{N_{part}}^4 + 2^4]^{1/4}.
\eea
Those relations establish a correspondence $N_{part} \leftrightarrow n_x$.

\begin{figure}[h]
  \includegraphics[width=1.63in]{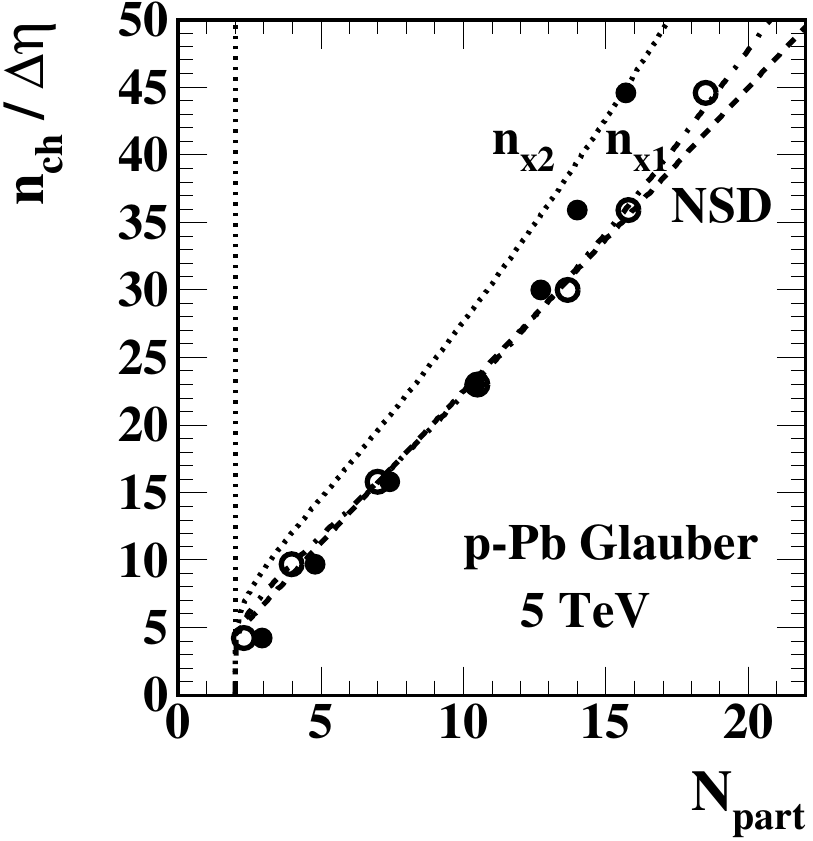}
  \includegraphics[width=1.67in]{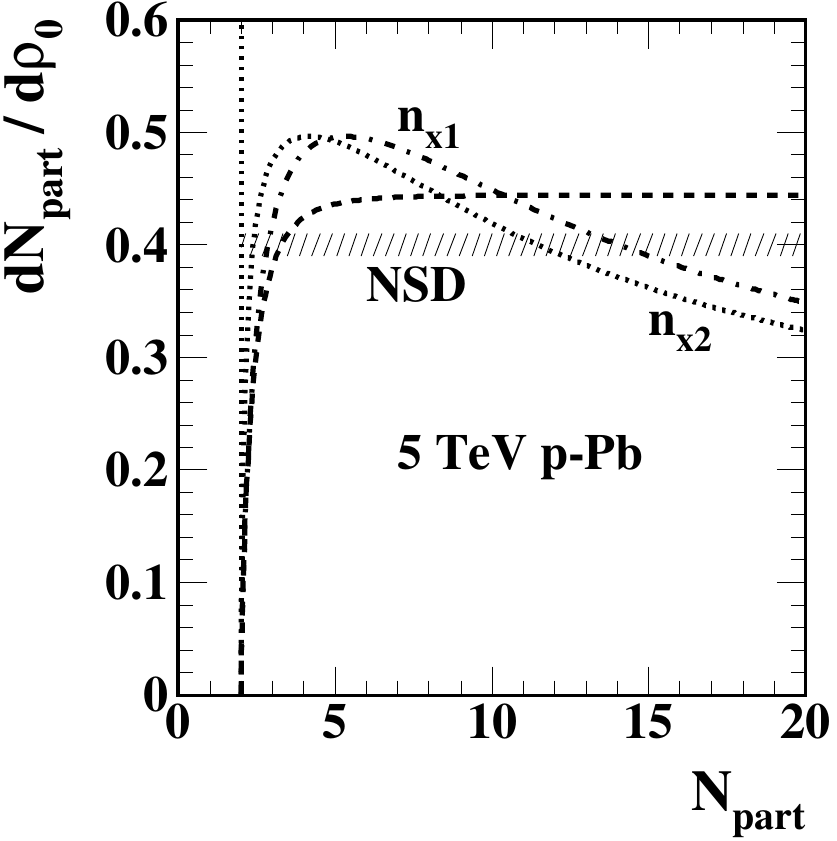}
  \caption{\label{nchnpart}
Left:  Relations between Glauber predicted $N_{part}$ and measured $\bar \rho_0$ from Table~\ref{ppbparams1} (solid and open points). The curves are explained in the text.
 Right: Jacobians relating $N_{part}$ to $\bar \rho_0$ derived from curves in the left panel.
   } 
 \end{figure}

Figure~\ref{v0anch} (right) shows pairs of bin edges on $n_x$ vs corresponding midrapidity centrality-mean values of $\bar \rho_0' \equiv n_{ch}' / \Delta \eta$ for seven V0A centrality bins from Table~\ref{ppbparams1} inferred from Fig.~16 (lower left) of Ref.~\cite{aliceppbprod}. The solid curve $n_{x1}$ defined by Eq.~(\ref{nchnx}) (first line) is
determined to pass between pairs of bin edges, closer to the lower edges per the V0A distribution on $n_x$. Final values for the two constants were established by accommodating data in Fig.~\ref{nchnpart} (left, open circles) for a self-consistent system.
\bea \label{nchnx}
n_{x1}(\bar \rho_0) &=& \int_0^{\bar \rho_0} d\bar \rho_0'\, 6.8 \tanh(\bar \rho_0'/10)
\\ \nonumber
n_{x2}(\bar \rho_0) &=& \int_0^{\bar \rho_0} d\bar \rho_0'\, 5.4 \tanh(\bar \rho_0'/12)
\eea
Equation~(\ref{nchnx}) (second line) defines the dotted curve $n_{x2}$ in the right panel and is derived by matching bin edges from $P(n_x)$ in Fig.~\ref{ppbvoa} (left) with bin edges from $P(n_{ch})$ obtained from \mmpt\ data in Ref.~\cite{alicempt} and appearing in Fig.~\ref{ppbmult4}. Since the same event ensemble is distributed on $n_x$ and \nch\ that is an apples-to-apples comparison.
Equations~(\ref{nchnx}) establish correspondence $n_x \leftrightarrow n_{ch}$ while Eq.~(\ref{npartnx})  establishes correspondence $N_{part} \leftrightarrow n_x$. The two relations can be combined to determine $N_{part} \leftrightarrow n_{ch}$.


Figure~\ref{nchnpart} (left) shows $\bar \rho_0' = n_{ch}' / \Delta \eta$ vs $N_{part}'$ from Table~\ref{ppbparams1} (solid dots). The open circles represent the same $\bar \rho_0'$ values vs corrected $N_{part}$ values from Fig.~\ref{glauber1} (right) corresponding to the solid triangles. The dash-dotted and dotted curves are Eqs.~(\ref{npartnx}), (\ref{npart2})  and (\ref{nchnx}) combined to yield $\bar \rho_0'(N_{part})$. The prime indicates that the charge density does not in fact correspond to the model $N_{part}$ values obtained from the \ppb\ Glauber analysis, as demonstrated below. The dashed  curve is $\bar \rho_0'  \approx (N_{part} / 2) 4.5$ per the assumption of proportionality in Ref.~\cite{aliceppbprod}.

Figure~\ref{nchnpart} (right) shows Jacobians $dN_{part} / d\bar \rho_0$ for density transformations from one variable to the other. The relation $\bar \rho_0 \approx \bar \rho_{0NSD} N_{part}/2$ is indicated by the hatched band for 5 TeV \pp\ and \ppb\ collisions with $\bar \rho_{0NSD} \approx 5$. The dash-dotted, dotted and dashed curves are derived from corresponding curves in the left panel inferred from the Glauber analysis. The Jacobians test the initial assumption(s) of Ref.~\cite{aliceppbprod} that \nch\ is linearly related to, proportional to or ``scales with'' $N_{part}$, in which case the Jacobian $dN_{part} / d\bar \rho_0$ should be approximately constant.
 
Interpreting the V0A probability distribution $P(n_x)$ as a differential cross-section distribution defines \ppb\ centrality in terms of observable $n_x$ and the Glauber model of \pa\ geometry, e.g.\ parameters $N_{part}$ and $b$. The centrality bins on $n_x$ then define bin-averaged $\bar \rho_0$ values at midrapidity. The combination of a Glauber model for \ppb\ centrality and a \pn\ NBD distribution with an assumption of proportionality between $N_{part}$ and $n_x$ leads to inference of a \pp\ NBD on $n_x$. The combination suggests approximate proportionality between $N_{part}$ and $\bar \rho_0$, seeming to close the circle. In the next section results of the \ppb\ Glauber analysis of Ref.~\cite{aliceppbprod} are compared to geometry information inferred from \mmpt\ data in Ref.~\cite{tommpt} derived from the same underlying particle data.

\section{$\bf p$-$\bf Pb$ Glauber model $\bf vs$ $\bf \bar p_t$ TCM} \label{glaubervstcm}

An analysis of \mmpt\ data for \pp, \ppb\ and \pbpb\ collisions for several LHC energies was reported in Ref.~\cite{tommpt}, and pertinent details are reviewed in App.~\ref{ppmptapp} and Sec. \ref{ppb}. A TCM for \mmpt\ data from each collision system relates charge multiplicity to system centrality (where relevant) via manifestations of MB dijet production. In this section the \ppb\ TCM \mmpt\ results are compared to the Glauber-model description of \ppb\ data. The principal issue is apparent contradictions between the $N_{part}(\bar \rho_0)$ trend inferred from the \mmpt\ TCM  and from the Glauber model of \ppb\ centrality.

\begin{figure}[h]
  \includegraphics[width=1.65in]{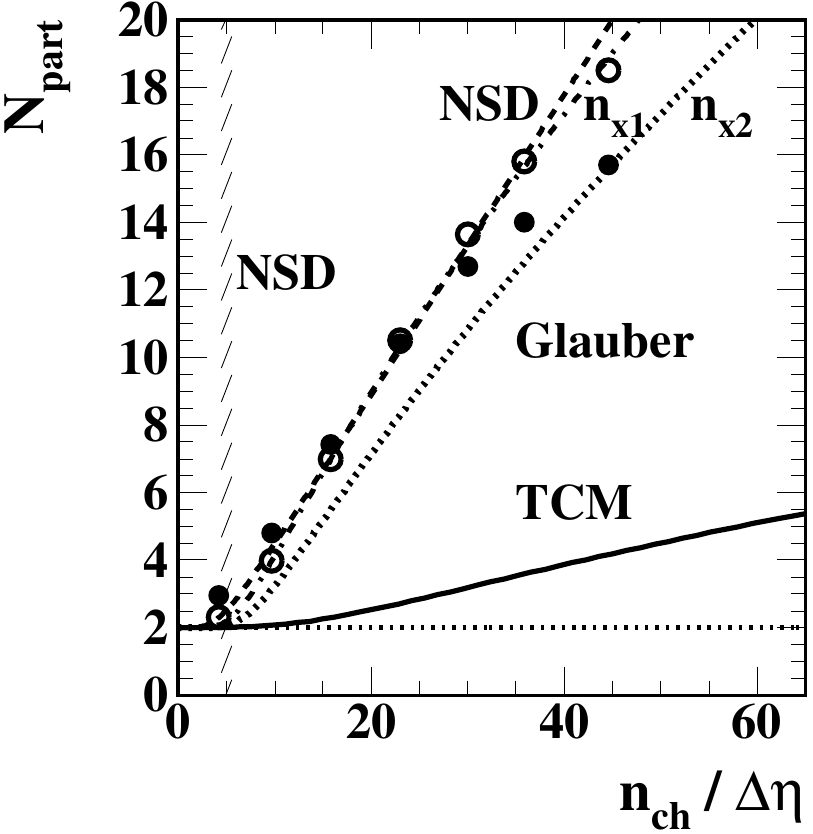}
  \includegraphics[width=1.65in]{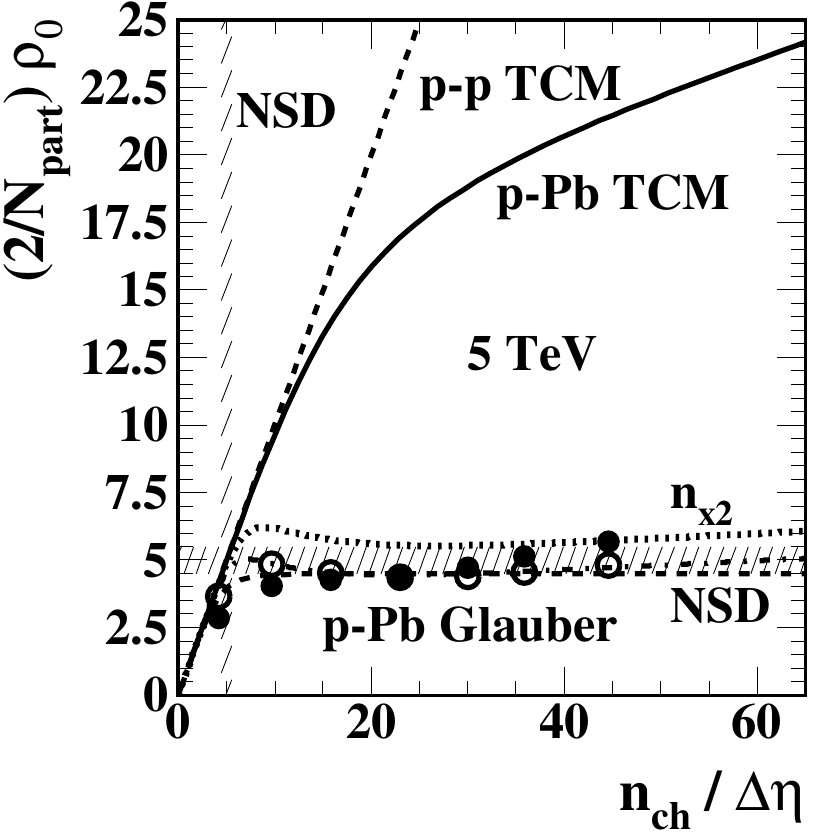}
 \caption{\label{tcmcomp1}
  Left: Comparison of $N_{part}(\bar \rho_0)$ trends from the Glauber study in Ref.~\cite{aliceppbprod} (points and three curves) and the \ppb\ \mmpt\ TCM (solid) from Ref.~\cite{tommpt}.
  Right: Hadron production per participant pair: trends from the Glauber study in Ref.~\cite{aliceppbprod} (points and three lower curves), from the  \ppb\ \mmpt\ TCM (solid) and from the \pp\ TCM (dashed).
 } 
 \end{figure}

Figure~\ref{tcmcomp1} (left) shows  the \ppb\ Glauber-model number of participants $N_{part}$ vs charge density $\bar \rho_0' = n_{ch}' / \Delta \eta$ (points) from Table~\ref{ppbparams1}. Solid and open points represent $N_{part}'$ and $N_{part}$ respectively. The dash-dotted ($n_{x1}$) and dotted ($n_{x2}$) curves are obtained from Eqs.~(\ref{npartnx}), (\ref{npart2}) and (\ref{nchnx}). The dashed curve represents the assumption that ``the number of participants is proportional to the number of charged hadrons,'' e.g.\ $n_{ch} \approx n_{chNSD} N_{part} /2$ [except $N_{part} \geq 2$ per Eq.~(\ref{npart2})]. The solid curve is from the \ppb\ TCM~\cite{tommpt}. The large difference is apparent.

Figure~\ref{tcmcomp1} (right) shows $(2/N_{part}) \bar \rho_0$ obtained from Table~\ref{ppbparams1} entries (solid dots, $N_{part}'$) and using corrected $N_{part}$ (open circles). The corresponding TCM \ppb\ trend is the solid curve, and the dash-dotted and dotted curves are obtained from the equivalent in the left panel. The dashed curve is $(2/ N_{part}) \bar \rho_0 = 4.4$ but with the constraint $N_{part} \geq 2$ imposed  per Eq.~(\ref{npart2}). The hatched band is $\bar \rho_{0NSD} \approx 5$ corresponding to 5 TeV \pp\ collisions~\cite{alicetomspec}, consistent with the basic Glauber assumption for \ppb\ analysis that ensemble-averaged \nn\ collisions are the same as \pp\ NSD for all \ppb\ centrality conditions and increasing \nch\ must be due entirely to increasing $N_{part}$.



\begin{figure}[h]
  \includegraphics[width=1.66in]{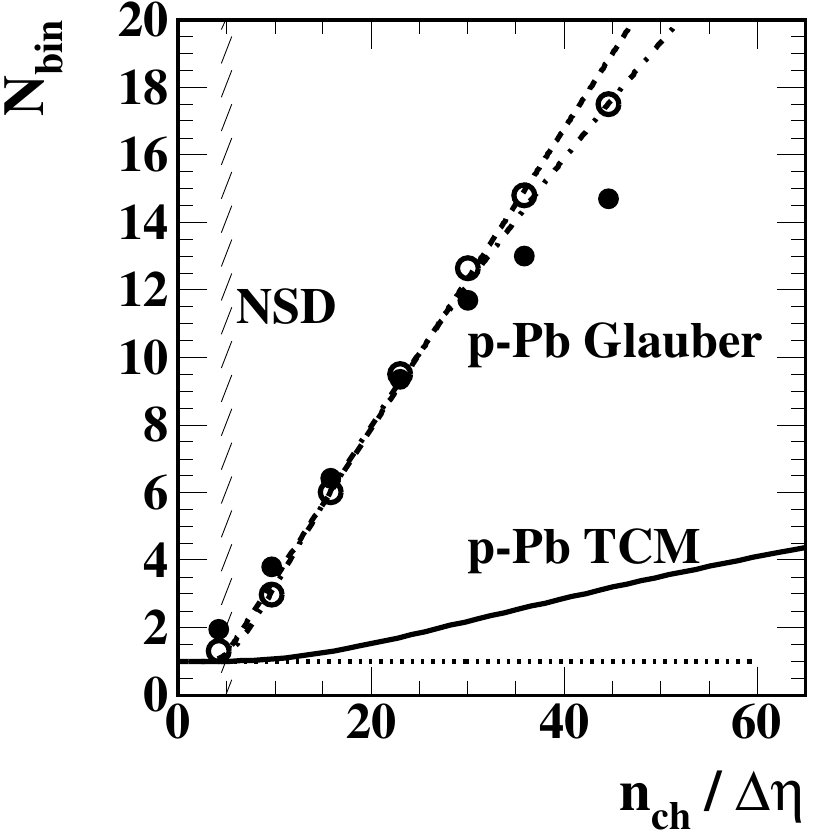}
 \includegraphics[width=1.64in]{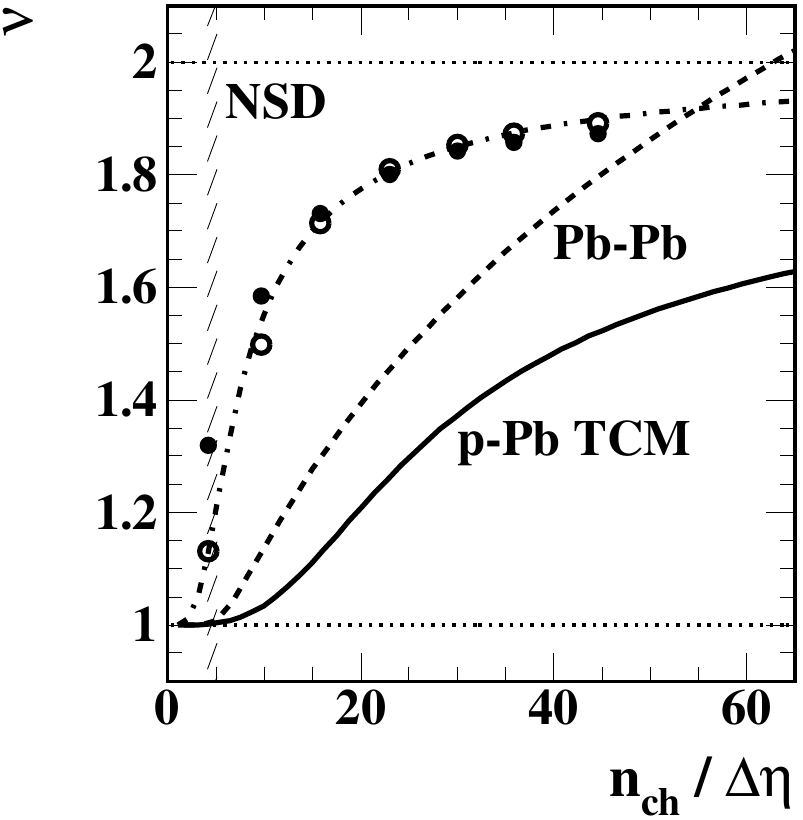}
 \caption{\label{ppbmult}
  Left: $N_{bin}$ vs $\bar \rho_0$ with $N_{bin} \equiv N_{part} - 1$ for \pa\ collisions.
  Right:  $\nu \equiv 2 N_{bin} / N_{part}$  obtained from Table~\ref{ppbparams1} values (solid dots, $N_{part}'$, $N_{bin}'$) and from corrected $N_{part}$ (open circles) compared to the \ppb\ TCM (solid curve).
   }  
 \end{figure}

Figure~\ref{ppbmult} (left) shows $N_{bin}$ vs $\bar \rho_0$ with $N_{bin} \equiv N_{part} - 1$ for \pa\ collisions. It is notable that the Glauber estimate $N_{bin} \approx 2$ applies to a centrality range where \ppb\ is dominated by \pn.
Figure~\ref{ppbmult} (right) shows $\nu \equiv 2 N_{bin} / N_{part}$  obtained from Table~\ref{ppbparams1} values (solid dots, $N_{part}'$, $N_{bin}'$) and from corrected $N_{part}$ (open circles) compared to the \ppb\ TCM (solid curve). The $\nu$ trend for \pbpb\ (dashed) is included for comparison.  It is notable that $\nu$ for \ppb\ from Ref.~\cite{aliceppbprod} {\em exceeds} that for \pbpb\ up to $\bar \rho_0 \approx 50$ whereas the \ppb\ TCM parameter values near  $\bar \rho_0 \approx 40$ are $N_{part} \approx 4$, $N_{bin} \approx 3$ and $\nu \approx 1.5$. In that case $(2/N_{part})\, \bar \rho_0 \approx 4.5 \bar \rho_{0NSD}$ implies approximately {\em 20-fold increase} in dijet production according to Ref.~\cite{ppprd}.


\begin{figure}[h]
 \includegraphics[width=1.65in]{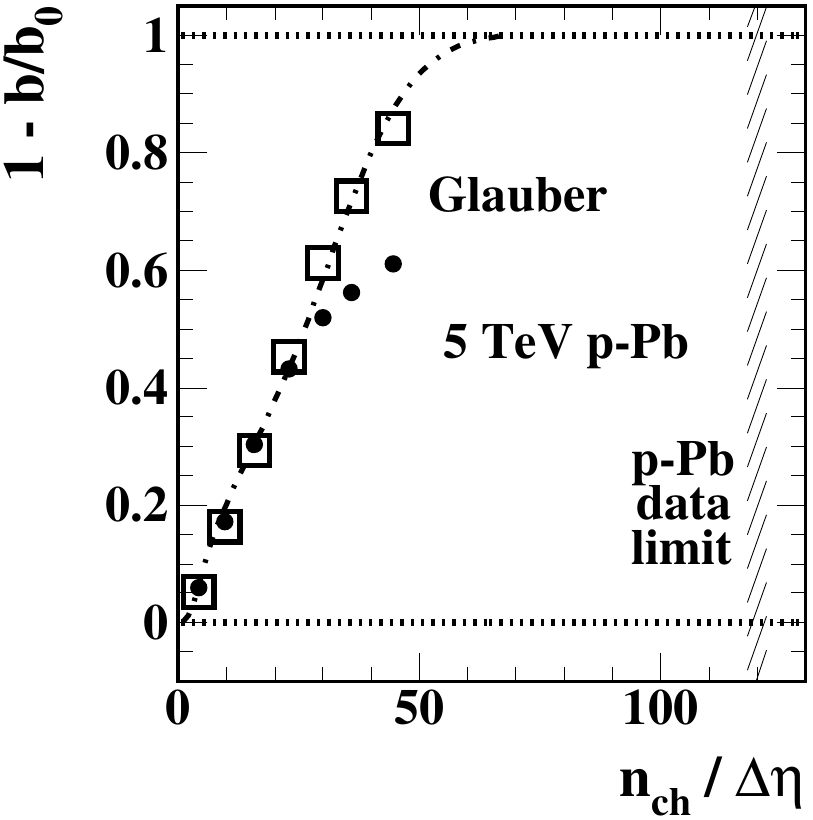}
  \includegraphics[width=1.65in]{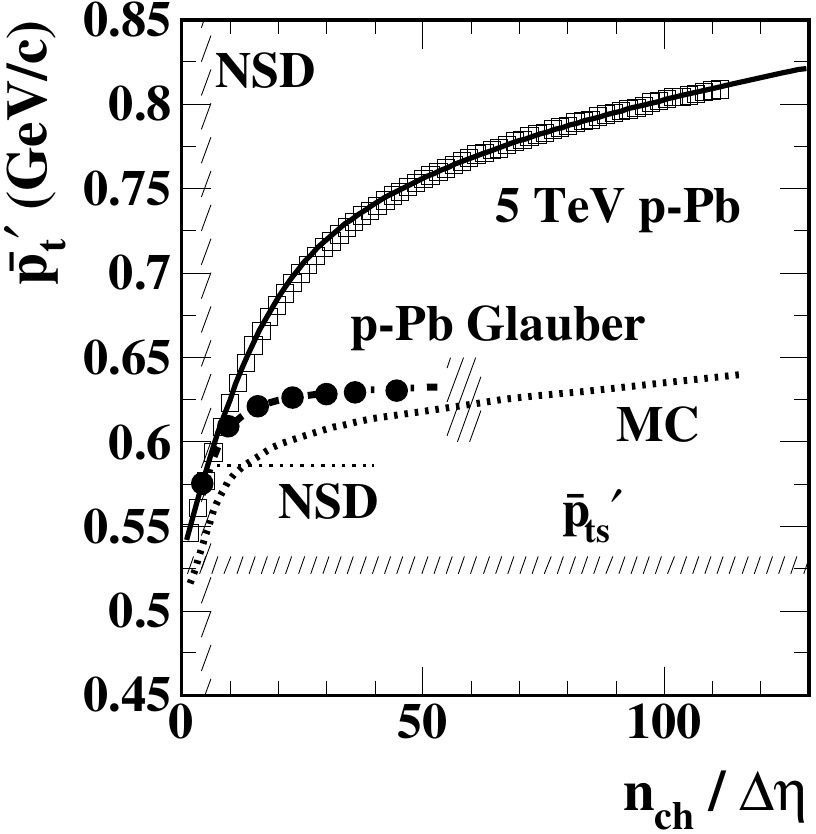}
  \caption{\label{ppbmult2}
  Left: Glauber-model \ppb\ centrality vs $\bar \rho_0$ from Ref.~\cite{aliceppbprod}  measured by impact-parameter  ratio $b / b_0$ (open squares) assuming that centralities in percent in Table~\ref{ppbparams1} represent $100 \sigma / \sigma_0$ and $\sigma / \sigma_0 \approx (b / b_0)^2$. The solid points are from Table~\ref{ppbparams1}.
  Right:  Uncorrected ensemble-mean \pt\ or $\bar p_t'$ vs corrected $\bar \rho_0 = n_{ch} / \Delta \eta$ for 5 TeV \ppb\ collisions from Ref.~\cite{alicempt} (open squares). The solid curve is the \ppb\ TCM. The solid points are predictions derived from the Glauber centrality analysis in Ref.~\cite{aliceppbprod}.  The Glauber MC curve (dotted) is taken from Fig.~3 of Ref.~\cite{alicempt}. The dotted line shows the $\bar p_t'$ estimate for NSD \pp\ collisions and all else the same.
   }  
 \end{figure}

Figure~\ref{ppbmult2} (left) shows Glauber-model \ppb\ centrality vs $\bar \rho_0$ from Ref.~\cite{aliceppbprod} where ``centrality'' is here measured by impact-parameter  ratio $b / b_0$ (open squares) assuming that centralities in percent in Table~\ref{ppbparams1} represent $100 \sigma / \sigma_0$ with $\sigma / \sigma_0 \approx (b / b_0)^2$.  The solid dots are obtained from the \ppb\ Glauber $b$ estimates in Table~\ref{ppbparams1} assuming $b_0 \approx 8$ fm (based on radii 7.1 fm and 0.85 fm for Pb and $p$). The dash-dotted curve is derived from fractional cross section  $1 - \sigma/\sigma_0$ shown as the Glauber running integral (dash-dotted curve) in Fig.~\ref{ppbmult4a} (left) below. The Glauber analysis suggests that fully-central \ppb\ collisions correspond to $\bar \rho_0 \approx 60$, whereas \mmpt\ data from Ref.~\cite{alicempt} extend out to $\bar \rho_0 \approx 115$ (hatched band).

Figure~\ref{ppbmult2} (right) shows uncorrected ensemble-mean $\bar p_t'$ vs corrected $\bar \rho_0 = n_{ch} / \Delta \eta$ data for 5 TeV \ppb\ collisions from Ref.~\cite{alicempt} (open squares). The solid curve is the corresponding TCM from Ref~\cite{tommpt}. The solid points and dash-dotted curve are \mmpt\ estimates based on results from the \ppb\ Glauber analysis of Ref.~\cite{aliceppbprod} using data from Table~\ref{ppbparams1} and Eq.~(\ref{pampttcm}) (first line) repeated here for convenience
\bea \label{cc}
  \bar p_t' &\approx& \frac{\bar p_{ts} + x(n_s) \nu(n_s) \, \bar p_{thNN}(n_s)}{\xi + x(n_s)\, \nu(n_s)}.
\eea
Consistent with the \ppb\ Glauber analysis $x(n_s) \approx \alpha \rho_{sNN}$ up to $\bar \rho_{sNSD}$ and then remains constant at the NSD value $\approx 0.06$. $\nu(n_s)$ is described by the dash-dotted curve in Fig.~\ref{ppbmult} (right) and $\bar p_{thNN} \rightarrow \bar p_{th0} = 1.3$ GeV/c with $\bar p_{ts} \approx 0.4$ and $\xi \approx 0.73$. Those trends are consistent with the assumption that average \nn\ collisions for any \ppb\ centrality should be equivalent to NSD \pp\ collisions. If that assumption were correct there is no possibility to match the published \ppb\ \mmpt\ data. According to the Glauber analysis \ppb\ data cannot extend beyond $\bar \rho_0 \approx 55$ (hatched band), yet the dotted  ``Glauber MC'' curve from Fig.~3 of Ref.~\cite{alicempt} extends out to $\bar \rho_0 \approx 70 / 0.6 \approx 115$. The MC vertical displacement from the other curves at small \nch\ is equivalent to a 5\% change in the inefficiency parameter $\xi$ in Eq.~(\ref{cc}).

To summarize, Glauber-model results from Ref.~\cite{aliceppbprod} appear to be inconsistent with  \mmpt\ data from Ref.~\cite{alicempt}, both from the same collaboration. There are three major issues: (a) The $N_{part}(\bar \rho_0)$ trend inferred from \mmpt\ data via the TCM analysis of Ref.~\cite{tommpt} is dramatically different from that assumed for the Glauber analysis. (b) The Glauber model suggests that most-central \ppb\ collisions correspond to $\bar \rho_0 < 55$ whereas \mmpt\ data extend to $\bar \rho_0 \approx 115$. (c) The \mmpt\ trend predicted by the Glauber analysis is very different from measurements in Ref.~\cite{alicempt} and described by the TCM in Ref.~\cite{tommpt}. In the next section possible sources of major differences are explored.


\section{$\bf p$-$\bf Pb$ Glauber-model discussion} \label{oops}



A description of centrality in A-B collisions includes several elements: (a) measured total cross section $\sigma_0 = \pi b_0^2$ distributed as differential cross section $d\sigma = \pi db^2$  manifesting as event frequency distributions on certain parameters; (b) A-B collision geometry measured by several internal geometry parameters, e.g.\ $N_{part}$; and (c) external or observable quantities that depend on hadron production, e.g.\ \nch. One then relates \nch\ to $N_{part}$ via $(\sigma,b)$. Several questions arise: (a) which nucleons in A and B are participants by what criteria, (b) what hadron production mechanism(s) apply, (c) is $d\sigma \propto dN_{evt}$ valid.

Reference~\cite{aliceppbprod} relates $N_{part}$ to $(\sigma,b)$ via a geometric Glauber MC,
and $d\sigma \propto dN_{evt}$ on \nch\ is assumed. The combination of $N_{part} \leftrightarrow \sigma$ and $\sigma \leftrightarrow n_{ch}$ then closes the circle with an inferred relation between $N_{part}$ and \nch.

\subsection{$\bf N_{part}$ vs $\bf n_{ch}$ scaling assumptions} \label{scaling}

A central element of the \ppb\ Glauber analysis of Ref.~\cite{aliceppbprod} is the assumed relation between Glauber $N_{part}$ and an observable quantity such as particle multiplicity within some $\eta$ acceptance, denoted by $n_x$ for simplicity.
The abstract includes ``Under the assumption that the multiplicity measured in the Pb-going rapidity region [i.e.\ V0A $n_x$] scales with the number of Pb-participants, an approximate independence of the multiplicity per participating nucleon measured at midrapitity [sic] of the number of participating nucleons is observed.'' 
The summary includes ``In particular,  we assume that the multiplicity at mid-rapidity is proportional to $N_{part}$.... We find... ii) that the multiplicity of charged particles at mid-rapidity scales linearly with the total number of participants....'' Similar statements appear elsewhere in the text.



Arguments for proportionality or ``scaling'' between $N_{part}$ and $n_{ch}$ based in part on the wounded-nucleon model of hadron production~\cite{bialas2} rely on fixed-target results at lower collision energies~\cite{wit4} or early data from RHIC (with large systematic uncertainties and limited centrality range)~\cite{phobosgeom}.  A description of \auau\ particle production vs centrality (PHOBOS) includes the statement ``However, within the systematic errors, the total [i.e.\ $4\pi$] yield per participant pair is approximately constant (within 10\%) over the measured centrality range, $65 < N_{part} < 358$, which corresponds to $3 < \bar \nu < 6$, where $\bar \nu$ is the average number of collisions undergone by each oncoming nucleon. [That interval includes only the most-central 40\% of the total cross section.] Thus, it appears that only the first few collisions have any appreciable effect on particle production...''~\cite{phobosauaunpart}.  The statement cautions however ``It should be noted that this simple scaling is not observed for [differential] particle yields measured in a limited pseudorapidity range near midrapidity''~\cite{phobosauaunpart}.
The same collaboration further states that ``...in d + Au collisions the total multiplicity of charged particles scales linearly with the total number of participants...''~\cite{phobosdaunpart}. But that conclusion depends critically on how $N_{part}$ is estimated. If the $N_{part}$ estimate is based on assumed proportionality (as in Ref.~\cite{aliceppbprod}) the quoted conclusion is trivial and the argument likely misleading.

\subsection{Alternative TCM $\bf p$-$\bf Pb$ centrality analysis} \label{details}

As an alternative to the Glauber analysis of Ref.~\cite{aliceppbprod} the following strategy is adopted. The TCM relation between $N_{part}$ and $\bar \rho_0$ near $\eta = 0$ inferred from \mmpt\ data in Ref.~\cite{tommpt} as in Fig.~\ref{tcmcomp1} (left, solid curve) is adopted as the starting point. A frequency distribution $P(n_{ch})$ on $\bar \rho_0$ near midrapidity is derived from \mmpt\ data reported in Ref.~\cite{alicempt}. A cross-section distribution on $N_{part}$ is then constructed by assuming that the geometric Glauber distribution in Fig.~\ref{glauber1} (left) may be a good approximation for peripheral collisions only and that $P(n_{ch})$ inferred from \mmpt\ data is determining for more-central collisions.

\begin{figure}[h]
   \includegraphics[width=1.62in]{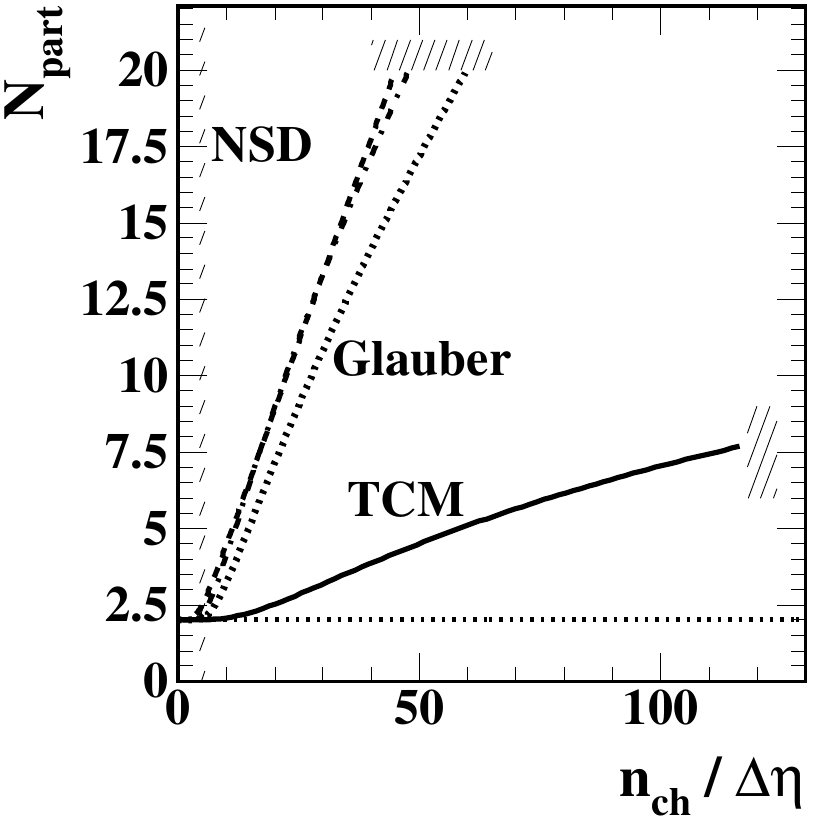}
 \includegraphics[width=1.68in]{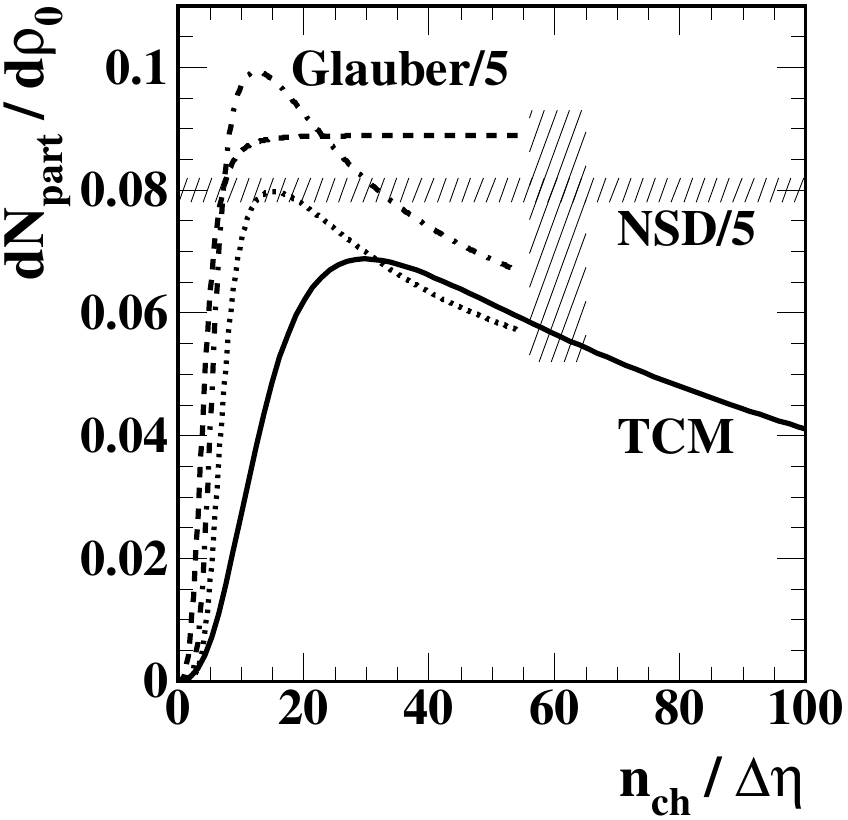}
  \caption{\label{ppbmult6}
Left: $N_{part}$ vs $\bar \rho_0$ for the Glauber analysis of Ref.~\cite{aliceppbprod} (dash-dotted and dotted), for ideal $N_{part}$ scaling (dashed) and for the \ppb\ \mmpt\ TCM (solid). 
Right:  Jacobians $dN_{part} / d\bar \rho_0$ derived from the left panel. An upper limit on attainable $\bar \rho_0$ implied by the Glauber analysis is denoted by the vertical hatched band. Glauber values are reduced by factor 1/5.
   } 
 \end{figure}

Figure~\ref{ppbmult6} (left) shows $N_{part}$ vs $\bar \rho_0$ for the Glauber analysis of Ref.~\cite{aliceppbprod} (dash-dotted and dotted), for ideal $N_{part}$ scaling (dashed) and for the \ppb\ \mmpt\ TCM (solid). 
The solid TCM curve is the starting point for this alternative analysis. Whereas the Glauber trends are consistent with numerous participants and smaller \pn\ multiplicities the TCM trend is consistent with fewer participants and larger  \pn\ multiplicities.

Figure~\ref{ppbmult6} (right) shows Jacobians $dN_{part} / d\bar \rho_0$ derived from curves in the left panel. Note that Glauber curves are scaled down by 1/5. The obvious large disagreement between TCM Jacobian and Glauber versions is a central issue for the present study: Assumption of $N_{part}$ ``scaling'' with some $n_x$ would result in a nearly constant Jacobian proceeding from NSD \nch, which is quite inconsistent with \mmpt\ data~\cite{tommpt} and MB dijet results~\cite{mbdijets}. 

The Jacobians in Figs.~\ref{ppbmult6} (right) and \ref{ppmult} (right) can be used to transform probability distributions and differential cross sections on $N_{part}$ or $n_x$ to common parameter \nch\ or $\bar \rho_0$.  Note that for an effective maximum value $N_{part} \approx 20$ from the differential cross section in Fig.~\ref{glauber1} the corresponding maximum charge density is $\bar \rho_0 \approx 20 / 0.4 = 50$ as indicated by the vertical hatched band in the right panel. The TCM value of $N_{part}$ for $\bar \rho_0 = 115$ is just less than 8 as indicated in the left panel.

Reference~\cite{aliceppbprod} does not provide an event frequency distribution on charge density $\bar \rho_0$ near midrapidity that can relate directly to Fig.~\ref{ppbmult6}, but such a distribution can be derived from \mmpt\ data in Ref.~\cite{alicempt}. Assume $\delta \bar p_t / \bar p_t \approx \delta N_{ch} / N_{ch} \approx 1/ \sqrt{\Delta \eta \bar \rho_0 N_{evt}}$ for Poisson statistics, where $N_{ch}$ is a multiplicity-bin total charge integrated over all events and within acceptance $\Delta \eta = 0.6$. Given reported statistical errors $\delta \bar p_t$ on \mmpt\ data vs $\bar \rho_0$  solve for $N_{evt}(n_{ch})$ to obtain event-frequency distribution $P(n_{ch})$ for $\eta = 0$.

\begin{figure}[h]
  \includegraphics[width=1.65in]{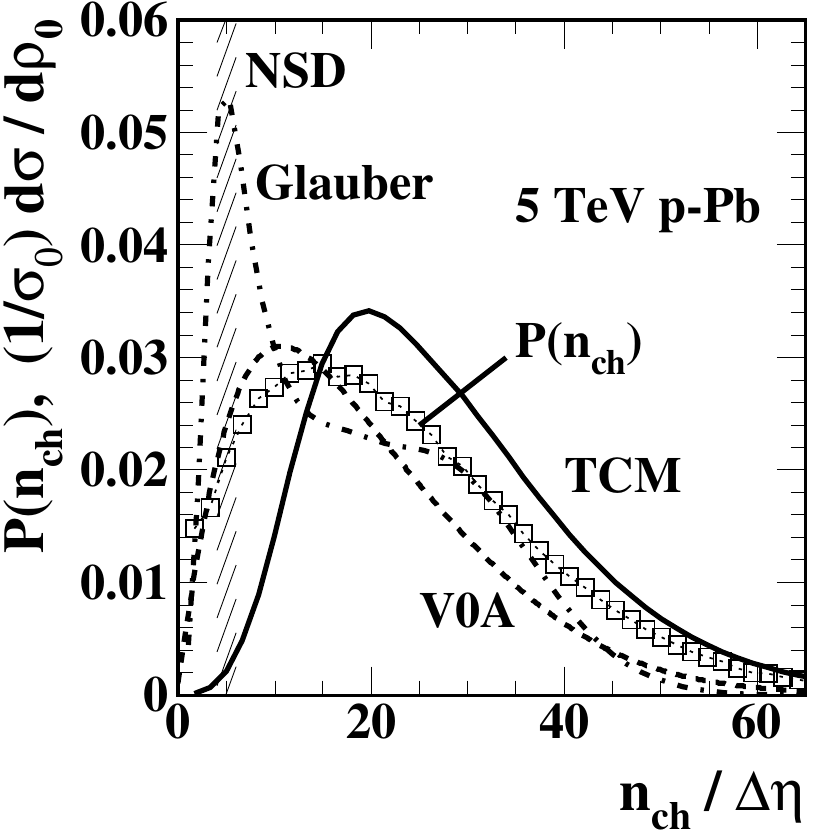}
  \includegraphics[width=1.65in]{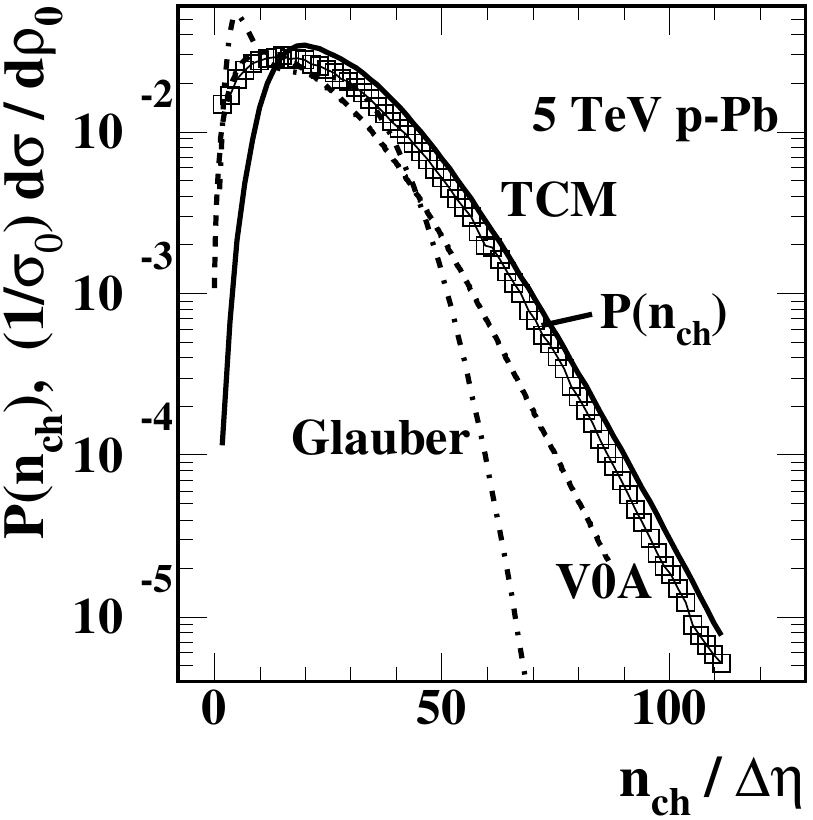}
  \caption{\label{ppbmult4}
Left:  Comparison of differential cross sections and probabilities obtained by transformations with relevant Jacobians and plotted within the same context.
 Right: The same curves in a semilog format. Curves are described in the text.
   } 
 \end{figure}

Figure~\ref{ppbmult4} (left) compares several curves of differing origins (within the same plot context) obtained by transformations with relevant Jacobians.  The dash-dotted curve is the Glauber MC differential cross-section on $N_{part}$ in Fig.~\ref{glauber1} (left) transformed with the dash-dotted Jacobian in Fig.~\ref{ppbmult6} (right) and is then {\em nominally} $(1/\sigma_0) d\sigma / d\bar \rho_0$, a differential cross-section distribution on $\bar \rho_0$.  The dashed curve is the V0A distribution $P(n_x)$ in Fig.~\ref{ppbvoa} transformed with the dashed Jacobian in Fig.~\ref{ppmult} (right) to obtain a normalized event distribution on  $\bar \rho_0$. The open squares represent the normalized event distribution $P(n_{ch})$ derived from \mmpt\ statistical errors $\delta \bar p_t$ reported in Ref.~\cite{alicempt} as described above. The solid curve is an inferred TCM differential cross section.

Figure~\ref{ppbmult4} (right) shows the same results in a semilog format. The TCM solid curves are derived as follows: The geometric Glauber MC cross-section distribution $(1/\sigma_0) d\sigma /dN_{part}$ in Fig.~\ref{glauber1} (left) is  assumed as a starting point, applying at least to peripheral \ppb\ collisions. The TCM Jacobian $dN_{part}/ d\bar \rho_0$ in Fig.~\ref{ppbmult6} (right, solid) is used to transform to $(1/\sigma_0) d\sigma /d\bar \rho_0$ which then greatly overshoots the mid-rapidity $P(n_{ch})$ (open squares) at larger $\bar \rho_0$ in the right panel. A function is applied to Glauber $(1/\sigma_0) d\sigma /dN_{part}$ in Fig.~\ref{glauber1} (left) such that the transformed distribution on $\bar \rho_0$ follows the form of $P(n_{ch})$ at larger $\bar \rho_0$ (see Sec.~\ref{laplace}). The result is the TCM solid curves in the two panels. The applied function is 
\bea \label{fac}
f &=&\exp\{ -[ (|N_{part} - 3| + N_{part} - 3)/2]^2 / 3 \}.
\eea
The dash-dotted Glauber curve drops off rapidly at larger \nch\ whereas from  Sec.~\ref{laplace} close correspondence with dashed  V0A $P(n_{ch})$ is expected, similar to the relation between the TCM solid curve and  $P(n_{ch})$ open boxes.

\begin{figure}[h]
  \includegraphics[width=1.65in]{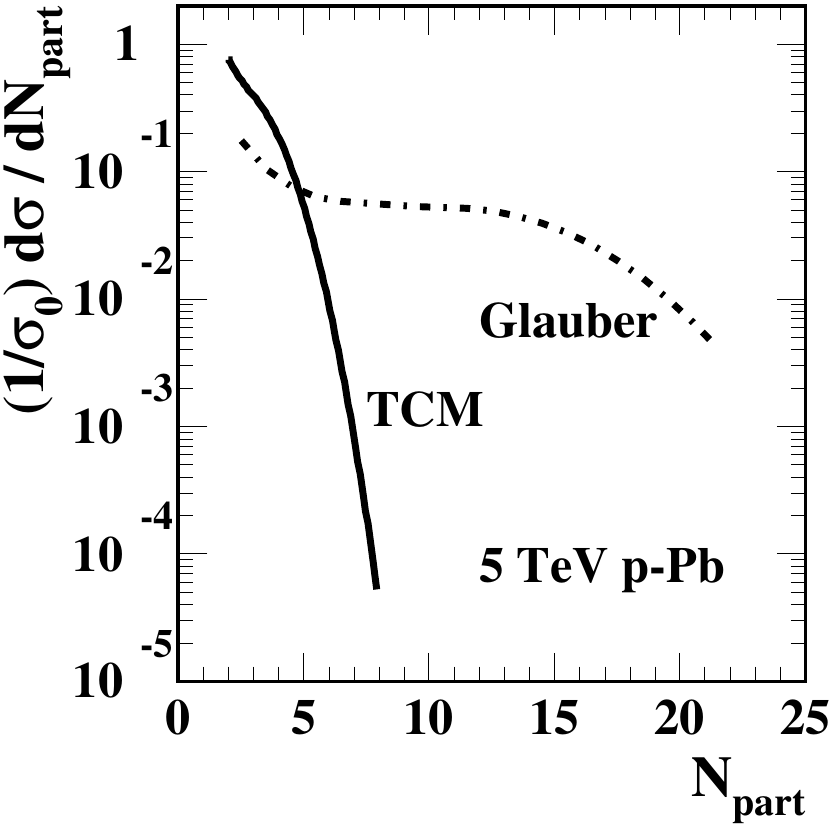}
 \includegraphics[width=1.65in]{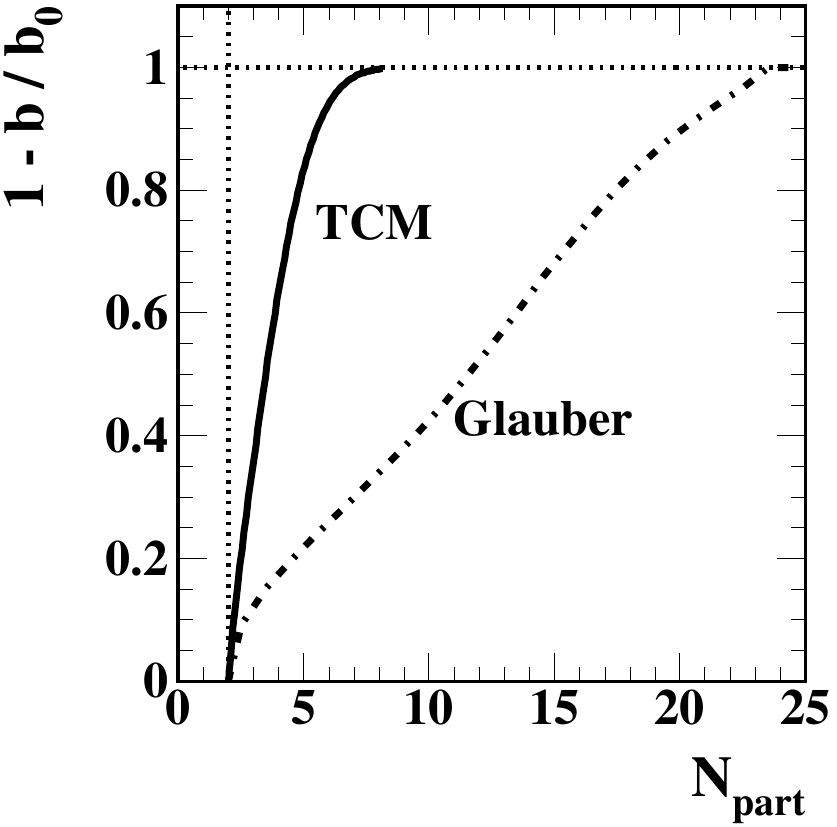}
  \caption{\label{glauber2}
  Left: Differential cross sections on $N_{part}$ derived from the geometric Glauber MC (dash-dotted) and as modified with Eq.~(\ref{fac}) as a factor to represent the TCM (dashed).
  Right: Running integrals $1 - \sigma / \sigma_0$ of curves on the left converted to $1 - (b/b_0 = \sqrt{ \sigma / \sigma_0})$.
   } 
 \end{figure}

Figure~\ref{glauber2} (left) repeats Fig.~\ref{glauber1} (left) now including the TCM cross-section distribution on $N_{part}$ (solid) as modified by Eq.~(\ref{fac}). Transformation of that curve with the Jacobian in Fig.~\ref{ppbmult6} (right, solid) gives the TCM solid curves in Fig.~\ref{ppbmult4}, suggesting that the true maximum participant number in central \ppb\ collisions is about 8.

Figure~\ref{glauber2} (right) shows running integrals of the differential distributions in the left panel transformed to fractional impact parameter $b/b_0 = \sqrt{\sigma / \sigma_0}$. Whereas the Glauber trend (dash-dotted) extends out to $N_{part} \approx 24$ the TCM trend (dashed), consistent with \mmpt\ data, extends to less than 8 (or $N_{bin} < 7$). For central \pbpb\ or \auau\ collisions the mean $N_{bin}$ per participant nucleon (i.e.\ in each projectile nucleus) is $\nu < 6$~\cite{anomalous}.

\begin{figure}[h]
  \includegraphics[width=1.65in]{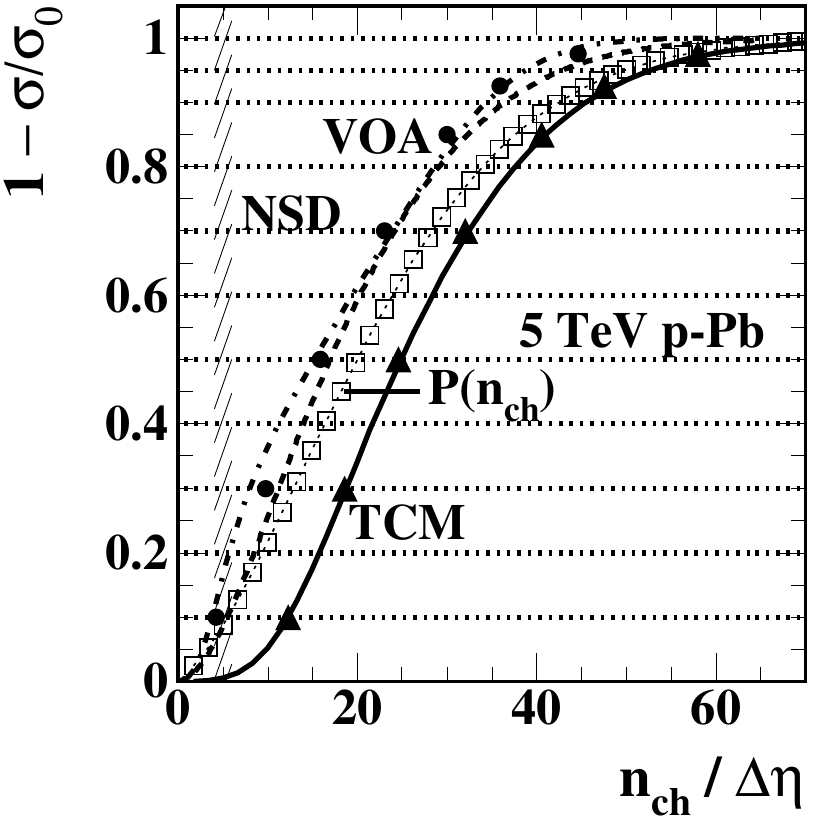}
  \includegraphics[width=1.65in]{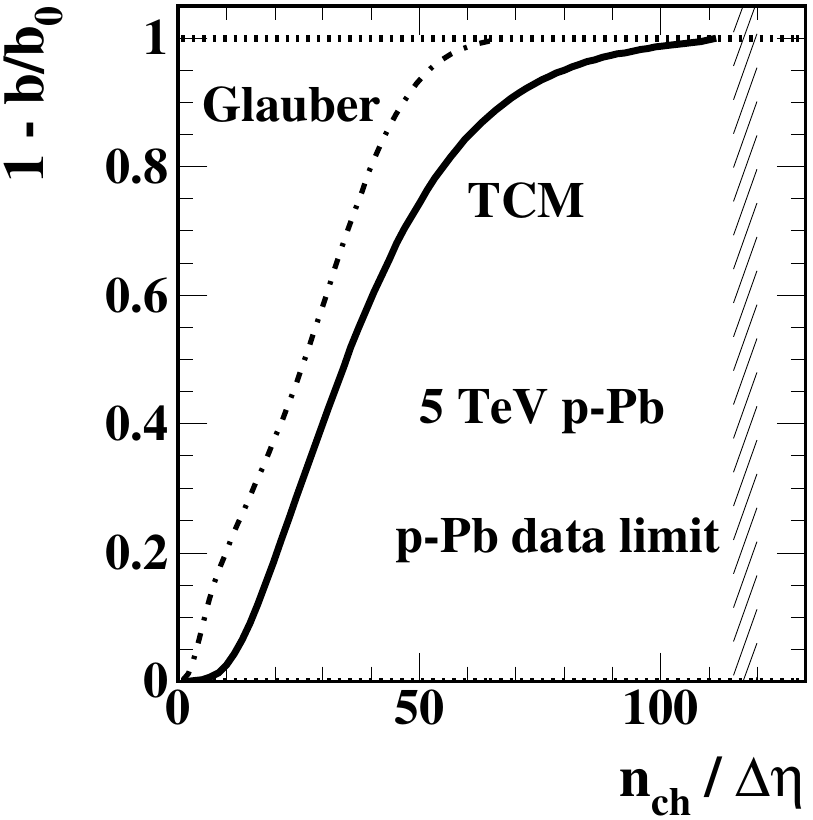}
  \caption{\label{ppbmult4a}
Left:  Running integrals of $d\sigma/d\bar \rho_0$ and $P(n_{ch})$ distributions in Fig.~\ref{ppbmult4} with corresponding line styles. Solid points are from Table~\ref{ppbparams1}. Solid triangles are explained in the text.
 Right: Impact-parameter $b / b_0 = \sqrt{\sigma / \sigma_0}$ trends on $\bar \rho_0$ illustrating extension of TCM centrality variation out to the limits of \mmpt\ data from Ref.~\cite{tommpt} (hatched band).
   } 
 \end{figure}

\begin{figure}[h]
  \includegraphics[width=1.63in]{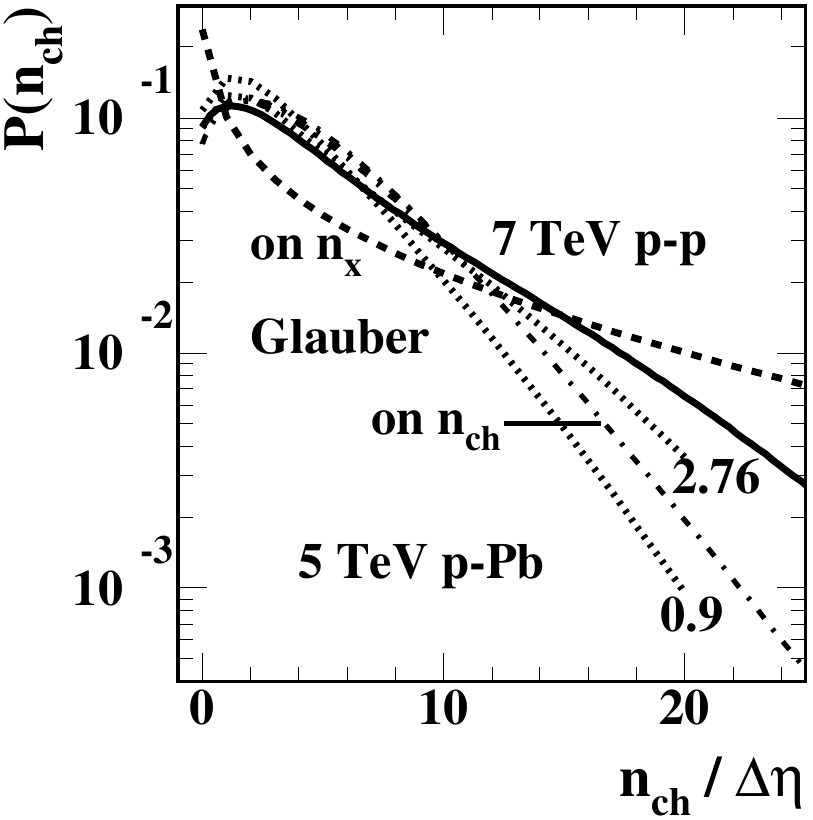}
 \includegraphics[width=1.67in]{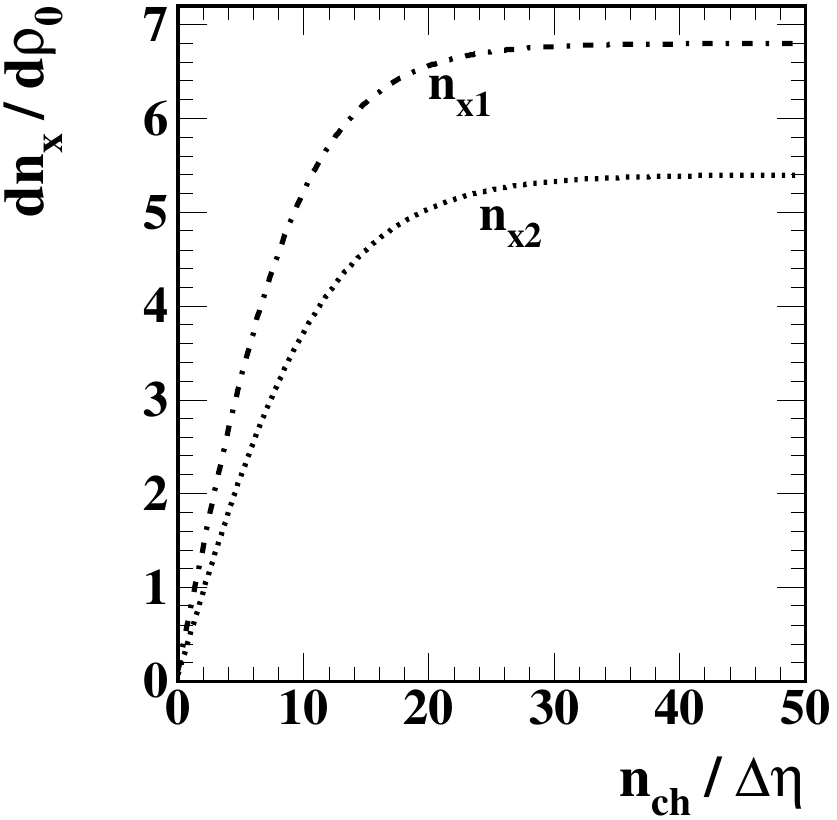}
  \caption{\label{ppmult}
  Left:  \pp\ multiplicity distributions shown in  Fig.~\ref{ppbmult1} (left) above. The dash-dotted curve on \nch\ is the V0A curve on $n_x$ (dashed) transformed with the Jacobian labeled $n_{x1}$ in the right panel.
Right: Jacobians derived from Eqs.~(\ref{nchnx}).
   } 
 \end{figure}

Figure~\ref{ppbmult4a} (left) shows running integrals of $d\sigma/d\bar \rho_0$ and $P(n_{ch})$ distributions in Fig.~\ref{ppbmult4} with corresponding line styles. The dashed V0A and dash-dotted Glauber MC trends contrast with the TCM trend (solid). Differences among V0A, Glauber, $P(n_{ch})$ and TCM distributions are discussed further in Sec.~\ref{laplace}. The horizontal dotted lines are centrality values defined in Ref.~\cite{aliceppbprod}. The solid dots are Glauber $\bar \rho_0'$ data from Table~\ref{ppbparams1} consistent with Glauber and V0A curves, although there are substantial differences between the two differential curves in Fig.~\ref{ppbmult4} (left). The triangles are corresponding TCM  $\bar \rho_0$ values lying on the solid curve. The TCM centrality bin centers are systematically shifted to larger $\bar \rho_0$.

Figure~\ref{ppbmult4a} (right) shows impact-parameter $b / b_0$ trends on $\bar \rho_0$. Whereas the Glauber cross section is already fully integrated on \nch\ ($b/b_0 \rightarrow 0$) near $\bar \rho_0 = 60$ the TCM centrality trend attains  its $b/b_0 \rightarrow 0$ limit only  near $\bar \rho_0 = 115$ (the upper limit for \mmpt\ data from Ref.~\cite{alicempt}). 

Figure~\ref{ppbmult4a} demonstrates that TCM \ppb\ centrality solves two problems arising from the Glauber analysis: (a) In an interval about $\bar \rho_{0} \approx \bar \rho_{0NSD}$ where \ppb\ $\approx$ \pn, $1 - \sigma / \sigma_0 \approx 0$ should persist as indicated by the TCM  trend in the left panel, but the Glauber equivalent rises rapidly. (b) Significant centrality variation should extend out to $\bar \rho_0 \approx 115$ consistent with \mmpt\ data from Ref.~\cite{alicempt}, as indicated by the TCM $b/b_0$ trend in the right panel, whereas the Glauber trend terminates near 60.

The structure of the \pn\ $P(n_{ch})$ distribution within \ppb\ collisions is critical for understanding \ppb\ centrality. In the Glauber study of Ref.~\cite{aliceppbprod} NBD parameters $(\mu,k) $ were inferred by fitting a convolution integral to the V0A $N_{evt}$ distribution on $n_x$ $P(n_x)$. Using the appropriate Jacobian that distribution can be transformed to \nch\ for direct comparison with measured \pp\ data.

Figure~\ref{ppmult} (left) repeats \pp\ multiplicity distributions shown in  Fig.~\ref{ppbmult1} (left) above. The solid curve is a double-NBD on mid-rapidity \nch\ fitted to 7 TeV \pp\ data with accuracy at the percent level~\cite{aliceppmult}. 
The value $\mu = 6$ is consistent with the NSD $\bar \rho_{0}$ value for 7 TeV. The dashed curve is the single NBD on $n_x$ resulting from Glauber analysis of 5 TeV \ppb\ V0A data described in Sec.~\ref{glaubresults}, with parameters $(\mu,k) = (11,0.44)$. The two distributions as plotted are very different, although Ref.~\cite{aliceppbprod} concludes that ``The values of the parameters $\mu$ and $k$ are similar to those obtained by fitting the corresponding multiplicity distributions in pp collisions at 7 TeV'' (text quoted in Sec.~\ref{details} above). But that statement contradicts the numbers above. Proper comparison requires transformation of the fitted density on $n_x$ to a density on midrapidity $\bar \rho_0$ with the correct Jacobian.

Figure~\ref{ppmult} (right) shows Jacobians $dn_x / d\bar \rho_0 $ obtained from Eqs.~(\ref{nchnx}). The bold dash-dotted curve on $\bar \rho_0$ in the left panel results from transforming the dashed curve on $n_x$ with the $n_{x1}$ Jacobian in the right panel. The transformed NBD is qualitatively similar to the 7 TeV \pp\ solid curve, and a comparison is now legitimate. The bold  curve terminates at $\bar \rho_0 \approx 10$ ($n_x \approx 30$) due to limitations in a routine to compute gamma functions. The thinner dash-dotted curve is a single NBD on $\bar \rho_0$ fitted to the bold curve with parameters  $(\mu,k) = (5,2)$. The value $\mu = 5$ is consistent with NSD  $\bar \rho_{0} \approx 5$ for 5 TeV \pp\ collisions.

The important difference lies with the $k$ values. Multiplicity fluctuations can be measured by relative variance {\em excess} (compared to Poisson) $\sigma_n^2 / \bar n - 1 = \mu / k$ for an NBD distribution. Detailed measurements at LHC energies~\cite{aliceppmult} reveal that \pp\ collisions exhibit large \nch\ variations. Charge densities $\bar \rho_0 \rightarrow 10\, \bar \rho_{0NSD} \approx 60$ were reported in Ref.~\cite{aliceppmult}. The 7 TeV NBD describing data has $\mu / k \approx 6 / 1.1 = 5.5$ whereas the NBD inferred from the Glauber analysis has $\mu / k \approx 5 / 2 = 2.5$, a factor 2.3 smaller although the collision energies are close. The large difference is manifested in the tails of the distributions: the Glauber fit is an {\em order of magnitude lower} than the measured \pp\ NBD tail near $\bar \rho_0 = 25$ (a region quite relevant to the \ppb\ analysis). The $n_{x2}$ Jacobian would result in greater reduction.
The assumptions in Ref.~\cite{aliceppbprod} noted in Sec.~\ref{scaling} are equivalent to assuming that average  \nn\ encounters for any \ppb\ \nch\ condition are equivalent to the \pp\ NSD mean. 

A consequence of that assumption is inference of a \pn\ multiplicity distribution [from the fit to V0A $P(n_x)$] with suppressed large-\nch\ tail sharply deviating from \pp\ data. Double-NBD curves (dotted) fitted to 0.9 TeV and 2.76 TeV \pp\ data~\cite{aliceppmult} are shown for comparison. The Glauber-inferred NBD for 5 TeV \ppb\ collisions is similar at larger $\bar \rho_0$ to a NBD  for \pp\ collisions near 1 TeV. The suppressed tail on \pn\ $P(n_x)$  is a consequence of overestimating the \ppb\ large-$N_{part}$ tail in Fig.~\ref{glauber2} (left).

\subsection{Joint probability distribution $\bf P(n_{ch},N_{part})$}

The differences between Glauber and TCM analyses and proposed mechanisms can be further explored by plotting joint probability distributions (normalized event-frequency distributions) $P(N_{part},\bar \rho_0)$ factorized as
\bea
P(N_{part},\bar \rho_{0NN}) &=& P(\bar \rho_{0NN}|N_{part}) (1/\sigma_0) d\sigma / dN_{part}~~~~~
\eea
with $\bar \rho_{0NN} = (2/N_{part}) \bar \rho_0$.  To simplify the exercise it is assumed that $P(\bar \rho_{0NN}|N_{part}) \approx P(\bar \rho_{0})$ is described by Glauber-inferred (dash-dotted) or \pp-data (solid) curves on $\bar \rho_0$ in Fig.~\ref{ppmult} (left)  and $(1/\sigma_0) d\sigma / dN_{part}$ is described by Glauber or TCM curves on $N_{part}$ in Fig.~\ref{glauber2} (left).

\begin{figure}[h]
  \includegraphics[width=3.3in]{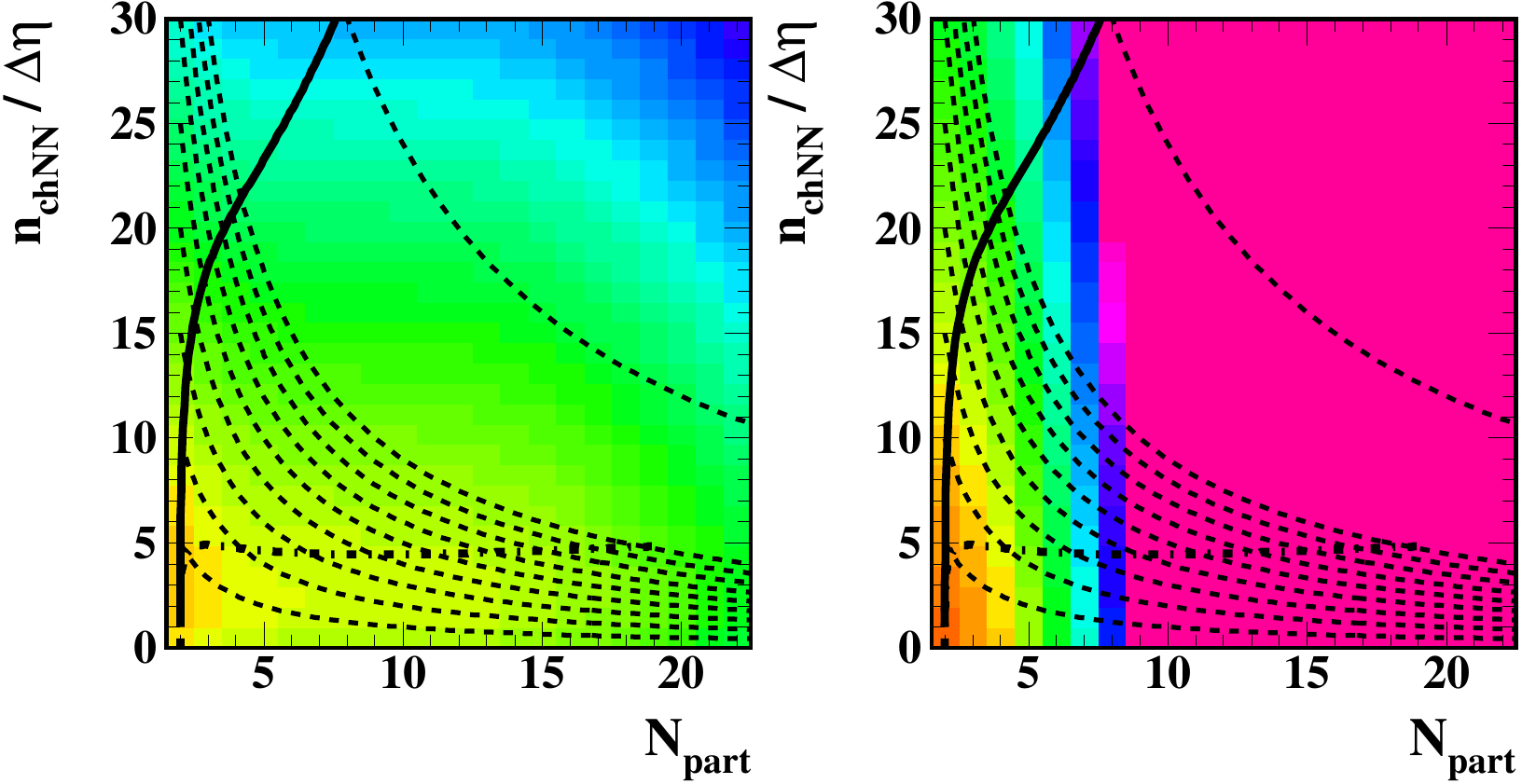}
  \caption{\label{jointprob}
 (Color online) Left:  Joint probability distribution $P(N_{part},\bar \rho_{0NN})$ representing the Glauber analysis of Ref.~\cite{aliceppbprod}. The $z$ axis is logarithmic over interval ($10^{-8}$,0.1).
  Right:  $P(\bar \rho_{0NN},N_{part})$ constructed with TCM versions of $P(\bar \rho_{0})$ and $d\sigma / dN_{part}$.  The $z$ axis is logarithmic over the same interval as the left panel.
   } 
 \end{figure}

Figure~\ref{jointprob} (left) shows joint probability distribution $P(N_{part},\bar \rho_{0NN})$ representing the Glauber analysis. The $z$ axis is logarithmic over interval ($10^{-8}$,0.1). The dashed curves are constraint loci defined by $\bar \rho_{0} = \bar \rho_{0NN} N_{part}/2$ with $\bar \rho_0$ increasing  in integral multiples of $\bar \rho_{0NSD}$ from 5 to 45. The right-most dashed curve corresponds to $\bar \rho_0 = 115$, the limiting value for the \mmpt\ analysis reported  in Ref.~\cite{alicempt} and appearing in Figs.~\ref{padata} and \ref{ppbmult2} (right). 
The solid curve is the TCM curve in Fig.~\ref{tcmcomp1} (right) plotted on $N_{part}$ rather than $\bar \rho_0$. The dash-dotted curve for the Glauber analysis has the same relation to Fig.~\ref{tcmcomp1} (right). 

The solid (TCM) and dash-dotted (Glauber) curves predict the most probable values or  {\em locus of modes} (approximately the mean values) of fluctuating $\bar \rho_{0NN}$ and $N_{part}$ given the ``centrality'' constraint $\bar \rho_0$ on the \ppb\ collision system. Up to $\bar \rho_0 \approx 10$ the actual locus appears to follow the most-probable point for each dashed curve, which corresponds to $N_{part} \approx 2$. Beyond that point it could be argued that probability values at larger $N_{part}$ and smaller $\bar \rho_{0NN}$ (the Glauber locus) may be greater. 

Figure~\ref{jointprob} (right) shows  $P(\bar \rho_{0NN},N_{part})$ constructed with TCM versions of $P(\bar \rho_{0})$ and $d\sigma / dN_{part}$.  The $z$ axis is logarithmic over the same interval as the left panel. The dramatic difference from the left panel is apparent and the Glauber locus is rejected here. The TCM locus, inferred from \mmpt\ data as in Ref.~\cite{tommpt}, follows the most probable points up to large $\rho_0$ and  intercepts the curve for $\bar \rho_0 = 115$ at $N_{part} \approx 7.5$ and $\bar \rho_{0NN} \approx 32$.  Note that \pp\ data for $P(n_{ch})$ in Ref.~\cite{aliceppmult} extend out to $\bar \rho_{0NN} \approx 60$ with $N_{part} \equiv 2$.  The large range of the TCM joint distribution on $\bar \rho_{0NN}$ is thus consistent with \pp\ data, whereas the much smaller range in the left panel is a direct consequence of the  broad distribution on $N_{part}$ required by the geometric Glauber MC. \ppb\  \mmpt\ data and \pp\ multiplicity distributions thus provide compelling evidence to support the alternative TCM-based centrality analysis.

\subsection{$\bf d\sigma$ vs $\bf dN_{evt}$ and Laplace's approximation} \label{laplace}

The distributions in Fig.~\ref{ppbmult4} are in principle not compatible within the same format. Normalized event-frequency distributions $P(n_{ch}) \rightarrow P(\bar \rho_0)$ (e.g.\ dashed curve, open squares) should be distinguished from differential cross sections $(1/\sigma_0)d\sigma / d\bar \rho_0$ (dash-dotted and solid curves). The difference requires distinction between event frequency $N_{evt}$ and centrality-related cross section $\sigma$: proportionality $d\sigma \propto dN_{evt}$ is not generally valid.


In Fig.~\ref{glauber2} (left) cross section $\sigma(N_{part})$ depends solely on $N_{part}$. Its relation to observed density $\bar \rho_0$ results only indirectly from the relation $N_{part}(\bar \rho_0)$ describing the  locus of modes for the $N_{evt}$  density on the space $(N_{part},\bar \rho_0)$.  The cross section is simply transformed to a density on $\bar \rho_0$ with a Jacobian (in Fig.~\ref{ppbmult6}, right, solid curve) as
\bea \label{sigpart}
\frac{d\sigma}{d\bar \rho_0} &=&
 \frac{dN_{part}}{d\bar \rho_0}
 \frac{d\sigma}{dN_{part}} 
\eea
which gives the solid curves in Fig.~\ref{ppbmult4}.  With \nch\ fixed or integrated the relation $d\sigma \propto dN_{evt}$ is valid on $N_{part}$. 

In contrast, $N_{evt}(N_{part},\bar \rho_0)$ is a function of (at least) two variables. The $N_{evt}$ {\em marginal} density $P(\bar \rho_0) = (1/N_0)dN_{evt}/d\bar \rho_0$ is given by the convolution integral
\bea \label{fold}
P(\bar \rho_0) &=& \int dN_{part} P(\bar \rho_0|N_{part}) P(N_{part}),
\eea
where $P(N_{part}) \equiv (1/\sigma_0) d\sigma / dN_{part}$. Equation~(\ref{fold}) has the same structure as  evidence $E$ in Bayesian inference defined by the first line of Eq.~(5) in Ref.~\cite{tombayes}. The second line of  Eq.~(5) introduces Laplace's approximation and may be reformulated in the present context as
\bea \label{poisson}
P(\bar \rho_0)  &\approx & \int_0^{\bar \rho_0} d\bar \rho_0' P(\bar \rho_0'|\tilde N_{part}) \left[\frac{dN_{part}}{d\bar \rho_0} P(N_{part})\right]_{\tilde N_{part}}~~
\eea
where $P(\bar \rho_0'|\tilde N_{part})$ is equivalent to the Bayesian {\em likelihood}, and the quantity in square brackets is the differential cross section on $\bar \rho_0$ evaluated at $\tilde N_{part}(\bar \rho_0)$ on the locus of modes for a given $\bar \rho_0$. The first factor is effectively a running integral of hadron production model $P(\bar \rho_0|N_{part})$ which distinguishes the form of $dN_{evt}/d\bar \rho_0$ from the form of $d\sigma / d\bar \rho_0$, why $dN_{evt} \propto d\sigma$ is not valid in that context. 

The various relations can be illustrated more simply by assuming that the locus of modes proceeds linearly down to lower limit $N_{part} = 2$ at $\bar \rho_{0NN} \approx 3\, \bar \rho_{0NSD} \approx 15$. As  $\bar \rho_{0}$ increases from zero to the endpoint of the locus of modes the running integral increases accordingly, with the factor in square brackets fixed at the value for lower limit $N_{part} \equiv 2$ and $\sigma/\sigma_0 \equiv 1$. Above that point $N_{part} \rightarrow \tilde N_{part}(\bar \rho_0)$  varies with $\bar \rho_0$ according to the locus of modes. The running integral should then remain at a fixed value $P[\bar \rho_0|\tilde N_{part}(\bar \rho_0)]$ while the quantity in square brackets varies in accord with  $\tilde N_{part}(\bar \rho_0)$. Within that interval $dN_{evt} \propto d\sigma$ is a good approximation.  Given that argument interpreting an event-frequency distribution, e.g.\ V0A distribution $P(n_x)$   in Fig.~\ref{ppbvoa}, as a differential cross section would produce systematic errors in the assigned cross-section intervals, as in Fig.~\ref{ppbmult4a} (left).



\begin{figure}[h]
   \includegraphics[width=1.65in]{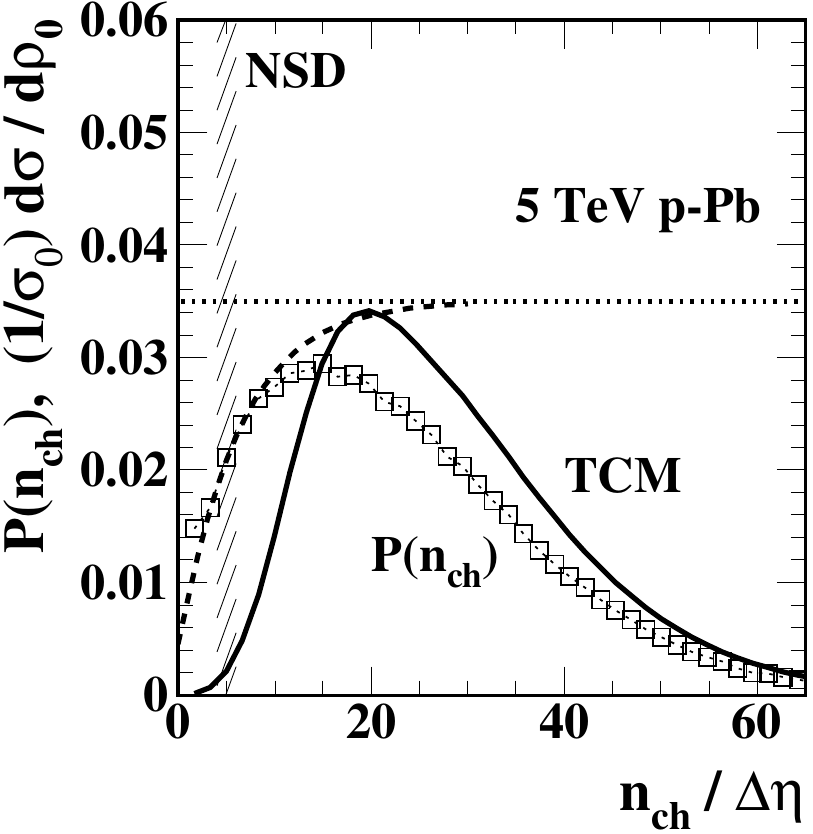}
 \includegraphics[width=1.65in]{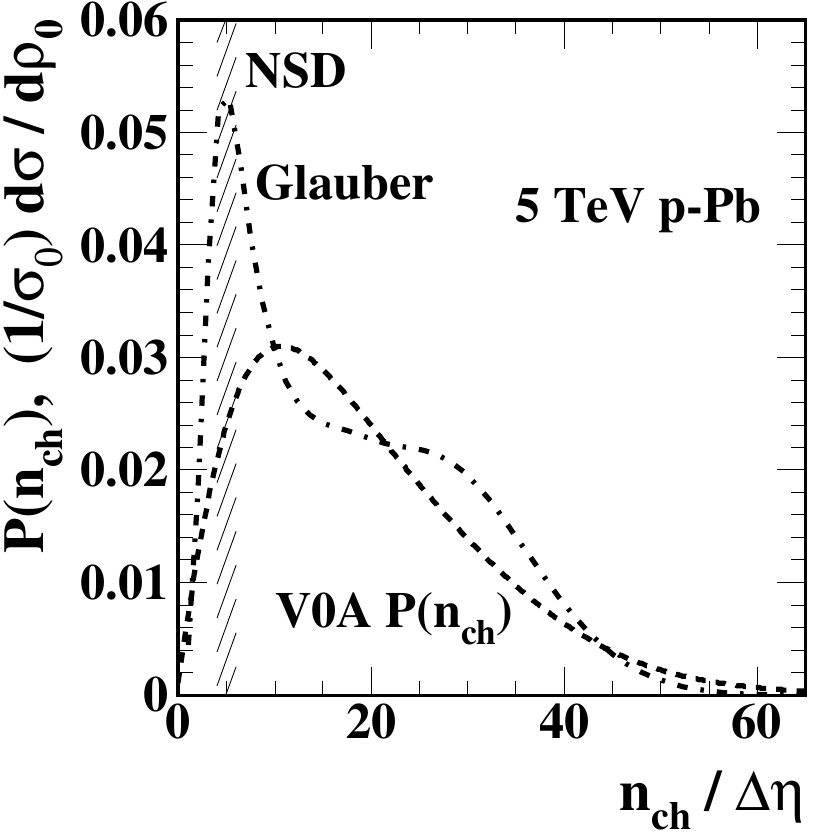}
  \caption{\label{invplot}
Left:  \ppb\ TCM differential cross section (solid curve) derived from Ref.~\cite{alicempt} \mmpt\ data and event distribution $P(n_{ch})$ derived from \mmpt\ statistical errors (open squares).
Right:  Glauber cross-section and V0A $P(n_{ch})$ trends.
   } 
 \end{figure}

Figure~\ref{invplot} (left) shows the \ppb\ TCM differential cross section (solid curve) obtained from Ref.~\cite{alicempt} \mmpt\ data and event distribution $P(n_{ch})$ derived from \mmpt\ statistical errors (open squares). Also plotted is a running integral (dashed curve) of the 7 TeV solid curve in Fig.~\ref{ppmult} (left) approximating 5 TeV \pp\ $P(n_{ch})$ multiplied by 0.035, the maximum value of the cross-section curve [corresponding to the limiting value of the term in square brackets in Eq.~(\ref{poisson})]. Correspondence with the $P(n_{ch})$ data is good.

Figure~\ref{invplot} (right) shows the Glauber cross-section and V0A $P(n_{ch})$ trends for comparison. The most important difference is between the TCM differential cross section (solid) and Glauber cross section (dash-dotted) at small \nch. The Glauber cross-section peak implies  that \ppb\ centrality is varying with \nch\ {\em most rapidly} near $\bar \rho_0 = \bar \rho_{0NSD}$ corresponding to {\em isolated} \pn\ collisions, whereas the TCM cross-section trend reasonably predicts little change in centrality in that most-peripheral region.

Equation~(\ref{fold}) can also be addressed as an inverse problem~\cite{inverse} wherein $dN_{evt} / d\bar \rho_0$ is the ``data,'' $d\sigma / dN_{part}$ is the ``model,'' and $P(\bar \rho_0|N_{part})$ relating $\bar \rho_0$ to $N_{part}$ is the ``kernel.'' An inverse problem is solved if given data and kernel the model can be inferred. For the analysis of Ref.~\cite{aliceppbprod} a \ppb\ geometry model was adopted from a Glauber MC based on certain assumptions. The relation of $\bar \rho_0$ to $N_{part}$ (hadron production model, the kernel) was also assumed and the resulting convolution integral was compared to V0A $P(n_{ch})$ data to ``validate'' the model. But such a solution is not unique. In the alternative TCM analysis of Sec.~\ref{details} the kernel was derived from \pp\ multiplicity data with the relation of $\bar \rho_0$ to $N_{part}$ determined by \ppb\ \mmpt\ data. The geometry model (cross-section distribution) was then inferred from the ``data'' by trial-and-error inversion. The TCM result is arguably unique because of the several direct contacts with measured data.

\section{$\bf p$-$\bf Pb$ $\bf p_t$-spectrum ratios} \label{ratios}

The abstract of Ref.~\cite{aliceppbprod} includes the statement ``Furthermore, at high-$p_T$ the p-Pb  spectra are found to be consistent with the pp spectra [when] scaled by $N_{coll}$ for all centrality classes...'' implying that defined spectrum ratio $Q_\text{pPb}(p_t)$ remains close to 1 above some \pt\ value, independent of \nch. The \ppb\ spectrum data are reconsidered here in the context of the present study.

\subsection{A TCM for spectrum ratios}

In the context of the \pt\ spectrum TCM of Ref.~\cite{alicetomspec} the spectrum ratio defined in  Ref.~\cite{aliceppbprod} can be expressed as 
\bea \label{qpa}
Q_\text{pPb}' &\equiv& \frac{1}{N'_{bin}} \frac{(N_{part}/2) \bar \rho_{sNN} \hat S_0(y_t) + N_{bin} \bar \rho_{hNN} \hat H_0(y_t) }{\bar \rho_{spp} \hat S_0(y_t) +  \bar \rho_{hpp} \hat H_0(y_t) }
 \nonumber\\
&=&  \frac{N_{bin}}{N'_{bin}} \frac{1}{\nu} \frac{\bar \rho_{sNN}\hat S_0(y_t) + \nu \bar \rho_{hNN} \hat H_0(y_t) }{ \bar \rho_{spp}\hat S_0(y_t) +  \bar \rho_{hpp} \hat H_0(y_t) }
\\ \nonumber 
&=&  \frac{N_{bin}}{N'_{bin}} \frac{1}{\nu} \frac{\bar \rho_{sNN}}{\bar \rho_{spp}} \frac{1 + \nu x_{NN} T_0(y_t) }{ 1 +  x_{pp}  T_0(y_t) }
\\ \nonumber 
&\rightarrow & \frac{N_{bin}}{N'_{bin}} \frac{1}{\nu} \frac{\bar \rho_{sNN}}{\bar \rho_{spp}}~~~(\text{low \pt\ $\Rightarrow$ LO})
\\ \nonumber 
&\rightarrow& \frac{N_{bin}}{N'_{bin}}  \frac{\bar \rho_{sNN}x_{NN}}{\bar \rho_{spp}x_{pp}} \approx \frac{N_{bin}}{N'_{bin}}  \frac{\bar \rho_{sNN}^2}{\bar \rho_{spp}^2}~(\text{high \pt\ $\Rightarrow$ HI})
\eea
where the first line is based on TCM \pt\ spectrum models in Eqs.~(\ref{ppspectcm}) and~(\ref{a1x}) and the last  two lines assume  $x_{xx} = \alpha \bar \rho_{sxx}$ for \pp\ and all \pn\  within \pa\ collisions. The unprimed parameters are values from the \ppb\ TCM that accurately describe \mmpt\ data as in Sec.~\ref{ppb}. The $N_{bin}'$ represents Glauber-model values taken from Table~\ref{ppbparams1}.

Spectrum ratios from Ref.~\cite{aliceppbprod} are taken from Fig.~19 (lower left, V0A). The accessible information is the limiting values in $p_t < 0.5$ (LO) and $p_t > 10$ (HI) GeV/c. Variation of $Q_\text{pPb}(p_t)$ in the transition region from LO to HI within 1-4 GeV/c is determined by ratio $T_0(y_t) = \hat H_0(y_y) / \hat S_{0}(y_t)$~\cite{alicetomspec} -- whether quantity $ \nu x_{NN} T_0$ is $ \gg 1$  (HI) or $\ll 1$ (LO). Details might permit reconstruction of spectrum hard components as in Ref.~\cite{alicetomspec}, but ratio data plotted on linear \pt\ (as opposed to logarithmic transverse rapidity \yt) do not provide access to that information.
If the assumptions invoked for the Glauber analysis  were correct $N_{bin} / N'_{bin} \rightarrow 1$ and $\bar \rho_{sNN}/\bar \rho_{spp} \approx 1$. According to Eqs.~(\ref{qpa}) LO should then vary as $1/\nu$ and HI remain near 1, as for \aa\ collisions with no jet modification. 

The following analysis is based on the assumption that $\bar \rho_0$ values in Table~\ref{ppbparams1} determined by V0A centrality bins correspond to spectrum ratios in Fig.~19 (V0A, lower left) of Ref.~\cite{aliceppbprod}. For each $\bar \rho_0$ value the TCM centrality is determined from Fig.~\ref{ppbmult4a} (left) and the $N_{bin}$ and $\nu$ values from Fig.~\ref{ppmult}. Values of $\bar \rho_s$ are obtained from $\bar \rho_0$ by back transforming Eq.~(\ref{nchppb}) (third line) using the relations in Fig.~\ref{paforms}. The values of $\bar \rho_{sNN}$ then follow from Eq.~(\ref{rhosnn}).

Table~\ref{rppbdata} shows nominal (primed) and  TCM (unprimed) fractional cross sections and  Glauber parameters, midrapidity charge density $\bar \rho_0$, \nn\ soft component $\bar \rho_{sNN}$ and TCM hard/soft ratio $x_{NN}$ required to evaluate Eqs.~(\ref{qpa}) for comparison with spectrum ratios from Ref.~\cite{aliceppbprod}.

\begin{table}[h]
  \caption{Nominal (primed) and  TCM (unprimed) fractional cross sections and  Glauber parameters, midrapidity charge density $\bar \rho_0$, \nn\ soft component $\bar \rho_{sNN}$ and TCM hard/soft ratio $x_{NN}$ used to evaluate 5 TeV \ppb\ spectrum ratios.
}
  \label{rppbdata}
\begin{center}
\begin{tabular}{|c|c|c|c|c|c|c|c|c|} \hline
 $\sigma' / \sigma_0$ &   $\sigma / \sigma_0$    & $N_{bin}'$ & $N_{bin}$ & $\nu'$ & $\nu$ & $\bar \rho_0$ & $\bar \rho_{sNN}$ & $x_{NN}$ \\ \hline
0.025         & 0.15 & 14.7  & 3.20 & 1.87  & 1.52 & 44.6 & 16.6  & 0.188 \\ \hline
0.075  & 0.24 &  13.0   & 2.59  &  1.86 & 1.43 & 35.9 &15.9  & 0.180 \\ \hline
  0.15  & 0.37 &  11.7 & 2.16 & 1.84 &  1.37 & 30.0  & 15.2  & 0.172 \\ \hline
  0.30 & 0.58 &  9.4 & 1.70 & 1.80  & 1.26  & 23.0  & 14.1  & 0.159  \\ \hline
0.50   &0.80  & 6.42  & 1.31 & 1.73  & 1.13 & 15.8 &   12.1 & 0.137  \\ \hline
 0.70  & 0.95 & 3.81  & 1.07 & 1.58  & 1.03  & 9.7  &  8.7 & 0.098 \\ \hline
 0.90 & 0.99 &  1.94 & 1.00 & 1.32  & 1.00  &  4.4  & 4.2 &0.047  \\ \hline
\end{tabular}
\end{center}
\end{table}

\subsection{Trends from measured spectrum ratios}

\begin{figure}[h]
  \includegraphics[width=3.3in]{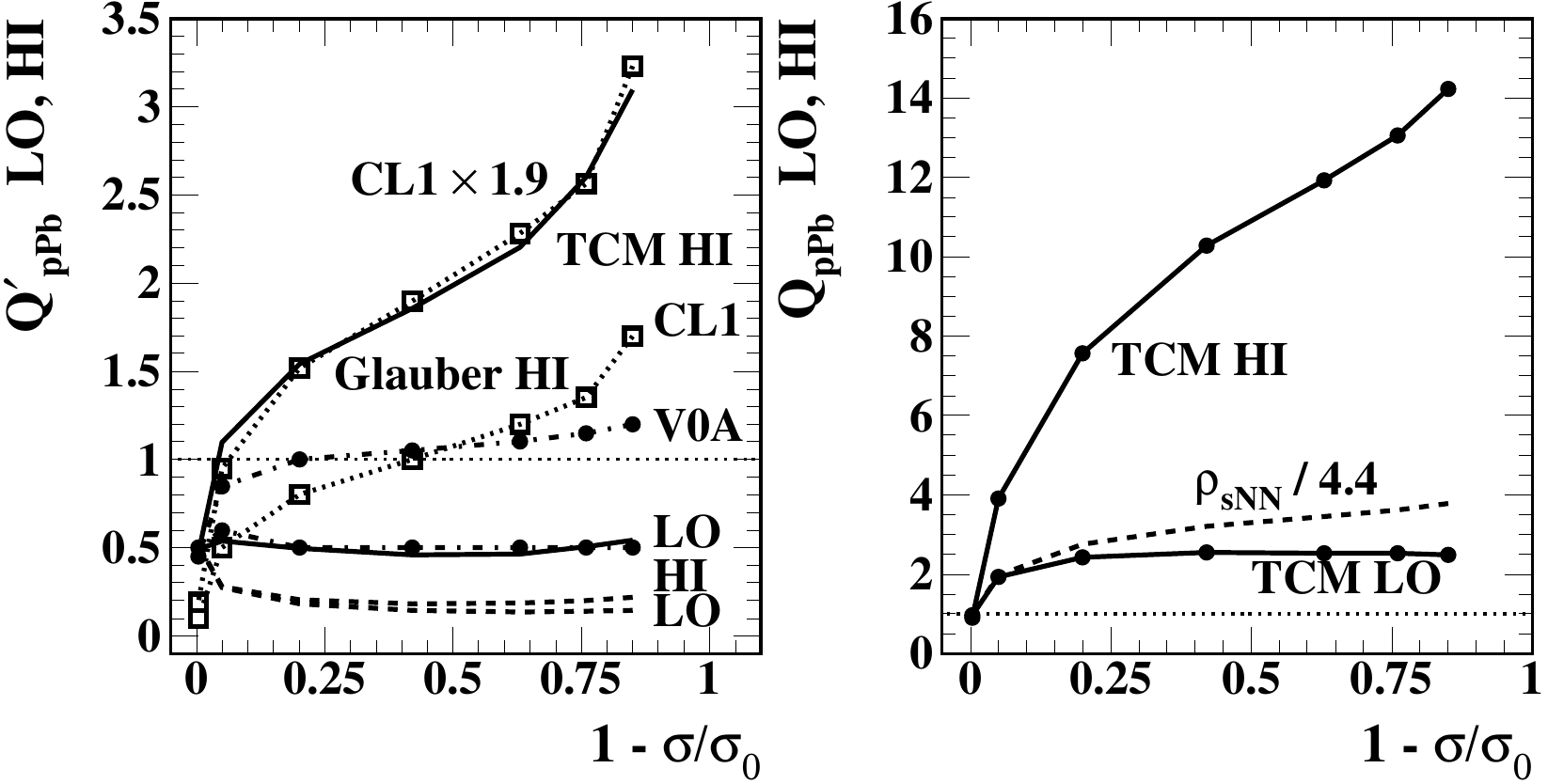}
  \caption{\label{glauber}
  Left:  $Q_\text{pPb}'$ (biased) LO and HI values (solid points, dash-dotted curves)  obtained from Fig.~19 (lower left) of Ref.~\cite{aliceppbprod} plotted vs unprimed (corrected) TCM centralities $\sigma / \sigma_0$ from Table~\ref{rppbdata}. Other curves are explained in the text.
  Right:  $Q_\text{pPb}$ trends (solid) that should be observed if $N_{bin}$ were obtained from the \ppb\ TCM as in Table~\ref{rppbdata} (unprimed).
   } 
 \end{figure}

Figure~\ref{glauber} (left)  shows $Q_\text{pPb}'$ (biased) LO and HI values (solid points, dash-dotted curves)  obtained from Fig.~19 (lower left) of Ref.~\cite{aliceppbprod} plotted vs unprimed (corrected) TCM centralities $\sigma / \sigma_0$ from Table~\ref{rppbdata}. The solid curves are corresponding TCM trends obtained from  Eqs.~(\ref{qpa}) using appropriate values from Table~\ref{rppbdata}. Whereas the LO values for TCM and Glauber correspond closely there is a large difference between HI values. The Glauber HI trend remains close to unity (as reported in Ref.~\cite{aliceppbprod}) but the TCM HI trend increases strongly up to $Q_\text{pPb}' \approx 3$. Lower open squares are CL1 (at midrapidity) HI values obtained from the upper-left panel in Fig.~19 of Ref.~\cite{aliceppbprod}. Those data, when multiplied by factor 1.9, correspond closely to the TCM values (also relevant to midrapidity). The dashed curves show the TCM result if $\bar \rho_{sNN} \approx \bar \rho_{spp}$ for all cases, as assumed for the Glauber \ppb\ analysis.

Figure~\ref{glauber} (right) shows Eqs.~(\ref{qpa}) with $N_{bin} ' \rightarrow N_{bin}$ and all else the same (solid curves); i.e.\ $Q_\text{pPb}$ trends that should be observed if $N_{bin}$ were inferred via the TCM. The dashed curve is ratio $\bar \rho_{sNN} / \bar \rho_{spp}$ appearing in Eqs.~(\ref{qpa})  with $\bar \rho_{spp} \rightarrow 4.4$ as in Fig.~\ref{tcmcomp1} (right) and $\bar \rho_{sNN}$ from Table~\ref{rppbdata}. Those results are compatible with the TCM \mmpt\ analysis of Ref.~\cite{tommpt} as in Sec.~\ref{ppb} where large \mmpt\ increases result from MB dijet production increasing quadratically with \pn\ (\nn) soft component $\bar \rho_{sNN}$.

Within the context of the Glauber \ppb\ centrality analysis and its assumptions one should expect $Q_\text{pPb} \approx 1$ for very peripheral \ppb\ (i.e.\ \pn\ $\approx$ \pp) collisions. LO should then decrease $\propto 1/\nu$ (i.e.\ toward 0.5 for \pa\ collisions) and HI should remain constant near unity with increasing centrality (e.g.\ V0A multiplicity). That the very peripheral values are near 0.5 rather than 1 is already notable in Fig.~19 of Ref.~\cite{aliceppbprod}. The TCM analysis reveals that the observed peripheral values correspond to $1/N_{bin}'  = 1/ 1.94  \approx 0.51$ -- the Glauber estimate for binary collision number  in {\em isolated} \pn\ collisions is  $N_{bin}\approx 2$. 

The \ppb\ TCM, in the form of Eqs.~(\ref{qpa}) and unprimed parameters in Table~\ref{rppbdata}, accurately {\em predicts} the observed Glauber LO trend in the left panel --  no parameters were adjusted to accommodate the $Q_\text{pPb}'$ data.  The large difference between TCM and Glauber HI trends is then difficult to explain, since according to Eqs.~(\ref{qpa}) the ratio HI/LO is simply the quantity $\nu\, \bar \rho_{sNN} / \bar \rho_{spp}$ and those factors, when included in  Eqs.~(\ref{qpa}), describe the measured LO trend accurately. It is also interesting to note that the CL1 HI trend from Fig.~19 of Ref.~\cite{aliceppbprod} follows the TCM HI trend closely when multiplied by constant factor 1.9.

The rapid increase of $Q_\text{pPb}$ HI for peripheral \ppb\ collisions is simply explained in terms of Fig.~\ref{ppbmult4a} (left): With an increasing \nch\ or $n_x$ ``centrality'' condition the multiplicity of very peripheral \ppb\ (i.e.\ individual \pn) collisions increases accordingly with little change in the actual \ppb\ centrality ($b/b_0 \approx 1$ persists). Peripheral \ppb\ basically follows the \pp\ \nch\ trend as in Sec.~\ref{ppb}. 

A general conclusion can be drawn from \ppb\ $Q_\text{pPb}$ data in the context of a TCM describing \mmpt\ data from the same collision system: In case jets are unmodified in \pa\ collisions and there {\em is} linear superposition of \pn\ collisions one should expect a large increase of $Q_\text{pPb}$ ($\gg 1$) in the high-\pt\ or HI region with increasing \nch\ due to quadratically increasing MB dijet production in \pn\ collisions. The LO region should increase modestly according to competition between decreasing $1/\nu$ and increasing (not static as assumed) $ \bar \rho_{sNN}$, both as in Fig.~\ref{glauber} (right). The unexpected results in Fig.~19 of Ref.~\cite{aliceppbprod} arise in part because of certain limitations in the Glauber analysis as described above, but other aspects remain unexplained.

\subsection{Identified-pion spectra for 5 TeV p-Pb collisions}

A critical test for the TCM description of $Q_\text{pPb}$ trends above can be established with published identified-pion spectra for 5 TeV \ppb\ collisions from Ref.~\cite{alicepionspec}.

\begin{figure}[h]
  \includegraphics[width=3.3in]{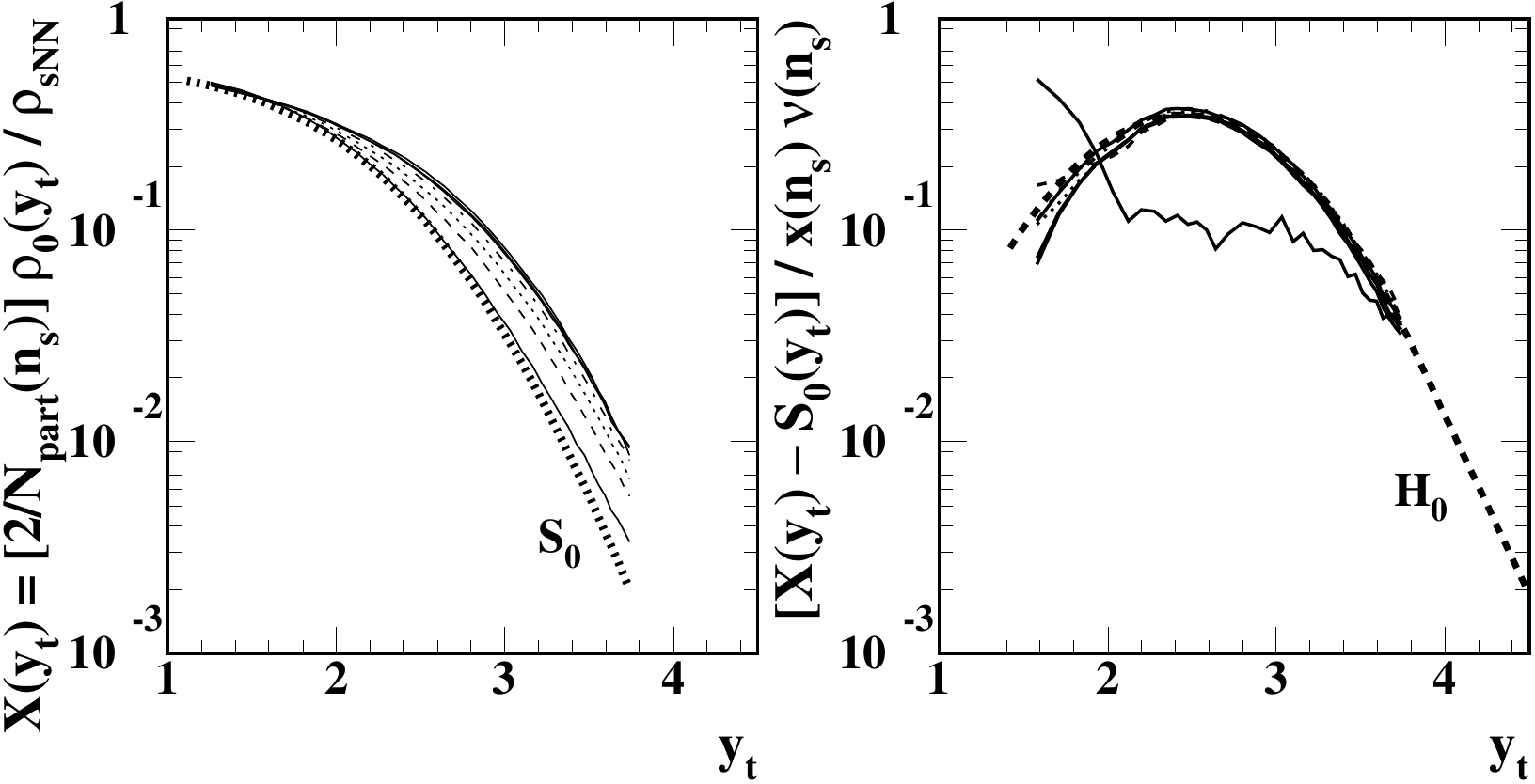}
  \caption{\label{qppb}
  Left: Identified-pion spectra for 5 TeV \ppb\ collisions from Ref.~\cite{alicepionspec} transformed to \yt\ with Jacobian $m_t p_t / y_t$ and normalized by TCM quantities in Table~\ref{rppbdata} (7 thinner curves of several styles). $\hat S_0(y_t)$ is the soft-component model.
  Right: Difference $X(y_t) - \hat S_0(y_t)$ normalized by $x(b) \nu(b) = \alpha \bar \rho_{sNN} \nu(b)$ using TCM values from Table~\ref{rppbdata} reported in Ref.~\cite{tommpt} (thinner curves). The dashed curve is hard-component model $\hat H_0(y_t)$ with exponential tail. The dotted curve is a Gaussian with the same parameters but no tail.
   } 
 \end{figure}

Figure~\ref{qppb} (left) shows identified-pion spectra from the same \ppb\ collision system considered throughout this article. The published spectra have been multiplied by $2\pi$ to be consistent with the $\eta$ densities used in this study and transformed to \yt\ with Jacobian $m_t p_t / y_t$. The spectra are then normalized by soft-component density $\bar \rho_s = (N_{part}/2) \bar \rho_{sNN}$ as reported in Table~\ref{rppbdata} following Eq.~(\ref{ppspectcm}) (third line) except that an additional factor 0.8 is applied to $\bar \rho_s$ values in Table~\ref{rppbdata} to reflect the pion fraction of soft hadrons. The normalized spectra $X(y_t)$ can then be compared with spectrum soft-component model $\hat S_0(y_t)$ shown as the bold dotted curve: a L\'evy distribution with parameters $T = 145$ MeV and $n = 8.3$ appropriate for 5 TeV \pp\ collisions as reported in Ref.~\cite{alicetomspec}.

Figure~\ref{qppb} (right) shows the difference $X(y_t) - \hat S_0(y_t)$ normalized by $x(n_s) \nu(n_s) = \alpha \bar \rho_{sNN} \nu(n_s)$ with TCM values reported in Ref.~\cite{tommpt} and Table~\ref{rppbdata}. There are no adjustments to accommodate the data. Given Eq.~(\ref{ppspectcm}) the result should be directly comparable to the \pp\ spectrum hard-component model in the form $\hat H_0(y_t)$ with model parameters  $(\bar y_t,\sigma_{y_t},q) = (2.65,0.59,3.9)$ for 5 TeV \pp\ collisions as reported in  Ref.~\cite{alicetomspec}. The dashed curve is $\hat H_0(y_t)$ with $(\bar y_t,\sigma_{y_t},q) \rightarrow (2.45,0.605,3.9)$. A shift to lower fragment momenta for pions is expected based on  Fig.~7 (left) of Ref.~\cite{eeprd}: pion FFs are softer than kaon FFs are softer than proton FFs.
The overall TCM spectrum description is well within point-to-point data uncertainties except for the lowest centrality class (solid curve) where the large deviation is expected based on Ref.~\cite{alicetomspec}. One may then conclude that the predictions for quantity $Q_\text{pPb}$ in Fig.~\ref{glauber} (right) are generally consistent with published \ppb\ pion spectrum data. The TCM description of \ppb\ spectra {\em assumes linear superposition} of \pn\ collisions within \ppb\ collisions. However, it also describes realistically the changing properties of \pn\ collisions depending on applied \ppb\ \nch\ condition. Those two panels can be compared with Figs.~1 in Refs.~\cite{ppquad,alicetomspec}. 

\subsection{Spectrum ratios for p-Pb identified pions}

The TCM HI and LO trends in Fig.~\ref{glauber} (right) describe limiting values. Using the pion \pt\ spectra in Fig.~\ref{qppb} data-model  comparisons for differential $Q_\text{pPb}(p_t)$ are possible.

\begin{figure}[h]
  \includegraphics[width=3.3in]{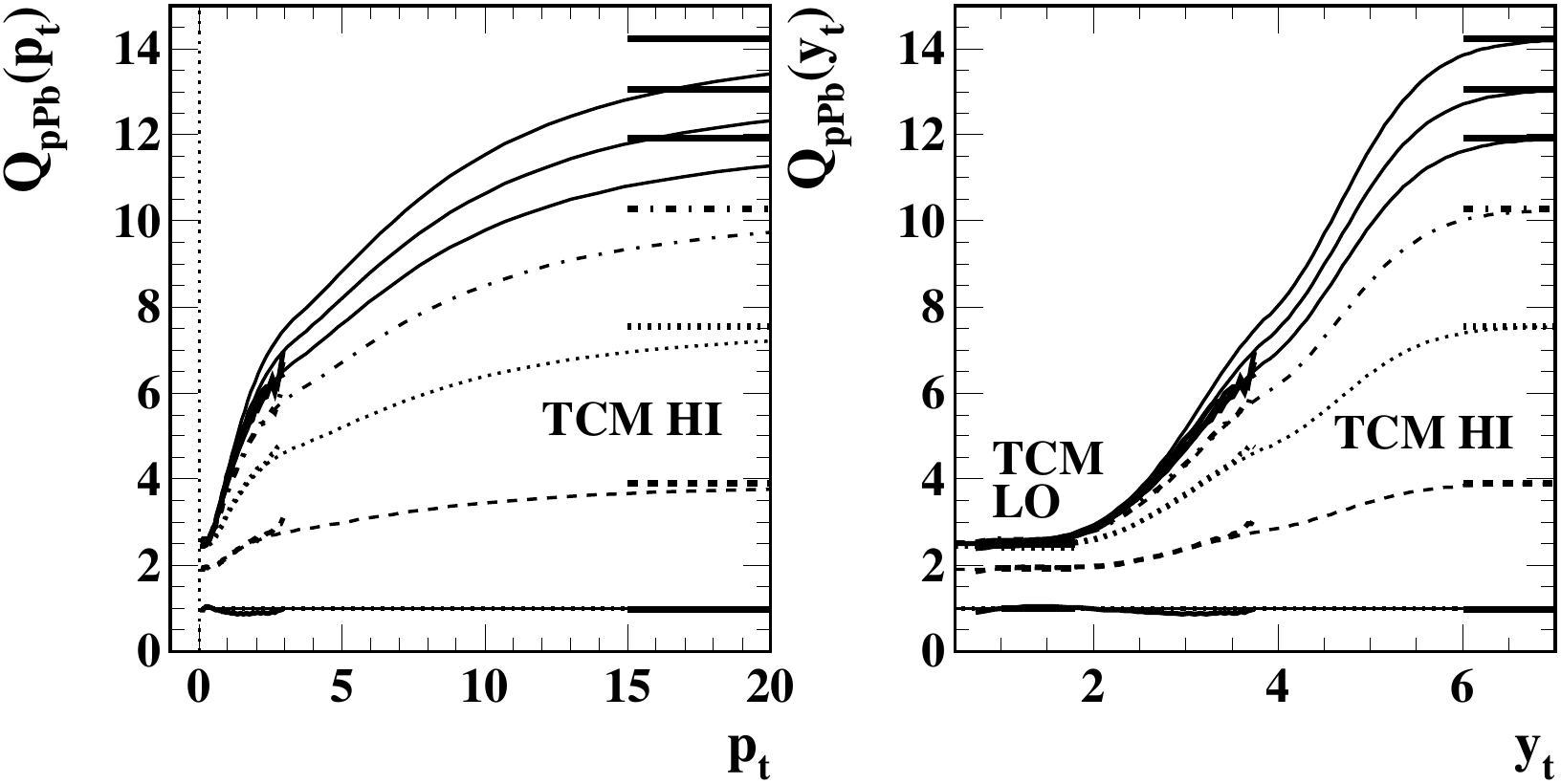}
  \caption{\label{qppb2}
  Left: Identified-pion spectrum ratios $Q_\text{pPb}(p_t)$ (bold curves of several line styles to 3 GeV/c). The most-peripheral TCM spectrum is adopted as the \pp\ reference. Bold horizontal lines indicate TCM HI values from Fig.~\ref{glauber} (right). Thinner curves extending to 20 GeV/c are TCM model ratios.
  Right:  The same data and curves plotted on logarithmic transverse rapidity \yt. Details in the LO region are now resolved. Note that $y_t = 6 \Rightarrow p_t \approx 28$ GeV/c and $y_t = 7 \Rightarrow p_t \approx 76$ GeV/c.
   } 
 \end{figure}

Figure~\ref{qppb2} (left) shows identified-pion spectrum ratios $Q_\text{pPb}(p_t)$ (bold curves of several styles to 3 GeV/c). The most-peripheral TCM spectrum is adopted as the \pp\ reference. Unprimed TCM $N_{bin}$ values from Table~\ref{rppbdata} are used for all ratios. 
For the most-central 5 TeV \ppb\ spectrum the ratio  $\alpha \rho_{sNN} \nu T_0(p_t)$ of hard/soft spectrum components  crosses 1 near $p_t = 1$ GeV/c and reaches only 5 at $p_t = 3$ GeV/c, the end of the published pion spectra. The soft component therefore contributes substantially at that endpoint and asymptotic HI limits are not attained.  The LO limits are not resolved in this plot format. The thinner TCM curves are discussed below.

Figure~\ref{qppb2} (right) shows the same data and curves plotted on logarithmic transverse rapidity \yt. The LO trends below 0.5 GeV/c ($y_t \approx 2$) are then clearly resolved, and the correspondence with the TCM LO predictions (bold lines with same line styles) is evident. The thinner curves are TCM predictions based on fixed TCM model functions $\hat S_0(y_t)$ and $\hat H_0(y_t)$ in Eq.~(\ref{qpa}) that appear in Fig.~\ref{qppb}. There is good agreement with the spectrum data, and the TCM ratios extrapolate to the predicted HI values

The systematic behavior of these spectrum ratios with increasing \ppb\ centrality is formally equivalent to the variation of \pp\ spectrum ratios with increasing \nch~\cite{alicetomspec}. That similarity is further evidence that $\bar \rho_{sNN}$ increases strongly with \ppb\ centrality as revealed by the TCM.

\section{Systematic uncertainties} \label{sys}

The main subject of this study is inference of \ppb\ centrality from certain aspects of measured data. This section considers systematic uncertainties for two competing methods. Some issues to consider: Are TCM results relating to \ppb\ collisions unique and reliable? Are Glauber results unique given related physical assumptions. Are those assumptions realistic? Do the results from either method ``make sense'' in relation to other data and collision systems? Are certain conjectured biases considered in the Glauber study relevant to data?

\subsection{p-Pb TCM uncertainties}

TCM uncertainties have been estimated for spectra and correlations in several previous studies~\cite{ppprd,hardspec,anomalous,alicetomspec,ppquad} and especially for the recent \mmpt\ analysis~\cite{alicetommpt}. TCM data descriptions are typically within data uncertainties.

The relevant issue for this study is the accuracy of hard/soft fraction $x(n_s)$ from which all other aspects of the \ppb\ TCM (e.g.\ $N_{part}$, $\nu$) are derived. According to \mmpt\ data from Ref.~\cite{alicempt} the \ppb\ TCM is essentially the \pp\ TCM below the transition point $\bar \rho_s \approx 3\, \bar \rho_{sNSD} \approx 15 \equiv \bar \rho_{s0}$ and inherits \pp\ uncertainties in  that interval which are described in Refs.~\cite{ppprd,ppquad,alicetomspec}. Within that interval $x(n_s) \approx \alpha \bar \rho_s$ is accurate to a few percent over a {\em 10-fold increase} in MB dijet production, implying that $N_{part} \approx 2$ within the same interval. Above $\bar \rho_{s0}$ the $x(n_s)$ model is a conjecture based on (a) continued monotonic increase with (b) the simplest form. The TCM then has two adjustable parameters with which to accommodate \mmpt\ data and does at the percent level as shown in Fig.~\ref{padata}.

In all cases, elements of  the TCM are derived based on data trends, simplicity and self-consistency. The TCM must describe or predict data from all available data systems or be falsified. No {\em ad hoc} elements are invoked. As noted, a key issue is the precise coincidence of \pp\ and \ppb\ \mmpt\ data over a  substantial $\bar \rho_0$ interval, implying that \ppb\ centrality does not change over that interval (wherein $b/b_0 \approx 1$ and $N_{part} \approx 2$). Whatever centrality method is invoked should arrive at that common result. 

\subsection{Geometric Glauber model -- general issues}

Estimation of systematic {uncertainties} relating to the Glauber analysis is problematic given the large deviations from published data and other analyses as demonstrated in the present study. Within the context of the Glauber analysis the key assumptions (a) applicability of a geometric Glauber model to \pa\ and (b) validity of the $n_{ch} \propto N_{part}$ assumption are of critical importance. 

Assumption (a) represents a limiting case: Any target nucleon within an eikonal corridor defined by the in-vacuum $\sigma_{pp}$ inelastic cross section {\em must} suffer a collision with the single projectile proton. But no data support that assumption. In symmetric \aa\ collisions there are many projectile nucleons, and the mean number of binary collisions per nucleon is less than 6. Most participants may be created by a projectile with no prior collisions. This assumption should be considered quite uncertain.

Assumption (b) is generally inconsistent with observations for more-central \aa\ collisions. Reference~\cite{phobosauaunpart} cited in support of that assumption warns  ``It should be noted that this simple scaling [$n_{ch} \propto N_{part}$] is not observed for [differential $dn_{ch}/d\eta$] particle yields measured in a limited pseudorapidity range near midrapidity'' as noted in Sec.~\ref{scaling}. The $n_{ch} \propto N_{part}$ trend, which describes the TCM soft component only, is violated by 80\% for more-central \pbpb\ collisions at 2.76 TeV as in Ref.~\cite{tommpt}, Sec.~V-A. The large deviation results from the MB dijet contribution with $n_{ch} \propto N_{bin}$. The validity of  this assumption should therefore be considered unlikely.  Further qualitative issues (a) - (f) are itemized in Sec.~\ref{summ}.

\subsection{Statistical bias and collision models}

A statistical quantity or statistic is biased if it returns a mean or median that is different from the true value for a given data population. In the case of centrality determination for instance values of $N_{part}$ or $b$ different from the true value for a given event sample would represent such a bias. A related aspect is internal consistency wherein statistical analysis of a given data population returns values consistent with one another whether biased or not, for instance mean values of $N_{part}$ for different subsets of an event ensemble following an expected trend.

Reference~\cite{aliceppbprod} refers to ``dynamical'' bias of centrality estimates based on particle multiplicities \nch\ due to large fluctuations (e.g.\ of \nch\ for fixed $N_{part}$). ``...centrality classification...based on [charge] multiplicity may select a sample of [\nn] collisions which is biased...'' (p.~16) and ``...generate a dynamical bias in centrality classes'' (abstract). A proposed measure of such bias as presented in Fig.~8 (left) of Ref.~\cite{aliceppbprod} is the quantity $R = \langle \text{Multiplicity} \rangle / [\langle N_\text{ancestor} \rangle \mu]$, where $N_\text{ancestor}$ (clan model) is apparently equivalent to $N_{part}$ according to the caption, and the data are obtained from the geometric Glauber MC coupled with NBD and fitted to various $n_x$ (e.g.\ V0A) distributions. Deviations of $R$ from unity are expected to indicate dynamical bias.

\begin{figure}[h]
  \includegraphics[width=1.65in]{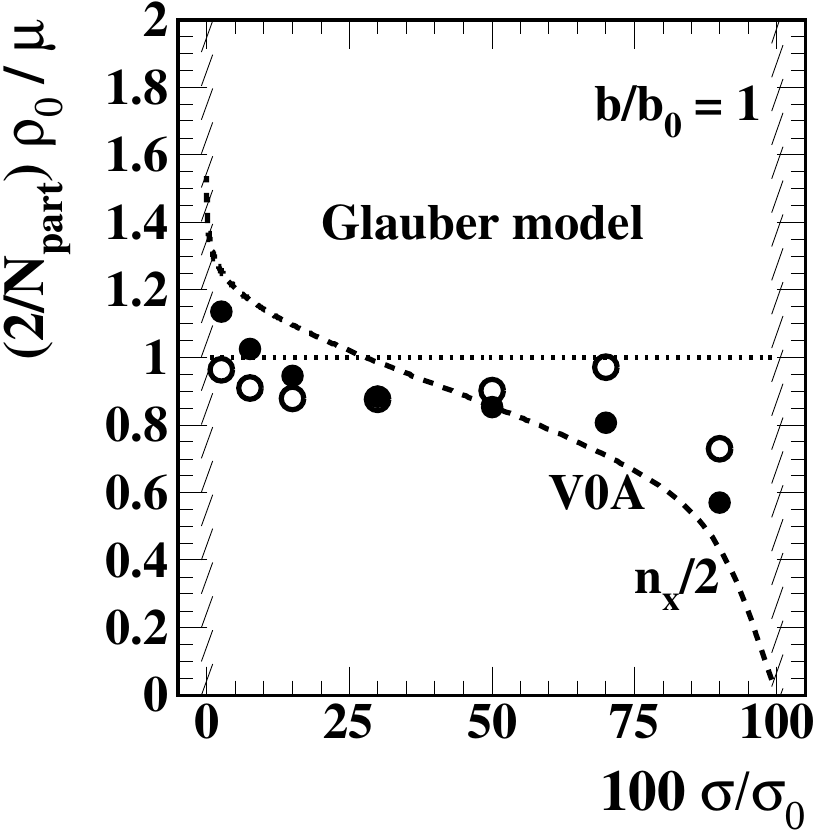}
  \includegraphics[width=1.65in]{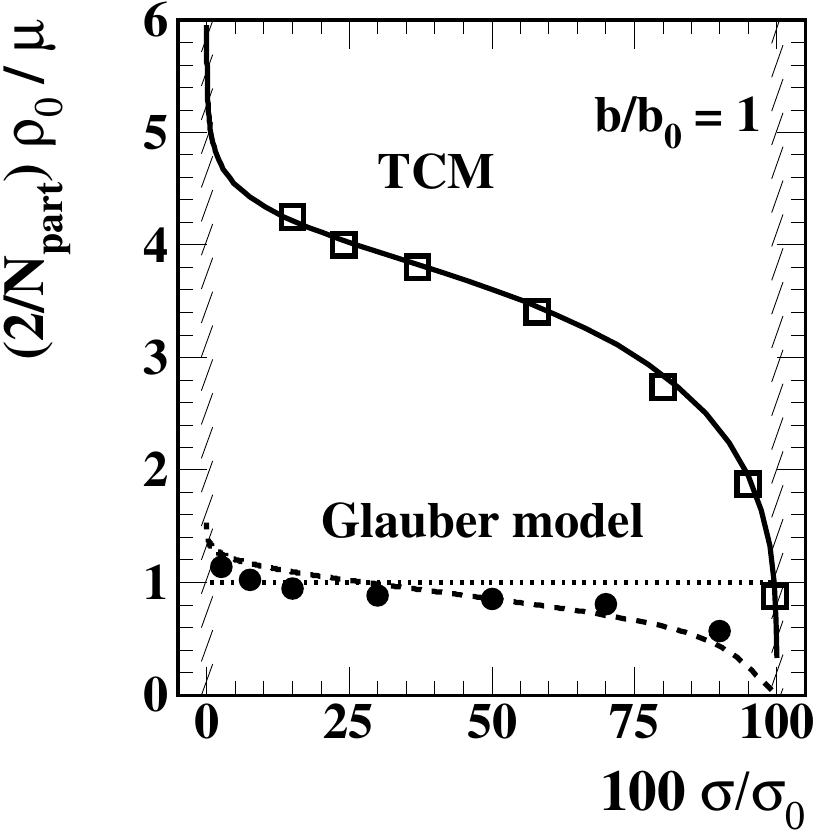}
  \caption{\label{bias}
  Left: Midrapidity charge density per participant pair (points) compared to NBD mean $\mu = \bar \rho_{0NSD} = 5$. $N_{part}$ and $N_{part}'$ values are from Table~\ref{ppbparams1}. For the V0A curve (dashed) $\bar \rho_0 \rightarrow n_x/2$ and $\mu =11$ are assumed so as to match Fig.~8 (left) of Ref.~\cite{aliceppbprod}.
  Right: Similar to the left panel but the added TCM trend is as described in the  text. The open squares are data from Ref.~\cite{aliceppbprod} with TCM Table~\ref{rppbdata} values for centralities and $N_{part}$. The close agreement with the TCM is notable.
   } 
 \end{figure}

Figure~\ref{bias} (left) shows a ratio equivalent to R above for seven multiplicity bins (points) based on values in Table~\ref{ppbparams1}. The open circles correspond to unprimed $N_{part}$ in that table. The data points are assigned $\mu = \bar \rho_{0NSD} = 5$ from 5 TeV \pp\ collisions. The dashed curve represents V0A data with $\mu = 11$ per  the NBD obtained from fits to those data and is derived from Eqs.~(\ref{nxprob}), (\ref{npartnx}) and (\ref{npart2}), assuming as in Ref.~\cite{aliceppbprod} that $P(n_x) \approx (1/\sigma_0)d\sigma/dn_x$. For that curve $\bar \rho_0 \rightarrow n_x/2$ is used to match Fig.~8 (left) of Ref.~\cite{aliceppbprod}, but note  that $\Delta \eta = 2.3$ for the V0A detector. The correspondence between this dashed curve and the V0A data points (solid squares) in Ref.~\cite{aliceppbprod} is good.

Figure~\ref{bias} (right) shows the TCM equivalent (solid curve) derived from Eqs.~(\ref{xmodel}) for $x(n_s)$, (\ref{nparmodel}) for $N_{part}(n_s)$, (\ref{nchppb}) to determine $\bar \rho_0(n_s)$ and the solid curve in Fig.~\ref{glauber2} (left) to determine $\sigma / \sigma_0$ vs $N_{part}$, with $\mu = \bar \rho_{0NSD} = 5$ as in the left panel. The open squares are the same $\bar \rho_0$ values from Ref.~\cite{aliceppbprod} and Table~\ref{ppbparams1} but with the other parameters in $R$ given unprimed TCM values from Table~\ref{rppbdata}. The Glauber results in the left panel are repeated for comparison.  Whereas the data points in the left panel seem to deviate significantly from a hypothesis based on $\bar \rho_{0NSD} \approx 5$, a property of 5 TeV \pp\ collisions reported in Ref.~\cite{alicetomspec}, the same data in the right panel are in good agreement with a TCM trend based on the same value.

The dotted lines represent the relation $n_{ch} \propto N_{part}$ assumed for the Glauber analysis in the form $(2/N_{part})\bar \rho_0 = \bar \rho_{0NSD}$. If that were a true property of the \ppb\ data ensemble then significant deviations might indeed represent statistical bias. But the \mmpt\ analysis of Ref.~\cite{tommpt} reveals that \ppb\ data are better represented by the TCM solid curve in the right panel as the ``true'' data trend. In fact, individual \pn\ collisions within \ppb\ collisions {\em are} strongly biased relative to ensemble-averaged isolated \pp\ collisions depending on the imposed \ppb\ \nch\ condition, just as isolated \pp\ collisions are biased as to the mix of hadron production mechanisms depending on an imposed \pp\ \nch\ condition. That bias is accurately described within the TCM, e.g.\ by the solid curve in Fig.~\ref{bias}.

The contrasting trends for peripheral collisions are notable. As the multiplicity \nch\ condition increases from zero the TCM curve remains within the $b/b_0 \approx 1$ hatched band with $N_{part} \approx 2$ for peripheral \ppb, dominated by isolated \pn\ collisions. Above $\bar \rho_0 \approx 2\, \bar \rho_{0NSD}$ \ppb\ centrality begins to increase ($\sigma / \sigma_0$ decreases below 1) and the \pn\ mean multiplicity increases toward a limiting value $\bar \rho_{0NN} \approx 6\, \bar \rho_{0NSD} \approx 30$ corresponding to upper limit $N_{part}  \approx 8$ as in Fig.~\ref{glauber2} (left). The product $(N_{part}/2) \bar \rho_{0NN}$ then corresponds to the upper limit $\bar \rho_0 \approx 115$ for the \ppb\ event sample reported in Ref.~\cite{alicempt}. In contrast, the V0A dashed curves immediately deviate from the $b/b_0 \approx 1$ hatched bands as \nch\ increases from zero because the assumption $P(n_x) \approx (1/\sigma_0)d\sigma/dn_x$ is incorrect, as discussed in Sec.~\ref{laplace} and demonstrated in Fig.~\ref{invplot} (left). One result is the Glauber MC estimate $N_{bin} \approx 2$ for $\bar \rho_0 \approx \bar \rho_{0NSD}$ and isolated \pn\ collisions.

\subsection{Internal consistency and centrality strategies}

The internal consistency of the Glauber implementation in Ref.~\cite{aliceppbprod} may be questioned. Glauber MC estimates for $N_{part}$ in Table~\ref{ppbparams1} are inconsistent with a running integral of the Glauber differential cross section in Fig.~\ref{glauber1} (left) as shown in the right panel (solid dots vs solid curve). The Glauber MC estimates for $b$ in Table~\ref{ppbparams1} (solid dots) are inconsistent with the $b/b_0$ trend (open boxes and dash-dotted curve) in Fig.~~\ref{ppbmult2} (left) representing defined centralities. The \ppb\ Glauber MC estimate for $\nu$ (derived from Glauber $N_{part}$) in Fig.~\ref{ppbmult} (right) exceeds the \pbpb\ trend over a substantial $\bar \rho_0$ interval.

Reference~\cite{aliceppbprod} describes several strategies for determining \ppb\ centrality depending on the $\eta$ acceptance and nature of detectors used to define centrality classes. The several detectors include the V0A detector as described above and a zero-degree neutron calorimeter (ZNA). Reference~\cite{aliceppbprod} asserts that whereas V0A and other charge-multiplicity detectors may suffer from bias due to \pn\ multiplicity fluctuations (see previous subsection) no bias is expected for estimates based on the ZNA assumed to be causally disconnected from midrapidity \nch. However, different $\eta$ intervals involve different combinations of soft and hard components of hadron production which are not well defined. The TCM hard component, representing MB dijets arising from low-$x$ gluons, is strongly peaked at midrapidity~\cite{wolschin} where the TCM is well-established.

\begin{figure}[h]
  \includegraphics[width=1.64in]{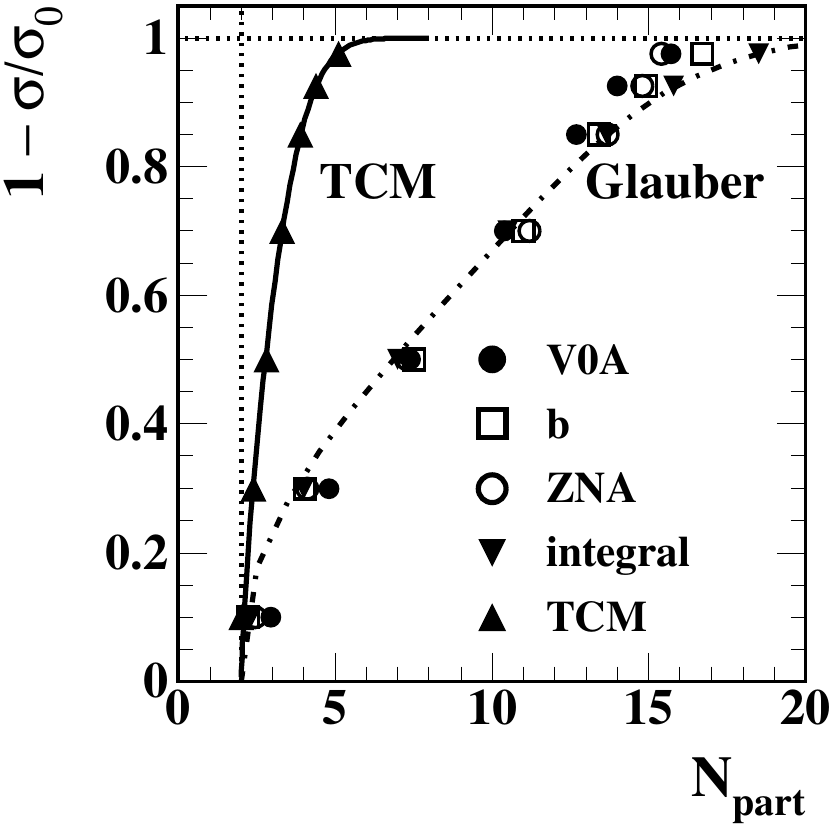}
  \includegraphics[width=1.66in]{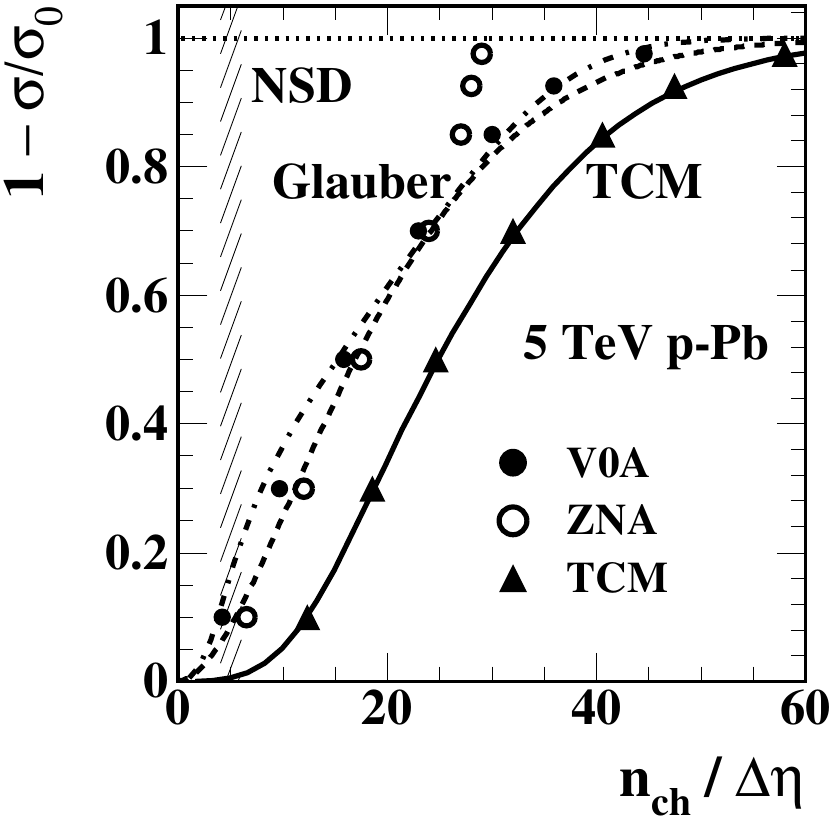}
  \caption{\label{ppbmult9}
  Left: Fractional cross section $\sigma / \sigma_0$ vs $N_{part}$ comparing several centrality strategies. The points and curves are described in the text.
  Right: Similar to the left panel but with midrapidity $\bar \rho_0$ instead of $N_{part}$. The differences between Glauber MC and TCM in the two panels are complementary.
   } 
 \end{figure}

Figure~\ref{ppbmult9} (left) shows fractional cross sections in the form $1 - \sigma / \sigma_0$ vs participant number $N_{part}$. Several centrality strategies are represented, including the V0A detector emphasized in this study, the ZNA neutron detector at zero degrees and the known impact parameter $b$ within the Glauber MC, whose results cluster around the Glauber trend (dash-dotted curve) derived as a running integral of Eq.~(\ref{glauberdata}). The inverted triangles are points on that curve corresponding to standard centrality values. The large disagreement with the TCM curve (solid) inferred from \mmpt\ data has been noted above.

Figure~\ref{ppbmult9} (right) shows fractional cross section vs midrapidity charge density $\bar \rho_0$. The dash-dotted curve is the corresponding Glauber MC curve in the left panel transformed to $\bar \rho_0$ via the Jacobian in Fig.~\ref{ppbmult6} (right). The dashed curve is derived from V0A $P(n_{ch})$ assumed equivalent to a differential cross section. The ZNA $\bar \rho_0$ values are taken from Fig.~16 (lower right) of Ref.~\cite{aliceppbprod}. The TCM curve is a running integral of the TCM differential cross section in Fig.~\ref{ppbmult4} defined below that figure. The V0A data (solid dots) are consistent with the V0A curve or the Glauber MC curve. The ZNA data (open circles) favor the V0A curve except for the most-central points where there are large deviations. In effect, changes in the ZNA centrality condition there produce no corresponding change in the measured quantity. Again there is major disagreement with the TCM trend.
Generally, the systematic differences between alternative centrality detectors are minor compared to the systematic Glauber-TCM difference. 

In the left panel the TCM-Glauber difference arises from the preferred \pa\ collision model---what nucleons within an eikonal corridor are actually ``wounded''---which is an experimental issue. The TCM result is inferred inductively from \mmpt\ data whereas the geometric Glauber MC is an assumed model.
In the right panel, the V0A-TCM difference at small $\bar \rho_0$ arises from confusing the V0A $P(n_{ch})$ with a differential cross section. The result is a Glauber prediction that for $\bar \rho_0 \approx \bar \rho_{0NSD}$ the \ppb\ centrality is 90\%, whereas the TCM prediction is 100\% or $b \approx b_0$, i.e.\ isolated \pn\ collisions. At larger $\bar \rho_0$ the Glauber-TCM difference is complementary to the result in the left panel: The TCM describes larger charge multiplicities for smaller $N_{part}$ as a characteristic of \pn\ collisions within \pa\ collisions, consistent with \mmpt\ data.

\section{Discussion} \label{disc}

This section emphasizes four topics relating to \ppb\ centrality determination. (a) Any centrality model must be related to an observable quantity based on a model of hadron production within high-energy nuclear collisions. (b) A major information source for formulating hadron production models is \pt\ spectra and \mmpt\ data. (c) Any model for \pa\ collisions must rely on the nature of the proton projectile and its interaction with target nucleons within a dense nucleus. And (d) MB dijets should play a major role in the formulation of any collision model.

\subsection{TCM vs Glauber hadron production models}

Hadron production in elementary \pp\ collisions is observed to proceed via  two dominant mechanisms represented by the TCM soft and hard components. The TCM efficiently and accurately describes a broad array of yield, spectrum and correlation data. The hard component representing MB dijets plays a dominant role in evolution of data structures with charge multiplicity \nch\ and/or A-B centrality measured for instance by nucleon participant number $N_{part}$.  More-central \auau\ or \pbpb\ collisions are consistent with participant scaling for the \nch\ soft component, but the jet-related hard contribution varies with $N_{bin}$ in complex ways. The relation between $N_{part}$ and $n_{ch}$ for \pa\ collisions is thus an open question that should be resolved via quantitative analysis, for example as described in Refs.~\cite{alicetommpt,tommpt} and Sec.~\ref{ppb}.

The Glauber study relies on an assumed geometric Glauber MC and a basic assumption about hadron production: hadron multiplicity \nch\ is simply proportional to $N_{part}$. Based on that context larger values of $N_{part}$ are related to smaller \nch. But \ppb\ \mmpt\ data {\em require} smaller $N_{part}$ in combination with larger $\bar \rho_{0NN}$ to accommodate noneikonal dijet production per the TCM. The geometric Glauber MC is a common basis for centrality determination in \aa\ collisions and for several MC collision models such as HIJING~\cite{hijing} and AMPT~\cite{ampt}. These \ppb\ results motivate reconsideration of the validity of such models.

The data and TCM trends in Fig.~\ref{paforms} and \pa\ \mmpt\ data in Fig.~\ref{padata} are consistent with the following scenario: Increase of jet-related hadron production in \pa\ collisions may proceed via two mechanisms depending on control parameter $\bar \rho_s$: (a) increasing depth of splitting cascades on momentum fraction $x$ within {\em single} \pn\ collision partners that increases $\bar \rho_{sNN}(n_s) \approx \bar \rho_s$ for the most-peripheral \pa\ collisions or (b) increasing participant-nucleon number $N_{part}(n_s)$ with increasing \pa\ centrality and $\bar \rho_{sNN}(n_s) < \bar \rho_s$. The relative contributions depend on probabilities. 
Below transition point $\bar \rho_{s0}$ single \pn\ collisions dominate and the noneikonal $\bar \rho_h \approx \bar \rho_{hNN} \approx \alpha \bar \rho_{sNN}^2$  trend for dijet production observed in \pp\ collisions~\cite{ppquad} is determining.   Above $\bar \rho_{s0}$ \pa\ centrality dominates and increasing \pn\ binary-collision number  $N_{bin}$ plays the dominant role in dijet production with $\bar \rho_{h}(n_s) \approx N_{bin}(n_s) \alpha \bar \rho_{sNN}^2(n_s)$. Dijet manifestations provide an essential and effective probe of \pa\ centrality. A primary message from \ppb\ \mmpt\ data is the trend toward larger $\bar \rho_{0NN} = (2/N_{part})\bar \rho_0$ and smaller $N_{part}$ compared to expectations based on $\bar \rho_0 \propto N_{part}$ linear scaling as assumed for the Glauber study in Ref.~\cite{aliceppbprod}. 


\subsection{TCM for $\bf p_t$ spectra vs $\bf \bar p_t$ data}

The TCM for \pt\ spectra such as reported in Refs.~\cite{ppprd,hardspec,ppquad} inspired the TCM for \mmpt\ data as in Ref.~\cite{tommpt}. It is not surprising then to find structural equivalents. Normalized spectra $\bar \rho_0(y_t) / \bar \rho_s$ in Fig.~\ref{qppb} (left) relative to universal soft-component model $\hat S_0(y_t)$ are equivalent to $\bar P_t / \bar \rho_s$ in Eq.~(\ref{pampttcm}) (second line) relative to universal soft component $\bar p_{ts}$ which may be derived from $\hat S_0(y_t)$. 
Relative to the soft components the hard components then vary with common factors $x(n_s) \nu(n_s)$. Dividing by those factors leaves $\bar p_{thNN}$ for \mmpt\ data and $\hat H_0(y_t)$ for \pt\ spectra from which $\bar p_{thNN}$ may be derived. Spectrum ratios such as $Q_\text{pPb}$ are like $\bar p_t = \bar P_t / n_{ch}$ in that distinct soft and hard components and their corresponding hadron production mechanisms may be confused and obscured.

Hadron production in A-B collisions may be summarized as follows: The product $\bar \rho_s = (N_{part}/2) \bar \rho_{sNN}$ applies generally. In \pp\ collisions hadron production is controlled by $\bar \rho_{sNN} \rightarrow \bar \rho_s$ with noneikonal $\bar \rho_h \propto \bar \rho_s^2$ and $N_{part} \equiv 2$ fixed. In \aa\ collisions hadron production is controlled by $N_{part}$ with eikonal $N_{bin} \approx (N_{part}/2)^{4/3}$ and $\bar \rho_{sNN} \approx \bar \rho_{sNSD}$ approximately constant. Production in \pa\ collisions is intermediate and must be determined by experiment, for example by TCM analysis of \ppb\ \mmpt\ data where MB dijets dominate data variations. Figure~\ref{qppb} confirms that the $N_{part}$ and $\bar \rho_{sNN}$ trends in Table~\ref{rppbdata} inferred from the \mmpt\ study in Ref.~\cite{tommpt} {\em accurately predict} \ppb\ \pt\ spectrum data for pions in Ref.~\cite{alicepionspec}. It also validates the assumption in Ref.~\cite{tommpt} that jet production in \ppb\ is unmodified, since spectrum hard components $H(y_t)$ from \ppb\ spectra are consistent with \pp\ $\hat H_0(y_t)$.

\subsection{The proton projectile within p-A collisions}

A major issue arising from comparisons of the Glauber MC and TCM for \ppb\ collisions in Sec.~\ref{oops} is the large difference between the upper limit on participant number $N_{part} \approx 16$-19 for $b \approx 0$ estimated with the geometric Glauber MC and the much lower $N_{part} \approx 8$ inferred from TCM analysis of \mmpt\ data. The \ppb\ Glauber MC is based on the assumption that even for a central \ppb\ collision each of approximately 18 binary \pn\ encounters is equivalent to an {\em isolated} in-vacuum \pp\ collision. But in most-central \auau\ or \pbpb\ collisions the mean number of binary collisions per participant is less than 6, and 18 successive binary \pn\ {\em collisions} has never been otherwise observed.  How a proton projectile interacts with target nucleons after many \pn\ collisions {\em and} within a dense nuclear environment is an open question. \ppb\ data may provide a unique basis for its resolution.

A participant nucleon within the target nucleus as defined has been {\em effectively} struck by a projectile proton. The geometric Glauber MC assumes that any target nucleon residing within an ``eikonal corridor'' relative to the projectile-proton trajectory (defined by a \pn\ cross section $\approx \sigma_{pp}$) becomes a participant. However, the effective number of participants may be reduced in at least two ways: (a) The capacity of the projectile proton to ``wound'' a target nucleon may decrease along its trajectory after multiple prior \pn\ collisions. (b) The  capacity of a projectile proton to ``wound'' at any point along its trajectory within a dense nuclear environment may be substantially less than predicted via the  \pp\ cross section. Data from high-energy \pa\ collisions  such as \mmpt\ data from Ref.~\cite{alicempt}, spectrum data from Ref.~\cite{alicepionspec} and centrality data from Ref.~\cite{aliceppbprod} provide a unique opportunity to explore such novel possibilities. A follow-up study of that problem will be presented in a future article.

\subsection{MB dijets and p-Pb centrality determination}

The distinction between soft and hard components of yields, spectra and correlations from high-energy nuclear collisions is the basis for the TCM. That separation is the result of observed data structures rather than {\em a priori} assumptions. Attribution of the hard component to MB dijet production is also the result of comparisons between measured data trends and measured jet characteristics. MB dijets then provide the main source of information for centrality determination in \pa\ collisions, as in  Sec.~\ref{ppb}. Generally, the importance of MB dijets for understanding high-energy nuclear collisions should be emphasized~\cite{mbdijets}.

Measured MB jet trends indicate that \pp\ collisions are noneikonal based on the observed relation $\bar \rho_h \propto \bar \rho_s^2$. There is no restricted ``eikonal corridor'' depending on \pp\ impact parameter, all participant low-$x$ partons in each proton may interact within any collision. The same relation in \pa\ collisions may then be used to track the factorization $\bar \rho_s = (N_{part}/2) \bar \rho_{sNN}$ via the MB jet yield per $\bar \rho_{hNN} \propto \bar \rho_{sNN}^2$, e.g.\ the \mmpt\ analysis of Ref.~\cite{tommpt}, from which $N_{part}(\bar \rho_0)$ may be inferred as in Fig.~\ref{paforms2} (left).

Although \mmpt\ data may be very precise they retain limited spectrum information. Predictions of full spectra provide a more rigorous test of the TCM and the role of MB jets in hadron production. The validity of TCM-inferred $N_{part}(\bar \rho_0)$ is confirmed by accurate prediction of \ppb\ spectra and spectrum ratios in Figs.~\ref{qppb} and \ref{qppb2}. In contrast, the Glauber analysis apparently does not incorporate any aspect of jet production. The central assumption $\bar \rho_0 \propto N_{part}$ is consistent with a solitary soft component and no jet contribution, which contradicts a broad array of A-B data~\cite{ppprd,hardspec,anomalous,ppquad}. The geometric Glauber MC is based on the eikonal approximation applied at the \pn\ level as well as the composite A-B level, which contradicts an assortment of \pp\ data~\cite{ppprd,ppquad,alicetomspec,jetspec2}

\section{Summary}\label{summ}

This article reports a comparative study of two methods for estimating the centrality of \pa\ (specifically \ppb) collisions. One method is based on simulations of projectile-proton interactions with Pb target nucleons via a geometric Glauber Monte Carlo and an assumption that hadron production is proportional to number of participant nucleons as simulated. The other method is based on a two-component (soft + hard) model (TCM) of hadron production describing any A-B collision system.

The TCM has been applied successfully and accurately to yields, spectra and two-particle correlations from a variety of A-B collision systems over a broad range of collision energies. The basic elements of the TCM, soft (projectile dissociation) and hard (scattered-parton fragmentation to jets) hadron production mechanisms, reflect high-energy physics results over several decades. Most recently, the TCM was used to describe ensemble-mean \mmpt\ data from \pp, \ppb\ and \pbpb\ within their uncertainties. 
From the \mmpt\ study the inferred relation between \ppb\ participant-nucleon number $N_{part}$ and charge density $\bar \rho_0$ near midrapidity was obtained as a simple modification of the data trend for \pp\ collisions. According to \mmpt\ data the {\em noneikonal} nature of \pp\ collisions---each participant parton in one proton can interact with {\em any} participant in the partner proton---continues for \pn\ collisions within \pa\ collisions, and MB dijet production is then related quadratically to the \pn\ charge density.

The Glauber Monte Carlo (MC) study reports rapid increase of $N_{part}$ with \nch\ (as a centrality control parameter) to a mean value within 16-19 for most-central density  $\bar \rho_0 \equiv n_{ch}/ \Delta \eta \approx 45$, with fluctuations of $N_{part}$ to as high as 30. In contrast, the TCM describes slow increase of $N_{part}$, with maximum value near 8 for $\bar \rho_0  \approx 115$. The present study aims to determine the origin of those differences and which method, if either, may be correct.

Detailed comparison of the two methods leads to the following observations: (a) The Glauber MC predicts a differential cross section $d\sigma / dN_{part}$ describing maximum centrality variation for most-peripheral collisions (smallest \nch) where isolated \pn\ collisions should dominate, whereas the TCM predicts minimal centrality variation in the same interval. (b) The Glauber analysis locates most-central \ppb\ collisions below $\bar \rho_0 \approx 50$, whereas measured \ppb\ \mmpt\ data extend out to 115. (c) The Glauber study describes the number of hadrons per participant pair as remaining near the \pp\ mean value 5, whereas the TCM derived from 5 TeV \ppb\ \mmpt\ data describes rapid increase of that quantity for more-peripheral \ppb\ collisions, similar to \pp\ data, up to a maximum value near 30. (d) Trends derived from the Glauber MC can be combined to predict \ppb\ \mmpt\ data, but the predicted values are much smaller than measured \mmpt\ data and the TCM description.  (e) The tail of probability distribution $P(n_{ch})$ for \pn\ collisions within \ppb\ collisions inferred via the Glauber MC drops off much more rapidity than measured distributions for \pp\ collisions. (f) The $p$-Pb/$p$-$p$ spectrum ratio $Q_\text{pPb}(p_t)$ (inferred by scaling with \nn\ binary-collision number $N_{bin}$ from the Glauber MC) is said to remain near 1 for $p_t > 10$ GeV/c for all \ppb\ centralities, whereas the TCM derived from \mmpt\ data predicts rapid increase of that ratio to about 14. The TCM ratio prediction is found to be consistent with measured pion spectra from 5 TeV \ppb\ collisions. 

Those observations, taken together, suggest that the Glauber MC is not a valid description of \ppb\ collisions and/or that the basic assumption $\bar \rho_0 \propto N_{part}$ is not valid, whereas the TCM accurately describes a variety of data. One question that emerges concerning the basis for the Glauber MC: To what extent is a projectile proton able to ``wound'' a target nucleon and thereby produce a participant? The Glauber MC assumes that any {\em encounter} corresponding to a \pn\ cross section of 70 mb is a collision. TCM results suggest that the {\em effective} participant production rate may be only 1/3 of the Glauber estimate.

The geometric Glauber MC is a common basis for centrality determination in \aa\ collisions and for several MC collision models such as HIJING and AMPT. These \ppb\ results motivate reconsideration of the validity of such models. As to recent claims of collectivity in \pa\ systems, \ppb\ pion spectra and their ratios indicate no deviation from {\em linear superposition} of \pn\ collisions over a large \ppb\ \nch\ and centrality interval. Jet production, at least, appears to remain unmodified in \ppb\ collisions.

\begin{appendix}

\section{TCM for $\bf p$-$\bf p$ spectra} \label{specapp}

The TCM for hadron production in high energy nuclear collisions represents consistent observations~\cite{ppprd,hardspec,anomalous,jetspec,ppquad,alicetomspec,alicetommpt,tommpt} that hadron production near midrapidity is dominated by two mechanisms: (a) projectile-nucleon dissociation to charge-neutral hadron pairs (soft) and (b) scattered-parton fragmentation to correlated hadron jets (hard).  A TCM reference can be defined in terms of {linear superposition} of a basic process: low-$x$ parton-parton interactions within \pp\ collisions or nucleon-nucleon (\nn) interactions within \pa\ and \aa\ collisions. Deviations from a TCM reference may then reveal genuine novelty (e.g.\ nonlinearity) in a composite system. The TCM for \pt\ spectra and \mmpt\ in \ppb\ collisions can be used to determine centrality in that  system.



\subsection{Basic TCM description of p-p $\bf p_t$ or $\bf y_t$ spectra} \label{aa1}

The TCM for \pp\ collision data emerged from analysis of 200 GeV \pt\ spectrum data. Systematic analysis of the \nch\ dependence of \pt\ spectra from 200 GeV \pp\ collisions described in Ref.~\cite{ppprd} led to a compact phenomenological TCM with {\em approximate factorization} of multiplicity \nch\ and {transverse-rapidity} \yt\ dependence in the form
\bea \label{a1x}
\frac{d^2n_{ch}}{y_t dy_t d\eta} &=& S_{pp}(y_t,n_{ch}) + H_{pp}(y_t,n_{ch})
\\ \nonumber
&\approx& \bar \rho_s(n_{ch}) \hat  S_0(y_t) + \bar \rho_h(n_{ch}) \hat H_0(y_t)
\eea
with mean angular densities $\bar \rho_x = n_x / \Delta \eta$. Transverse rapidity $y_t \equiv \ln[(p_t + m_t)/m_h]$ with transverse mass defined by $m_t^2 = p_t^2 + m_h^2$ provides improved visual access to spectrum structure at lower \pt\ or \yt\ (for unidentified hadrons pion mass $m_h = m_\pi$ is assumed). Unit-integral soft-component model $\hat S_0(m_t)$ is consistent with a L\'evy distribution on \mt, while peaked hard-component model  $\hat H_0(y_t)$ is well approximated by a Gaussian on \yt\ centered near $y_t \approx 2.65$ ($p_t \approx 1$ GeV/c) with exponential (on \yt) tail reflecting an underlying power-law (on \pt) jet spectrum~\cite{jetspec2}. Conversion from \pt\ or \mt\ to \yt\ is accomplished with Jacobian $p_t m_t/  y_t$.

Integration of Eq.~(\ref{a1x}) over \yt\ results in the angular-density TCM $\bar \rho_0 = \bar \rho_s + \bar \rho_h$. Spectrum~\cite{ppprd,alicetomspec} and angular-correlation~\cite{ppquad} data reveal that soft and hard angular densities are related by $\bar \rho_h = \alpha \bar \rho_s^2$ with $\alpha \approx 0.006$ within $\Delta \eta = 2$ at 200 GeV. The two relations are equivalent to a quadratic equation that uniquely defines $\bar \rho_s$ and $\bar \rho_h$ in terms of $\bar \rho_0$ (when corrected for inefficiencies). That quadratic relation is valid over a $\bar \rho_s$ interval corresponding to {\em 100-fold variation} of MB dijet production.


\subsection{Recent TCM results for $\bf p$-$\bf p$ $\bf p_t$ spectra}

The dijet production trend $\bar \rho_h \propto \bar \rho_s^2$ inferred from \pp\ hadron spectra as described above combined with $\bar \rho_s \propto \ln(\sqrt{s}/\text{10 GeV})$~\cite{alicetomspec} describes jet-spectrum energy trends over large \pp\ collision-energy and jet-energy intervals~\cite{jetspec2}. Predicted jet-spectrum trends can then be combined with measured fragmentation functions (FFs)~\cite{eeprd} to predict hadron-spectrum hard component $H(y_t)$~\cite{fragevo}. 

\begin{figure}[h]
  \includegraphics[width=1.66in]{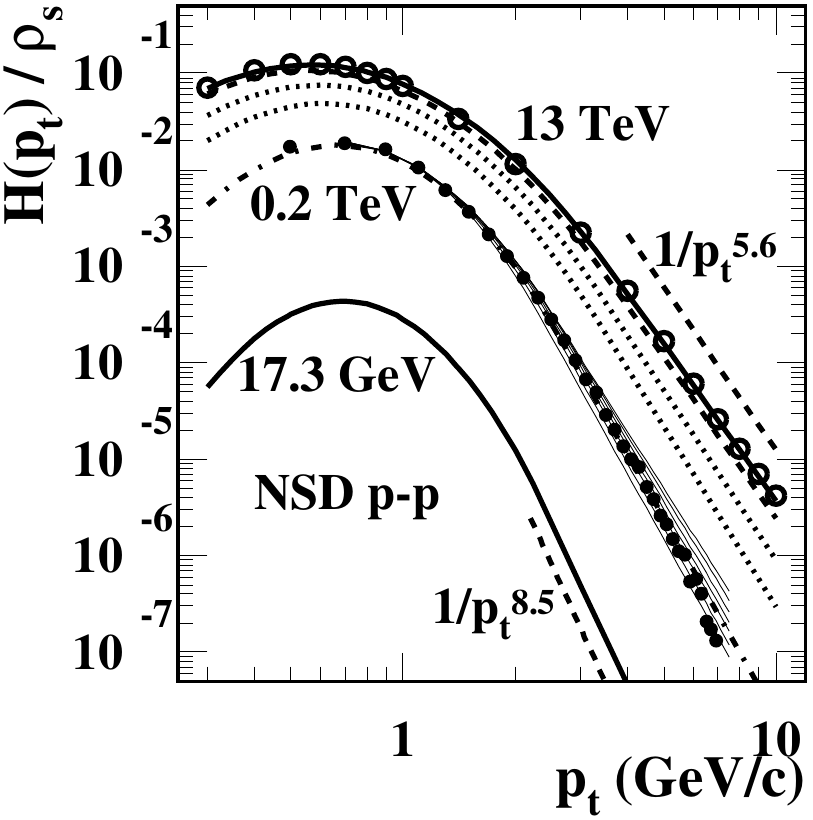}
  \includegraphics[width=1.60in]{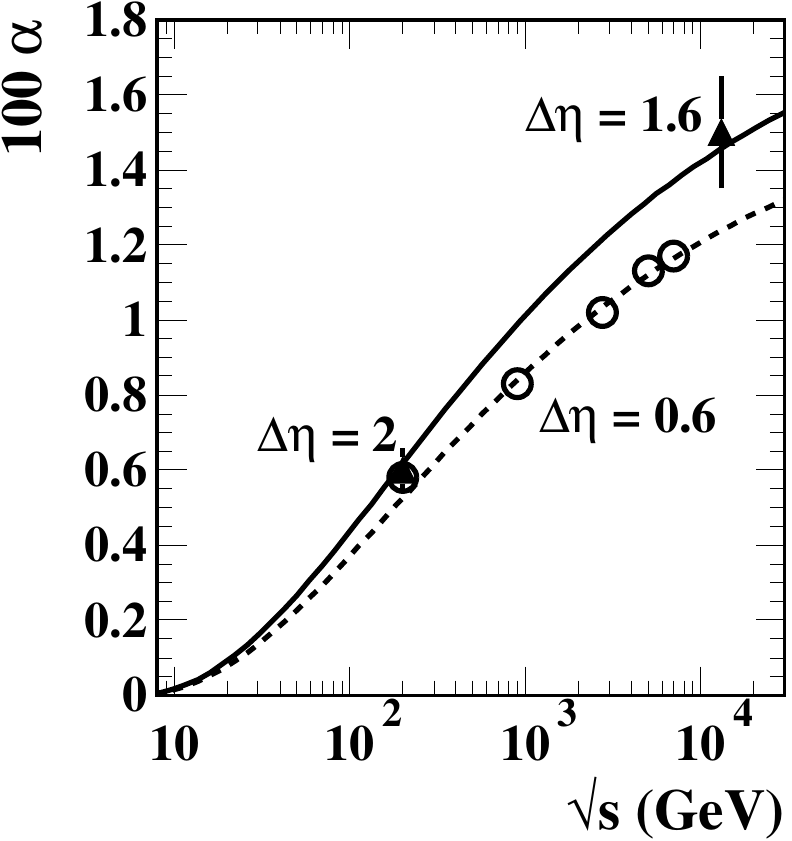}
\caption{\label{enrat3}
Left: A survey of spectrum hard components over the currently accessible energy range from threshold of dijet production (10 GeV) to LHC top energy (13 TeV). The curves are determined by TCM parameters for NSD \pp\ collisions from Ref.~\cite{alicespec}. The 200 GeV fine solid curves illustrate \nch\ dependence. The points are from Refs.~\cite{ppquad} (200 GeV) and \cite{alicespec} (13 TeV).
Right: TCM hard-soft ratio parameter $\alpha$ determined by analysis of spectrum-ratio data (solid points) from Ref.~\cite{alicespec}. The solid curve is defined in Ref.~\cite{alicetomspec}. The dashed curve is the solid curve reduced by factor 0.83 corresponding to the $\Delta \eta$ acceptance reduction. The open circles are derived in Ref.~\cite{tommpt} from \pp\ $\bar p_t$ data in Ref.~\cite{alicempt}.
 } 
\end{figure}

Figure~\ref{enrat3} (left) shows ratio $H(p_t,\sqrt{s}) / \bar \rho_s(\sqrt{s}) \approx \alpha(\sqrt{s}) \bar \rho_s(\sqrt{s}) \hat H_0(p_t,\sqrt{s})$ for NSD \pp\ collisions measuring the spectrum hard component  {\em per soft-component hadron} corresponding (by hypothesis) to dijet production per participant low-$x$ gluon. The 13 TeV TCM solid curve is compared to spectrum data (open points). The two dotted curves are for 0.9 and 2.76 TeV and the dashed curve is for 7 TeV. The 200 GeV summary includes parametric variation of $\hat H_0(y_t,n_{ch})$ for seven multiplicity classes (thin solid curves). Corresponding data (solid points) represent NSD \pp\ collisions. Isolated hard components clarify spectrum energy evolution and its relation to dijet production. The predictions for six collision energies (curves) are compared to data from four energies (13, 7, 0.9 and 0.2 TeV) in Ref.~\cite{jetspec2}. The overall result is a comprehensive description of dijet contributions to \pt\ spectra vs \pp\ collision energy over three orders of magnitude.

Parameter $\alpha$ connecting soft and hard components of \pp\ hadron yields is 
related to jet systematics by
\bea
\alpha \bar \rho_{sNSD}^2 &=& \bar \rho_{hNSD}  = \epsilon(\Delta \eta) f_{NSD}  2\bar n_{ch,j},
\eea
where $2\bar n_{ch,j}$ is the mean hadron fragment multiplicity per dijet averaged over a jet spectrum for given collision energy~\cite{eeprd} and $f_{NSD} = (1/\sigma_{NSD}) d\sigma_{jet} /  d\eta$ is the dijet frequency and $\eta$ density per NSD \pp\ collision~\cite{fragevo}. The energy trends for those quantities, inferred from isolated-jet data, can be used to predict an energy trend for $\alpha$. 



 Figure~\ref{enrat3} (right) shows values for $\alpha(\sqrt{s})$ (solid points) obtained for 200 GeV and 13 TeV from Refs.~\cite{ppquad,alicetomspec} respectively. The solid curve from Ref.~\cite{alicetomspec} is based on measured properties of isolated jets~\cite{jetspec2}. The open points are inferred from $\bar p_t$ trends in Ref.~\cite{tommpt} (described below). The dashed curve is the solid curve reduced by factor $\approx 0.83$ ($\approx 0.5/0.6$) corresponding to the reduced angular acceptance $\Delta \eta = 0.6$ in Ref.~\cite{alicempt} compared to previous results for $\Delta \eta = 2.0$~\cite{ppprd,ppquad}.

\section{$\bf \bar p_t$ TCM for $\bf p$-$\bf p$ collisions}  \label{ppmptapp}



The present study emphasizes comparisons between \mmpt\ analysis of \pp, \ppb\ and \pbpb\ collisions and a Glauber analysis of \ppb\ centrality. This appendix briefly reviews basic  \mmpt\ TCM analysis for elementary \pp\ collisions.

Reference~\cite{alicetommpt} established that a TCM for ratio $ \bar p_t = \bar P_t / n_{ch}$ constitutes a good description of LHC data from \pp\ and \pbpb\ collisions at several energies and  provided hints as to the mechanism of \ppb\ \mmpt\ variation -- evolving from a \pp\ trend for smaller \nch\ to a quantitatively different but similar trend above a transition point.
%
Reference~\cite{alicetomspec} presented a detailed TCM analysis of \pp\ \pt\ spectra for a range of energies from 17 GeV to 13 TeV. Soft component $\hat S_0(m_t,\sqrt{s})$ varies  weakly with energy and not at all with \nch, but hard component $\hat H_0(y_t,n_s,\sqrt{s})$ varies strongly with energy (consistent with jet properties) and significantly with \nch\ (as established with 200 GeV spectra). 
Those new spectrum results have been incorporated in a revised \mmpt\ analysis in Ref.~\cite{tommpt} summarized for \pp\ here and for \ppb\ in Sec.~\ref{ppb}.



Quantities $\bar p_{th}(n_s,\sqrt{s})$, $\alpha(\sqrt{s})$ and an effective detector acceptance ratio $\xi$ are used  to update results from Ref.~\cite{alicetommpt}. The TCM for charge densities averaged over some angular acceptance $\Delta \eta$ (e.g.\ 0.6 for Ref.~\cite{alicempt}) is 
\bea
\bar \rho_0 &=& \bar \rho_{s} + \bar \rho_{h}
\\ \nonumber
&=& \bar \rho_{s}[1+ x(n_s)],
\\ \nonumber
\frac{\bar \rho_0'}{ \bar \rho_s}&=& \frac{n_{ch}'}{n_s} ~=~ \xi+ x(n_s),
\eea
where $x(n_s) \equiv \bar \rho_{h} / \bar \rho_{s} \approx  \alpha \bar \rho_s$ is the ratio of hard-component to soft-component yields~\cite{ppprd} and $\alpha(\sqrt{s})$ is shown in Fig.~\ref{enrat3} (right). Primes denote uncorrected quantities. The TCM for ensemble-mean {\em total} $p_t$ integrated over some angular acceptance $\Delta \eta$ from \pp\ collisions for given $(n_{ch},\sqrt{s})$ is expressed as
\bea \label{mptsimple}
\bar P_t &=& \bar P_{ts} + \bar P_{th}
\\ \nonumber
&=& n_s \bar p_{ts} + n_h \bar p_{th}.
\eea
The conventional intensive ratio of extensive quantities
\bea \label{ppmpttcm}
\frac{\bar P_t' }{n_{ch}'} \equiv \bar p_t' &\approx & \frac{\bar p_{ts} + x(n_s) \bar p_{th}}{\xi + x(n_s)}
\eea
(assuming $\bar P_t' \approx \bar P_t$~\cite{tommpt}) in effect partially cancels MB dijet manifestations represented by ratio $x(n_s)$.  The alternative ratio
\bea \label{niceeq}
\frac{n_{ch}'}{n_s} \bar p_t'   \approx \frac{ \bar P_t}{n_s} &= & \bar p_{ts} + x(n_s) \bar p_{th}(n_s)
\\ \nonumber
&=& \bar p_{ts} + \alpha(\sqrt{s})\, \bar \rho_s \, \bar p_{th}(n_s,\sqrt{s})
\eea
preserves the simplicity of Eq.~(\ref{mptsimple}) and provides a convenient basis for testing the TCM hypothesis precisely.

\begin{figure}[h]
   \includegraphics[width=1.65in,height=1.6in]{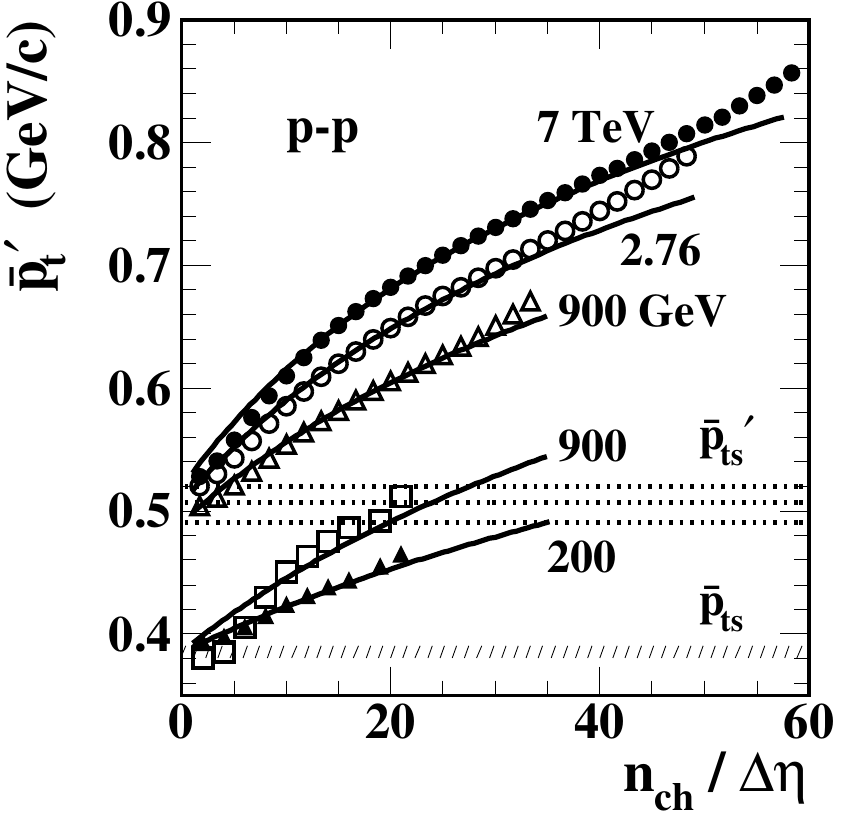}
   \includegraphics[width=1.65in,height=1.6in]{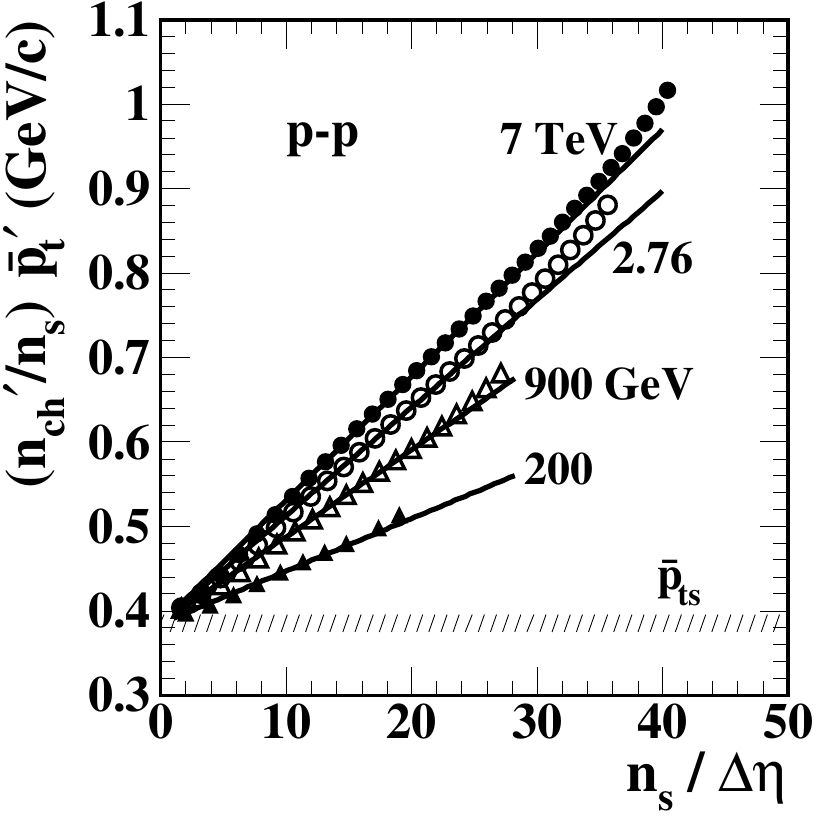}
 \caption{\label{alice5a}
 Left: \mmpt\ vs \nch\ for several collision energies. The upper group of points from Ref.~\cite{alicempt} are derived from particle data with a lower \pt\ cutoff. The lower 900 GeV data from Ref.~\cite{ua1mpt} and 200 GeV data from Ref.~\cite{ppprd} are extrapolated to zero \pt.
Right: Data from the left panel multiplied by factor ${n'_{ch}} / n_s$ that removes the jet contribution and the effect of the low-\pt\ cut on the soft component from the denominator of \mmpt. 
} 
  \end{figure}

Figure~\ref{alice5a} (left) shows $\bar p_t$ data for four \pp\ collision energies from 
the RHIC  (solid triangles~\cite{ppprd}),
the Sp\=pS  (open boxes~\cite{ua1mpt})%
and the LHC (upper points~\cite{alicempt}) increasing monotonically with charge density $\bar \rho_0 = n_{ch} / \Delta \eta$. The lower points and curves correspond to full \pt\ acceptance.  For acceptance extending down to zero ($\xi = 1$), $\bar p_t' \rightarrow \bar p_t$ in Eq.~(\ref{ppmpttcm}) should vary between the universal lower limit $\bar p_{ts} \approx 0.4$ GeV/c ($n_{ch} = 0$) and  $\bar p_{th}$ ($n_{ch} \rightarrow \infty$) as limiting cases. For a lower \pt\ cut $p_{t,cut} > 0$ the lower limit is $\bar p_{ts}' = \bar p_{ts} / \xi$ (dotted lines) and the data are systematically shifted upward (upper points and curves).  The solid curves represent the \pp\ $\bar p_t$ TCM from Ref.~\cite{tommpt}. Note  that  the 7 TeV \mmpt\ data extend to $\bar \rho_0 \approx 10\, \bar \rho_{0NSD} \approx 60$ and were derived from 150 million \pp\ collision events.

Figure~\ref{alice5a} (right) shows data on the left transformed via Eq.~(\ref{niceeq}) to $(n_{ch}' / n_s) \bar p_t' \approx  \bar P_t / n_s$ (points). The TCM curves undergo the same transformation and the slopes of the resulting straight lines are $\alpha(\sqrt{s}) \bar p_{th0}(\sqrt{s})$. The data deviate significantly from the straight-line TCM because of systematic variation with \nch\ of the \pt\ spectrum hard-component shape as reported in Refs.~\cite{alicetomspec,tommpt}. However, those details are beyond the scope of the present study.


The success of the \pp\ \mmpt\ TCM confirms that variation of \pp\ \mmpt\ is dominated by jet fragments from large-angle-scattered low-$x$ gluons. 
The hard yield or angular density $\bar \rho_{h} \approx \alpha(\sqrt{s})\, \bar \rho_s^2$ represents the dijet fragment density determined precisely by soft component $\bar \rho_s$. 
The \pt\ spectrum TCM hard component and underlying jet energy spectrum evolve according to the same rules~\cite{jetspec2}. 
%
The quadratic relation $\bar \rho_h \propto \bar \rho_s^2$ implies that \pp\ collisions are {\em noneikonal} (compared to the eikonal trend $\bar \rho_h\propto \bar \rho_s^{4/3}$). 
The quadratic trend (each participant gluon in one proton can interact with {\em any} participant gluon in the partner proton) implies that \pp\ collisions with large \nch\ are very jetty.
Reference~\cite{tommpt} demonstrates a direct connection between \mmpt\ hard component $\bar p_{th}(n_s)$, \pt\ spectrum hard component $H(p_t,n_s)$~\cite{alicetomspec} and jet spectra as in Ref.~\cite{jetspec2}. 
Thus, a variety of \pp\ data provide strong evidence that MB dijets dominate \pp\ collisions and $\bar p_t(n_{ch},\sqrt{s})$ trends.

\end{appendix}


\end{document}